\documentclass[aps,prb,twocolumn,superscriptaddress,citeautoscript,floatfix,showpacs,longbibliography]{revtex4-1}

\usepackage{graphicx}
\usepackage{dcolumn}
\usepackage{bm}
\usepackage{natbib}
\usepackage{amsmath}
\usepackage{subfigure}
\usepackage{dsfont}
\usepackage{eufrak}
\usepackage{tikz}
\usepackage{appendix}
\usepackage{makecell}
\usepackage[margin=0.8in]{geometry}
\usepackage[normalem]{ulem}
\usepackage{comment}

\newcommand{\n}{\notag \\}

\newcommand{\half}{\frac{1}{2}}

\newcommand{\dd}{\text{d}}

\newcommand{\tr}{\text{tr}}

\begin{document}

\preprint{APS/123-QED}

\title{Simulating Large One-Dimensional Neutral Atom Quantum Systems}

\def\urbana{
Institute for Condensed Matter Theory and IQUIST and NCSA Center for Artificial Intelligence Innovation and Department of Physics, University of Illinois at Urbana-Champaign, IL 61801, USA
}

\author{James Allen}
\email{jamesza2@illinois.edu}
	\affiliation{\urbana}
\author{Matthew Otten}
\affiliation{HRL Laboratories LLC, 3011 Malibu Canyon Road, Malibu, CA 90265, USA}
\author{Stephen K. Gray}
\affiliation{Center for Nanoscale Materials, Argonne National Laboratory, Lemont, IL 60439, USA}

\author{Bryan K. Clark}
 \email{bkclark@illinois.edu}
\affiliation{\urbana}

\date{\today}

\begin{abstract}
While abstract models of quantum computation assume a closed system of two-level states, practical quantum devices inevitably couple to the environment in some way, creating sources of noise. 
Understanding the tolerance to noise of specific quantum algorithms run on specific devices is important for determining the feasibility of quantum computing in the current noisy intermediate scale quantum era. Of particular interest is understanding the noise sensitivity of these devices as more qubits are added to the system. Classical simulations are a useful tool to understand the effects of this noise, but direct classical simulations of open quantum systems are burdened by an exponentially growing cost in the number of qubits and a large local Hilbert space dimension. For one dimensional, shallow circuits, using tensor networks can replace this exponential cost with a linear one and simulate far wider systems than what would normally be available. In this paper, we describe a tensor network simulation of a neutral atom quantum system under the presence of noise, while introducing a new purity-preserving truncation technique that compromises between the simplicity of the matrix product state and the positivity of the matrix product density operator. We apply this simulation to a near-optimized iteration of the quantum approximate optimization algorithm on a transverse field Ising model in order to investigate the influence of large system sizes on the performance of the algorithm. We find that while circuits with a large number of qubits fail more often under noise that depletes the qubit population, their outputs on a successful measurement are just as robust under Rydberg atom dissipation or qubit dephasing as smaller systems. However, such circuits might not perform as well under coherent multi-qubit errors such as Rydberg atom crosstalk. We also find that the optimized parameters are especially robust to noise, suggesting that a noisier quantum system can be used to find the optimal parameters before switching to a cleaner system for measurements of observables.  

\end{abstract}

\maketitle

\renewcommand{\thesection}{\Roman{section}}

\section{Introduction}
An ideal quantum computer is decoupled from the environment so as to minimize the effects of noise coming from sources such as dephasing and dissipation. Unfortunately, in practice such separation is difficult because quantum circuit operations necessarily couple the system with an outside source. Realistic quantum devices are best thought of as open quantum systems that interact with various sources of environmental noise. This limits the extent to which quantum circuits can be operated, both in terms of circuit size and depth, before breaking down. 

Given the current generation of noisy quantum computers, a wide array of quantum algorithms have been developed, such as the variational quantum eigensolver (VQE) and the quantum approximate optimization algorithm (QAOA) \cite{Graham2022,Farhi2014,Endo2021,Bharti2022}, which need low circuit depth and hopefully limited coherence.  It is still unclear, though, how even in this shallow depth circuit regime, the effect of realistic noise influences the output, such as variational energies and optimized parameters, of these algorithms, especially as we scale circuits to larger system sizes. Therefore, to facilitate our understanding and characterization of quantum devices, we would like to simulate them as best we can using  classical algorithms.

Simulations can play an important role in determining the effect of noise on quantum devices. Unfortunately, simulating quantum devices classically is exponentially difficult in general, a feature which is essential to the algorithmic strength of quantum computing.  These simulations are even more difficult in the case of open quantum systems where a single wavefunction cannot represent the full state. A traditional approach which directly represents the entire density matrix of the system becomes inefficient very quickly for state-of-the-art sizes such as Sycamore's 53 qubit system \cite{Arute2019}, and stochastic approaches are limited by a poorly scaling signal-to-noise cost\cite{Berquist2022}. For a quantum system based on neutral atom arrays, the qubit count can reach even higher, with systems up to 100 qubits being implemented\cite{Canonici2023}. 

Due to their ability to simulate circuits with a very large number of qubits (albeit at low depth), tensor network states (TNS) are particularly well equipped to study the scaling of noise-based errors with system size \cite{Noh2020}. 
Tensor networks are most frequently used to represent a wavefunction.
However, with an open quantum system we need to create a TNS that represents the density matrix. Besides the increased computational burden from the extra tensor indices, this introduces another problem: enforcing the positivity of the density matrix with a TNS is nontrivial.

In this work, our focus is two-fold.  First, we develop and validate a new tensor network approach  to approximately enforce positivity of the density matrix when simulating open quantum systems. The most naive representation of the density matrix -- a vectorized Matrix Product Operator (MPO) -- can be modified to enforce its positivity, following the Matrix Product Density Operator (MPDO) scheme\cite{Verstraete2004} 
(also known as a locally purified density operator \cite{Werner2016})
. While this type of tensor network is well-behaved for simple channels, it struggles to implement circuit operations that combine the channels of multiple time steps together. 
This limits the ability of the truncation algorithm to find the most accurate approximate forms -- in fact, we found that it did not perform as well as the naive MPO in our simulations.  Instead, we have devised an efficient compromise, the Purity-Preserving Truncation (PPT) algorithm, which keeps the purity of the density matrix constant after each truncation, limiting the lowest negative eigenvalue of the system.

Second, we have used the PPT, combined with an efficient massively parallel code (see Appendix~\ref{app:distributed_memory}) to determine the effect of noise on a neutral atom simulation of the QAOA. We find that even in the face of non-trivial dissipation and dephasing noise, both the energy and optimized parameters found in a QAOA simulation of a Transverse Field Ising Model (TFIM) are quite accurate even as the system size grows to 80 qubits. 
However, this is conditioned on a successful measurement, i.e. one without a qubit in a dark state. While the measured observable values when the system returns a result seem largely insensitive to noise, the probability of an unsuccessful measurement increases with both noise and system size resulting in the need for many more shots to achieve a similar result. We also find that certain coherent effects, such as crosstalk between qubits,  can influence the circuit in a way that creates compounding errors over system size, making it difficult to operate large, accurate circuits under these errors.

The rest of the paper is as follows.
In Section II, we introduce the dynamics of the neutral atom array, the specific quantum system that we will be simulating \cite{Saffman2020}. 
In Section III we outline our MPO simulation approach, explaining the PPT algorithm and showing that it performs better than the bare MPO and MPDO in a heavily truncated Random Circuit Sampling (RCS) algorithm. In Section IV, we apply our new machinery on a near-optimized iteration of a QAOA circuit, where the circuit parameters have been optimized to create a ground state wavefunction of the TFIM.  
In Section IVa, we demonstrate that under most sources of error, the VQE iteration's final evaluated energy depends only on error strength and not on system size, although coherent errors caused by Rydberg atom crosstalk might create a system size dependence. In Section IVb, we also consider the possibility of the algorithm selecting a spurious value due to errors, and find that this is also mostly independent of system size in the same way as the energy measurement. Moreover, we find that the parameter optimization tends to be more robust to the noise than the energy measurement. Our work opens a new approach for the simulation of large, open quantum systems and validates the efficacy of QAOA algorithms on neutral atom devices at the noisy intermediate scale.

\section{Lindblad Master Equation for Neutral Atom Arrays}
In this work, we focus on modeling a neutral atom array.  The neutral atom array is a system for implementing quantum circuits which is well suited for a large number of qubits, with some current systems composed of hundreds of qubits \cite{Henriet2020QuantumAtoms}.   
We focus in this work on a one-dimensional geometry, where each atom contains a two-level computational subspace $|0\rangle, |1\rangle$ and a high energy Rydberg state $|r\rangle$ as well as an additional set of dark states $|d\rangle$.

\begin{figure*}
\centering
    \begin{tikzpicture}
        \begin{scope}
            \node[anchor=north west,inner sep=0] (image_a) at (0,-0.9) {\includegraphics[width=0.99\columnwidth]{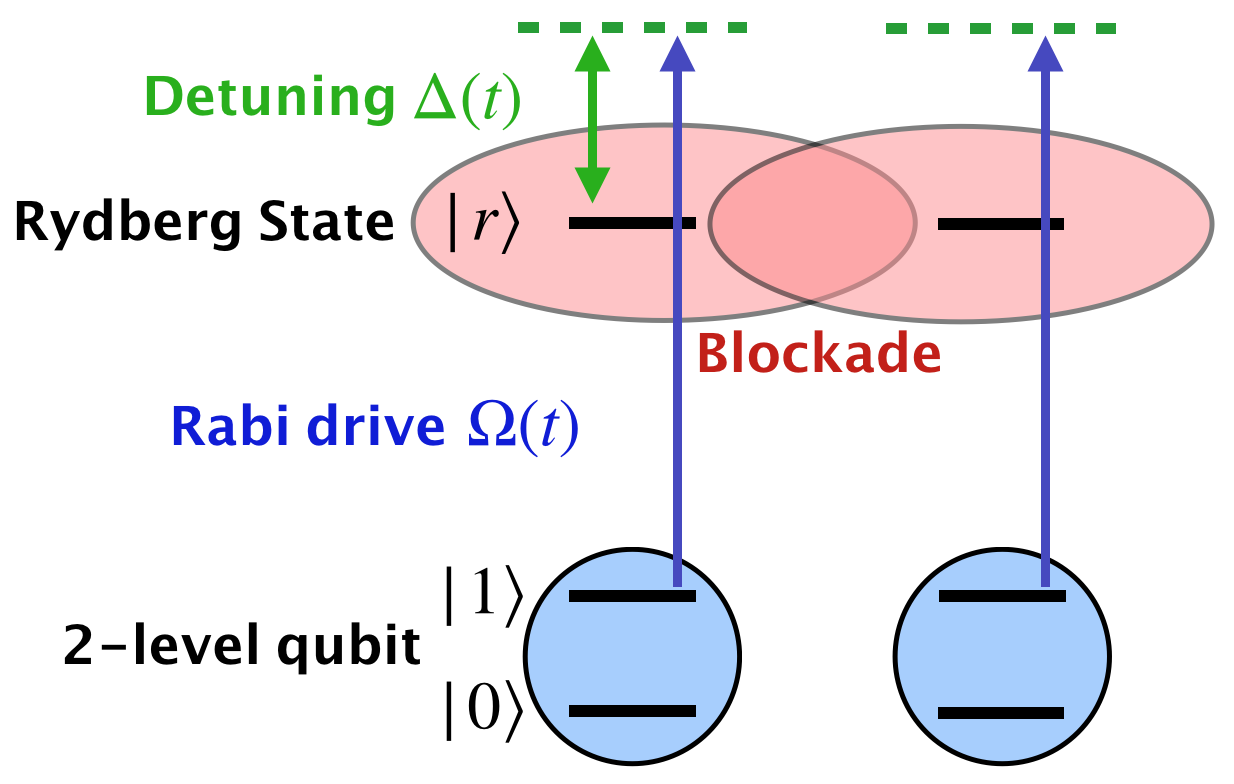}};
            \node [anchor=north west] (note) at (-0.1,-0.85) {\textbf{a)}};
        \end{scope}
        \begin{scope}[xshift=1.0\columnwidth]
            \node[anchor=north west,inner sep=0] (image_a) at (0,-0.5)
            {\includegraphics[width=0.9\columnwidth]{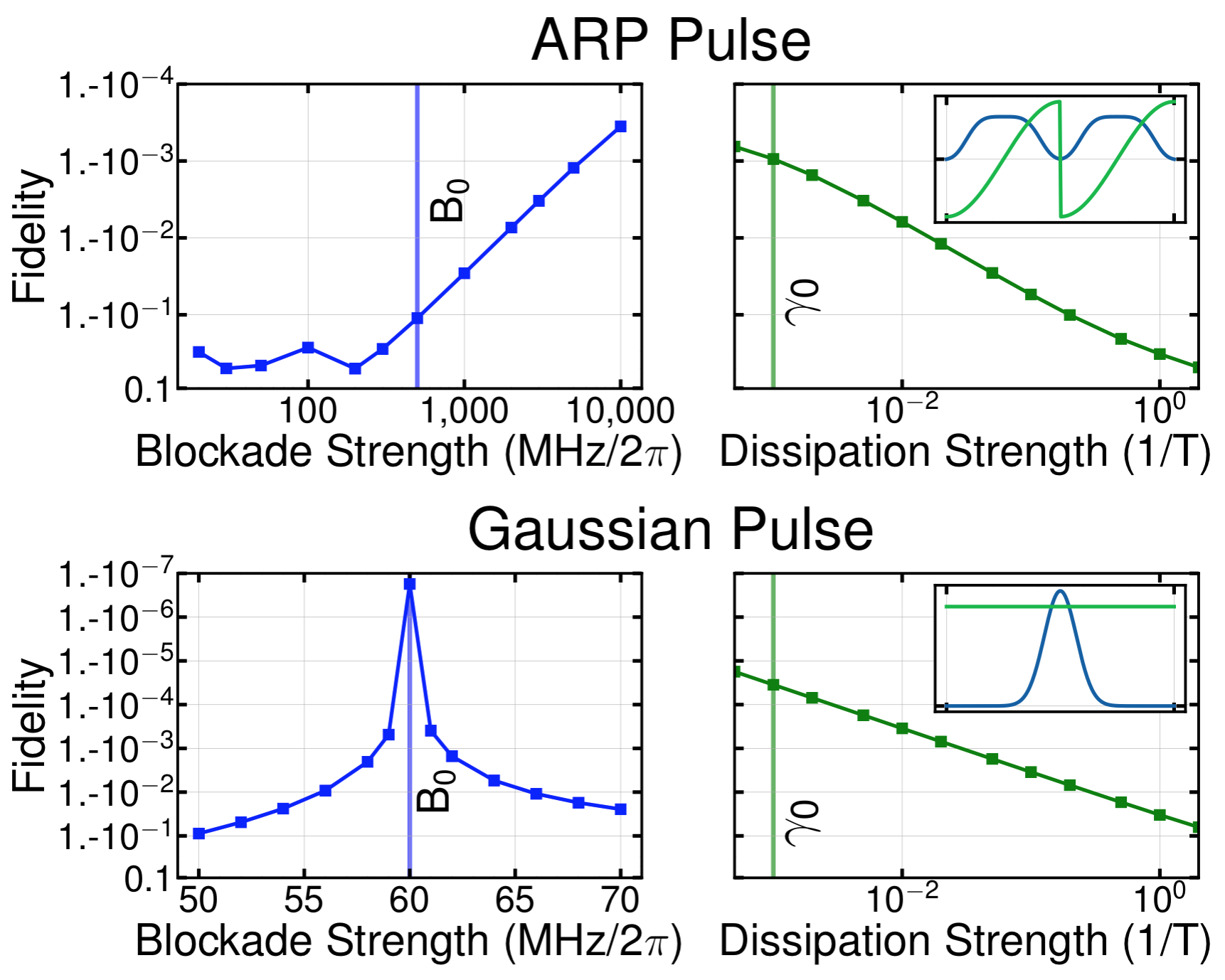}};
            \node [anchor=north west] (note) at (-0.2,-0.85) {\footnotesize{\textbf{b)}}};
            \node [anchor=north west] (note) at (4.05,-0.85) {\footnotesize{\textbf{c)}}};
            \node [anchor=north west] (note) at (-0.2,-3.9) {\footnotesize{\textbf{d)}}};
            \node [anchor=north west] (note) at (4.05,-3.9) {\footnotesize{\textbf{e)}}};
        \end{scope}
    \end{tikzpicture}
    \caption{(a) Rydberg blockade mechanism. A pulse $\Omega(t)$ promotes a $|1\rangle$ qubit state to a Rydberg state $|r\rangle$ with detuning $\Delta(t)$; the two neighboring Rydberg states experience an extra blockade interaction $B$.
    (b,c) ARP gate fidelity as a function of (b) blockade strength with dissipation $\gamma_{diss}=0$, and (c) dissipation with blockade $B=2\pi\times10000$MHz. $\Omega(t)$ and $\Delta(t)$ follow the blue and green curves of the inset in (c) respectively: $\Omega(t) = \Omega_{max}\big[e^{-(t-t_0)^4/\upsilon^4}-a]/(1-a)$ with pulse width parameter $\upsilon=0.175T$, $t_0=T/4$ and $a=e^{-(t_0/\upsilon)^4}$, while $\Delta(t)$ follows a split cosine. We use, following Saffman et al\cite{Saffman2020}, $T=0.54\mu$s, $\Omega_{max} = 17$MHz and $\Delta_{max}=23$MHz. (d,e) Gaussian gate fidelity as a function of (d) blockade strength with no dissipation, and (e) dissipation with blockade $B=2\pi\times 60$MHz. $\Omega(t)$ and $\Delta(t)$ follow the blue and green curves of the inset in (e) respectively: $\Omega(t) = \Omega_{max}\big[e^{-(t-t_0)^2/\upsilon^2}-a]/(1-a)$ with pulse width parameter $\upsilon=0.1T$, $t_0=T/2$ and $a=e^{-(t_0/\upsilon)^2}$, while $\Delta(t)$ is a constant. We use, following Robicheaux et al\cite{Robicheaux2021}, $T=2.165\mu$s, $\Omega_{max}=17$MHz and $\Delta=-14.7$MHz. In both pulses, the branching ratios for dissipation are taken as $b_0 = b_1 = 1/16, b_d = 7/8$.} 
    \label{fig:rydberg_blockade}
\end{figure*}

In a neutral atom array, entanglement between nearest-neighbor sites can be created via the Rydberg blockade (Fig.~\ref{fig:rydberg_blockade}a) \cite{Graham2022}, where two neighboring qubits are temporarily promoted to Rydberg states that repulsively interact with each other. In one such scheme, the two active qubits experience simultaneous pulses under a Hamiltonian
\begin{gather}
    H_p(t) = H_1(t) \otimes I_2 + I_1 \otimes H_2(t) + B|rr\rangle\langle rr| \label{eq:LME_Hamiltonian}
\end{gather}
where $B$ is the Rydberg blockade strength, and $H_1, H_2$ are the single-qubit components of the Hamiltonian. These components include a Rabi frequency $\Omega_i(t)$ which promotes a $|1\rangle$ qubit to the Rydberg state $|r\rangle$ and a Rydberg detuning $\Delta_i(t)$,
\begin{gather}
    H_i(t) = \frac{\Omega_i(t)}{2}\big[|r\rangle \langle 1 | + \text{h.c.}\big] + \Delta_i(t)|r\rangle \langle r| \label{eq:LME_Single_Site}.
\end{gather}
For all the systems considered in this paper, each pulse is identical, so $\Omega_1(t) = \Omega_2(t)$ and $\Delta_1(t) = \Delta_2(t)$.

In any quantum system, coupling between the system and environment introduces noise degrees of freedom that must be accounted for.
Provided the coupling is weak enough and the environment is memoryless, we can model the evolution of the reduced density matrix of the system $\rho$ with the Lindblad Master Equation (LME), 
\begin{align}
    \frac{\dd \rho(t)}{\dd t} &\equiv -i\mathcal{L}[\rho] \n
    &= -i[H_p(t),\rho(t)] + \sum_i L_i \rho(t) L^\dag_i - \half \{ L^\dag_i L_i, \rho(t)\}. \label{eq:LME}
\end{align}
Here $L_i$ are jump operators representing different noise sources in the neutral atom array. One such source is Rydberg atom dissipation. The Rydberg states have a finite lifetime and can decay to either the qubit states or arbitrary dark states $|d\rangle$ that represent any reachable atomic levels. These dark states cease to interact with the rest of the system and we assume that measuring an atom in a dark state counts as a failure of the entire circuit.
The jump operator for this mechanism is
\begin{gather}
    L_j = \sqrt{\gamma_{diss} b_j} |j\rangle \langle r|
\end{gather}
for branching ratios $b_0, b_1, b_d$, and overall dissipation strength $\gamma_{diss}$. If the Zeeman shift between qubit energy levels fluctuates, there is another jump operator to represent qubit dephasing, \cite{Zhang2012,Robicheaux2021}
\begin{gather}
    L_{deph} = \frac{\gamma_{deph}}{\sqrt{2}}\big(|0\rangle \langle 0| - |1\rangle \langle 1|\big).
\end{gather}

The key operation in a quantum circuit is a universal two qubit gate -- we will focus on simulating a CZ gate in this paper. To measure the quality of these gates, we use a metric based on a combined arithmetic-geometric mean fidelity, as introduced in \cite{Reich2013}. 
We investigate two types of pulses. 
The first type of pulse we consider is an Adiabatic Rapid Passage (ARP) pulse (Fig.~\ref{fig:rydberg_blockade}c inset) \cite{Graham2022}. This pulse attempts to use the Rydberg blockade to prevent simultaneous excitations of both sites into the Rydberg states. Due to a $\pi$ phase shift acquired by the site after it enters and exits the Rydberg states, this pulse will create a CZ gate in the limit of infinite blockade strength and no noise. 

With a realistic\cite{Saffman2020} blockade strength $B_0 = 2\pi\times 500$MHz, time period $T=0.54\mu s$ and dissipation $\gamma_{diss} = 0.001 T^{-1}$, we calculated a Bell fidelity of 0.989, 
with most of the inaccuracy coming from phase errors in the unitary operator caused by finite blockade strength (Fig.~\ref{fig:rydberg_blockade}b).

The second type of pulse uses a Gaussian profile for $\Omega(t)$, and a $\Delta(t)$ which is constant as a function of time.
(Fig.~\ref{fig:rydberg_blockade}e inset) \cite{Robicheaux2021}. 
Unlike in the ARP pulse, the application of the Gaussian pulse to a single qubit does not complete a full $|1\rangle \rightarrow |r\rangle \rightarrow |1\rangle$ oscillation.  Therefore at infinite blockade strength, a $\pi$ phase shift doesn't happen; instead one must select a specific finite blockade strength which in combination with the single qubit rotation generates a CZ. 
The Gaussian unitary varies more rapidly over blockade strength compared to the ARP pulse, so fluctuations in the blockade strength create more significant errors. However, we no longer need large blockade strengths to create an accurate unitary operator. For blockade strengths as low as $2\pi\times 60$MHz (Fig.~\ref{fig:rydberg_blockade}d) we can create a unitary with a combined infidelity of $6\!\times\!10^{-8}$, provided there are no  additional sources of noise. 

\section{Matrix Product Operator Representation}

In this section we will describe how to efficiently represent a large number of neutral atom qubits using tensor networks. We will also introduce a new way of maintaining the physicality of an approximate representation of the system under the time evolution of a quantum circuit.

The wavefunction of a one-dimensional chain of $N$ qubits can be represented by a matrix product state (MPS)\cite{Cirac2021}, a string of rank 3 tensors (except for the edges which are rank 2), with each tensor representing an individual site entangled with its neighbors through virtual/bond dimensions (Fig.~\ref{fig:ansatzes}a),
\begin{gather}
    |\psi\rangle_{\sigma_1 ... \sigma_N} = A^{1}_{\sigma_1 i_1} A^{2}_{\sigma_2 i_1 i_2}... A^{N}_{\sigma_N i_{N-1}}
\end{gather}
where $\sigma_j$ is the index of site $j$ corresponding to the local Hilbert space, and $i_j$ is the virtual, or bond, index between sites $j$ and $j+1$.
Likewise, a density matrix can be represented as a matrix product operator (MPO)\cite{Saberi2011}, where each tensor contains two copies of the local Hilbert space (Fig.~\ref{fig:ansatzes}b),
\begin{gather}
    \rho_{\sigma_1 ... \sigma_N, \sigma_1'...\sigma_N'} = B^1_{\sigma_1 \sigma_1' i_1} B^2_{\sigma_2 \sigma_2' i_1 i_2}... B^N_{\sigma_N \sigma_N' i_{N-1}}.
\end{gather}
The MPS representation of the wavefunction requires $O(Nd D^2)$ values, where $d$ is the local Hilbert space dimension and $D$ is the bond dimension. Unlike the statevector representation, which requires $O(d^{N})$ values, the MPS representation only grows linearly in $N$, so it becomes the more tractable representation for any quantum state where the required bond dimension is not expected to be too high. This difference is even more important for the density matrix, where each site requires two local Hilbert space indices, giving a $O(d^{2N})$ cost in statevectors and $O(Nd^2 D^2)$ in MPOs. This harsher scaling makes open quantum system density matrix simulations difficult for systems beyond 12 sites\cite{Otten2019}: for example, a 15 site system with the same three levels as the neutral atom array requires over 3PB of RAM using a naive implementation, while an MPO on the same system with a bond dimension of 4096 (larger than we need to use) only requires 20 GB.

\subsection{MPO form of the Noisy CZ Gate}
We can convert the density matrix into an MPS by vectorizing \cite{Noh2020} the forward and backward local Hilbert space indices at each site, $\sigma_i \sigma_i' \rightarrow \eta_i$. After that, we want to evaluate the time evolution channel of a pulse on the neutral atom array,
\begin{gather}
    \rho(t+T) = \mathcal{C}_T(\Omega, \Delta, B)[\rho(t)]
\end{gather}
with duration $T$, Rabi frequency $\Omega(t)$, detuning $\Delta(t)$ and Rydberg blockade $B$ in terms of a vectorized MPO. We split the time evolution into small steps of duration $\tau \ll T$ and attempt to find an approximate form for the small time step channel
\begin{gather}
    \mathcal{C}_\tau(t) = e^{-i\int_{t}^{t+\tau}\!\!\mathcal{L}(t)\dd t}.
\end{gather}

We use a Trotterization process\cite{Hatano2005} to decompose the Linbladian (\ref{eq:LME}) into individual components and apply the time-evolved form of the components separately, with $O\left(\tau^2\right)$ errors. If $\rho$ is in a vectorized form, we can interpret channel actions $A\rho B$ as an operator $(A\otimes B)(\rho)$ acting on the forward and backward local Hilbert space indices of $\rho$. In this notation,
\small\begin{align}
    \mathcal{C}_\tau(t)[\rho(t)] =& e^{-i \tau H_p(t) \otimes I} e^{i \tau I \otimes H_p(t)} \n&\bigg(\prod_j e^{L_i \otimes L_i^\dagger - \half(L_i^\dagger L_i \otimes I + I \otimes L_i^\dagger L_i)}\bigg) (\rho(t)).
\end{align}\normalsize
Note that we wish to evaluate the exponential of each component analytically, instead of taking a Taylor series approximation like $e^{\tau \hat{O}} \approx I + \tau \hat{O}$. This is to make sure the approximate channel remains CPTP. The pulse Hamiltonian is further broken down into its respective components
\small\begin{align}
    e^{-i\tau H_p(t)} = &\left(e^{-i\tau \big[\half\Omega_1(t) (|r\rangle\langle 1| + \text{h.c.}) + \half\Delta_1(t) |r\rangle \langle r|\big]} \otimes I_2 \right)\n &\left(\text{site 1} \leftrightarrow \text{site 2}\right)\left(e^{-i\tau B|rr\rangle \langle rr|}\right)
\end{align}\normalsize
where $\left(\text{site 1} \leftrightarrow \text{site 2}\right)$ refers to the first factor on the RHS with sites 1 and 2 exchanged.
Each single site operator $\hat{O}_{\sigma_i \sigma_i'}$ can be represented as a rank 2 tensor acting on that site's local Hilbert space index $\sigma_i$. The two-site operator $e^{-i\tau B|rr\rangle \langle rr|}$ becomes a rank 4 tensor acting on both $\sigma_1$ and $\sigma_2$. This is the only component of the time evolution channel that entangles the sites.

We assume that the Rydberg blockade can be represented by a nearest-neighbor time evolution operator. There are two justifications for this assumption. Firstly, the nearest-neighbor atoms are the only ones where the pulse is being applied, so any interactions beyond nearest-neighbor needs to be due to residual Rydberg population from previous pulses, which tends to be very low. Secondly, the Rydberg blockade used in the Gaussian pulse is only 60 MHz, giving a next-nearest neighbor interaction of less than 1MHz, lower than the laser parameters.

None of the jump operators we cover in this paper operate on multiple sites at once. However, they do operate on the forward and backward local Hilbert space indices of the density matrix at the same time. We can write these channels as an operator that acts on the combined vectorized index $\eta_i$ at a particular site. With all of these channel components combined, the time evolution channel $\mathcal{C}_\tau(t)$ in the vectorized picture is a rank 4 tensor that acts on the vectorized index of two neighboring sites. 

\begin{figure}
    \centering
    \begin{tikzpicture}
        \begin{scope}
            \node[anchor=north west,inner sep=0] (image_a) at (0,0)
            {\includegraphics[width=0.85\columnwidth]{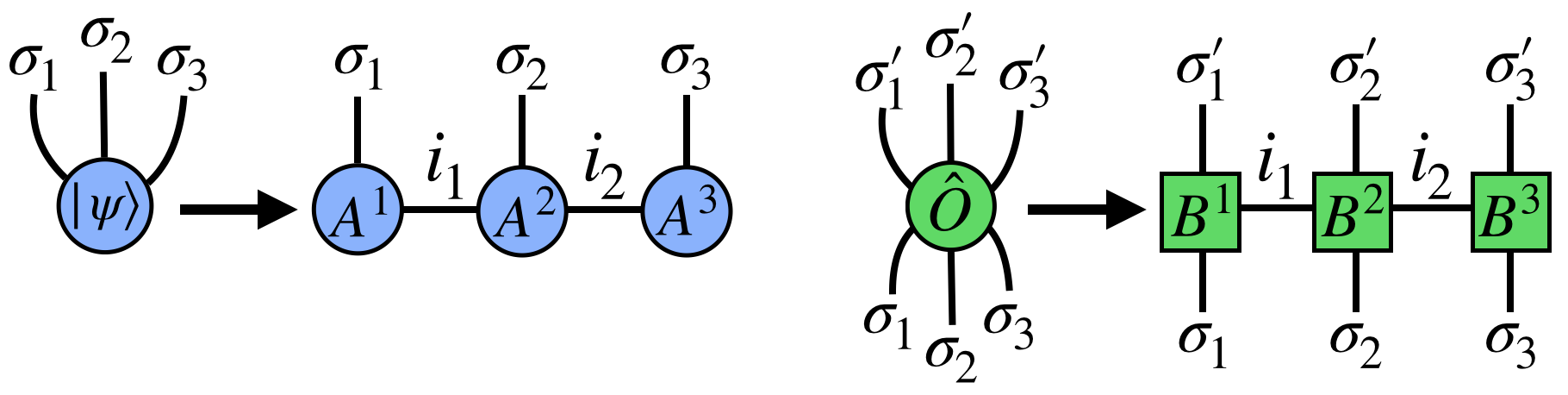}};
            \node [anchor=north west] (note) at (-0.4,0.1) {\small{\textbf{a)}} };
            \node [anchor=north west] (note) at (3.4,0.1) {\small{\textbf{b)} }};
        \end{scope}
    \end{tikzpicture}
    \begin{tikzpicture}
        \begin{scope}
            \node[anchor=north west,inner sep=0] (image_b) at (0,0)
            {\includegraphics[width=0.99\columnwidth]{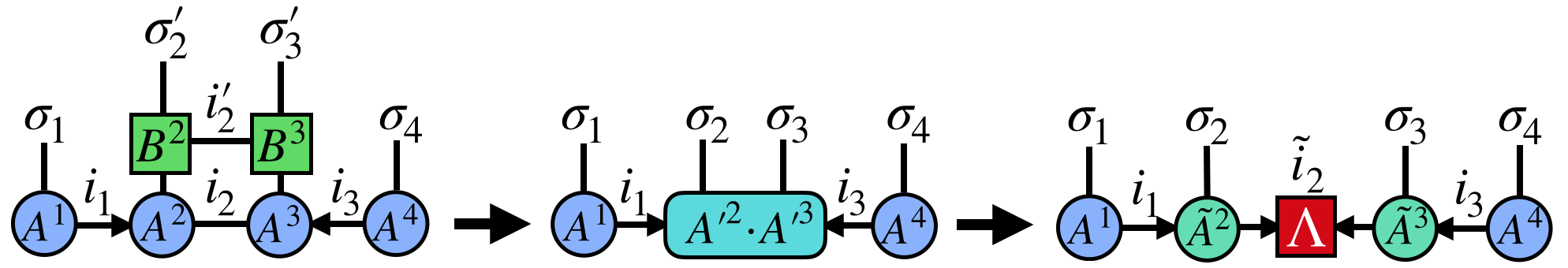}};
            \node [anchor=north west] (note) at (-0.1,0.1) {\small{\textbf{c)}}};
        \end{scope}
    \end{tikzpicture}
    \caption{a) Wavefunction/MPS. b) Operators or Density Matrix/MPO. c) A potential scheme for truncating an MPS while applying a two-site operator. Arrows on the MPS denote a canonization direction. All operator and active site tensors are first combined then resplit into two sites with SVD. This also produces a diagonal tensor $\Lambda$ between the sites containing the singular values of the squared density matrix. The truncated density matrix is then obtained by removing the lowest singular values of $\Lambda$.}
    \label{fig:ansatzes}
\end{figure}

At this point, we can time evolve a vectorized density matrix by applying this channel for each time step. This is accomplished by first multiplying together every tensor involved (Fig.~\ref{fig:ansatzes}c, first step). The resulting tensor now contains the local Hilbert space indices for two neighboring qubits. We use singular value decomposition (SVD, Fig.~\ref{fig:ansatzes}c second step) to split the two qubit degrees of freedom into separate tensors. This process also creates a new bond index between the tensors and a diagonal matrix $\Lambda$ on the bond. Generally, the dimension of this bond index will be larger than the previous bond dimension, but we can truncate it back to the original dimension by removing the least significant diagonal elements in $\Lambda$. This process is most efficient when the MPS has been canonized, with its center at one of the active sites of the gate \cite{Zhang2020}. Canonization is a gauge transformation that creates a specific center $c$ in the MPS such that the contraction of all sites around that center reduces to the identity,
\small\begin{align}
    &\sum_{i_k, k<c\text{-}1} \; \sum_{\eta_l, l < c} B^1_{\eta_1 i_1} ... B^{c\text{-}1}_{\eta_{c\text{-}1} i_{c\text{-}2} i_{c\text{-}1}} \!\!\! \sum_{i_k', k<c\text{-}1}\!\! B^{1 \dagger}_{\eta_1 i_1'} ... B^{c\text{-}1 \dagger}_{\eta_{c\text{-}1} i_{c\text{-}2}' i_{c\text{-}1}'} \n&\quad= \delta_{i_{c\text{-}1} i_{c\text{-}1}'}
\end{align}
\begin{align}
    &\sum_{i_k, k>c} \;\sum_{\eta_l, l > c} B^{N}_{\eta_N i_N} ... B^{c+1}_{\eta_{c+1} i_{c} i_{c+1}} \sum_{i_k', k>c} B^{N \dagger}_{\eta_N i_N'} ... B^{c+1 \dagger}_{\eta_{c+1} i_{c}' i_{c+1}'} \n&\quad= \delta_{i_{c} i_{c}'}.
\end{align}\normalsize
The argument in \cite{Zhang2020} is based on the L2-norm of the MPS being well-behaved, so for a density matrix MPS, as long as the purity is relatively close to 1 (which is the case for any light source of noise), canonization will yield a similar increase in truncation efficiency. 

Each application costs a time $O(d^3 D^3)$ where $d$ is the local Hilbert space dimension and $D$ the bond dimension of the vectorized density matrix MPS. As many of the frequencies used in the pulse are very large (particularly the Rydberg blockade), the required time step for accurate simulation is very small, requiring the application of thousands of time step channels for a single pulse. Applying each channel one-by-one to the MPS would be unnecessarily expensive. Instead, we integrate the time evolution channel for the entire pulse directly before applying it to the MPS. In a vectorized picture, this is simply a matter of multiplying all the time step channels together. Once the full time evolution channel of a CZ gate has been assembled, it does not have to be re-evaluated unless one of the dissipation or Hamiltonian parameters change and can be copied onto any CZ gate that appears in the quantum circuit we want to simulate.

The advantage of this method is that we can significantly reduce the amount of times we have to apply an operator directly onto the density matrix, but the disadvantage is that the form of the time evolution channel becomes more complicated to evaluate if it acts on more than two sites. This is the case for a circuit with significant global Rydberg atom crosstalk error. 

\subsection{MPO form of the Noisy CZ Gate with crosstalk}\label{sec:inf_crosstalk}

In this section we will cover a specific source of coherent error that will be introduced to some (but not all) of the circuit simulations in the rest of this paper. When a pulse excites a qubit to the Rydberg states, it leaves a residual population in that state. If the residual population does not fully decay before the next pulse, there can be unwanted crosstalk between the Rydberg population of the target sites of the pulse and residual populations in neighboring sites.
In order to add Rydberg atom crosstalk to our time evolution channel, we need to include the global blockade term
\begin{gather}
    H_b = \sum_{i=1}^{N-1} B |rr\rangle \langle rr|_{i,i+1} \negthickspace\negthickspace \underset{j\neq\{i,i+1\}}{\bigotimes_{j=1}^{N}} \negthickspace\negthickspace I_j
\end{gather}
to our time evolution channel. In all our simulations of unwanted crosstalk, we assume that there is no noise to combine with the crosstalk, so we time evolve under the Hamiltonian instead of the Lindbladian. 
All blockade terms commute with each other and the only blockade terms that do not commute with the rest of the Hamiltonian are those that have some overlap with the active sites, i.e. the active site blockade and the nearest neighbor blockade terms. All other terms can be applied as their own local operator onto the MPS, independent of the rest of the MPO. 

The blockade between an active site and its nearest non-active neighbor can be interpreted as a shift in the effective detuning of that site, $\Delta \rightarrow \Delta+B$, conditioned on whether the non-active neighbor is in a Rydberg state or not. Thus, we can write the gate with nearest-neighbor blockade $\mathcal{C}^{Xtalk}_{i,i+1}$ as a combination of unmodified two-site gates $\mathcal{C}_{i,i+1}(\Delta_i, \Delta_{i+1})$,
\small\begin{align} 
    \mathcal{C}^{Xtalk}_{i,i+1} = &(I-|r\rangle\langle r|)_{i-1} \otimes (I-|r\rangle\langle r|)_{i+2} \otimes \mathcal{C}_{i,i+1}(\Delta_i, \Delta_{i+1})\n
    +&(|r\rangle\langle r|)_{i-1} \otimes (I-|r\rangle\langle r|)_{i+2} \otimes \mathcal{C}_{i,i+1}(\Delta_i+B, \Delta_{i+1})\n
    +&(I-|r\rangle\langle r|)_{i-1} \otimes (|r\rangle\langle r|)_{i+2} \otimes \mathcal{C}_{i,i+1}(\Delta_i, \Delta_{i+1}+B)\n
    +&(|r\rangle\langle r|)_{i-1} \otimes (|r\rangle\langle r|)_{i+2} \otimes \mathcal{C}_{i,i+1}(\Delta_i+B, \Delta_{i+1}+B).
\end{align}\normalsize
This tensor gives us the form of the time evolution operator for the active sites and their nearest neighbors, which combined with the two-site blockade operators on the other sites gives the full time evolution of the system as a string of tensors. This assumes that each gate is being applied sequentially, which is not necessarily the case in a real circuit. We could also apply each gate in the same layer simultaneously. However, this would result in a far less tractable MPO, as we would no longer have commutativity of the blockade terms.

\subsection{Matrix Product Density Operators}

While the LME by itself preserves the positivity of the density matrix, positivity is not an inherent quality of the MPS. Therefore, if the density matrix is truncated, there is a possibility of introducing negative eigenvalues to the system. In the following sections we will describe two ways to alleviate this negative eigenvalue problem.

\begin{figure}
    \centering
    \begin{tikzpicture}
        \begin{scope}
            \node[anchor=north west,inner sep=0] (image_a) at (0,-0.5)
            {\includegraphics[width=0.48\columnwidth]{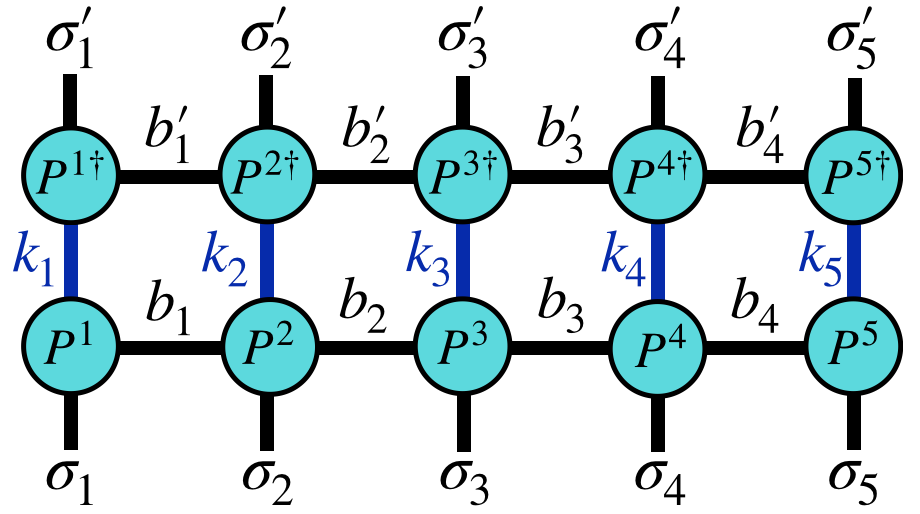}};
            \node [anchor=north west] (note) at (-0.4,0) {\small{\textbf{a)}}};
        \end{scope}
        \begin{scope}[xshift=0.5\columnwidth]
            \node[anchor=north west,inner sep=0] (image_a) at (0,0)
            {\includegraphics[width=0.48\columnwidth]{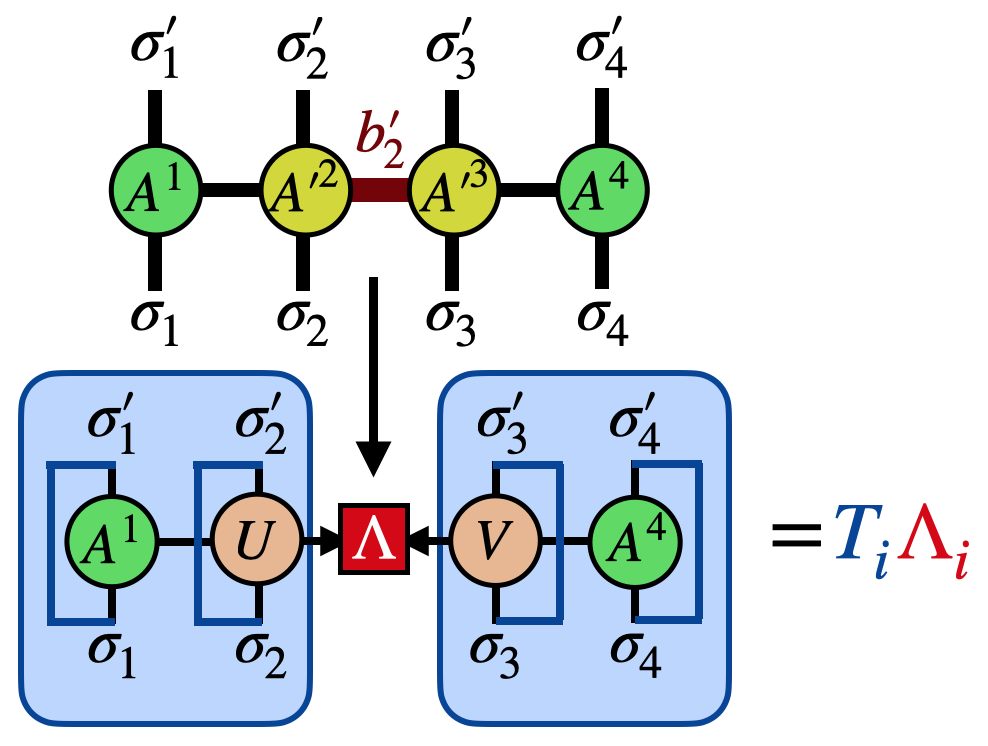}};
            \node [anchor=north west] (note) at (-0.15,0) {\small{\textbf{b)}}};
        \end{scope}
    \end{tikzpicture}
    \caption{(a) MPDO Ansatz. The $k_i$ indices are the inner dimensions of the tensor network, playing a similar role to the Kraus indices found in completely positive quantum channels. (b) Setup of the Purity-Preserving Truncation algorithm. Since the $\Lambda$ tensor is diagonal, the trace can be treated as the inner product of two vectors -- the $\Lambda$ tensor's diagonal entries and its environment.}
    \label{fig:ansatzes_positive}
\end{figure}

The traditional method, described in this section, is a tensor network representation of the density matrix that enforces its positivity at all times \cite{Verstraete2004, Werner2016}. This representation comes from the diagonalized form of the density matrix as a sum over wavefunction projectors,
\begin{gather}
    \rho = \sum_i p_i |\psi_i\rangle \langle \psi_i |.
\end{gather}
Each term in this sum can be represented as the outer product of an MPS and its complex conjugate. Instead of accumulating a potentially exponential amount of terms, we can distribute the summation over an extra internal dimension for every site (Fig.~\ref{fig:ansatzes_positive}a), creating the Matrix Product Density Operator. The drawback of this representation is the higher computational cost due to the extra index, and the increased difficulty of applying new gates to the density matrix and truncating it.  
Unlike in many other works\cite{Werner2016,Cheng2021}, the gate and noise operators are simulated at the pulse level, and occur simultaneously. While the individual time steps can be separated into entangling and noise components, the need to integrate each time step into a single channel, which was caused by the LME parameters requiring large frequencies and short time steps, prevents the gate from being separated this way. Instead, the tensor representing the integrated time evolution channel must be split along its bond and Kraus direction at the same time. In Appendix~\ref{app:4tsplit} we have constructed an algorithm to perform this simultaneous separation. This algorithm, which we call the Four-Tensor Split, is inspired by the Moses move used in isometric tensor networks \cite{Zaletel2020,Hauschild2018}, which also requires separating a tensor along multiple directions at once. In the Four-Tensor Split, the Kraus decomposition of the tensor is conducted first, then the Kraus components are adjusted to optimize its site-wise separation of information. However, such a splitting of information is inefficient when the operators responsible for entanglement and noise do not commute with the other terms of the LME, as is the case for all the noise models we consider in this paper.
In addition, applying the time evolution channel to an MPDO affects both its bond and inner dimension at the same time. The fact that four indices must be truncated simultaneously, as opposed to a conventional MPS truncation which only addresses one bond at a time, makes it difficult to concentrate information onto any individual site through a canonization-like scheme.

\begin{figure*}
    \centering
    \begin{tikzpicture}
        \begin{scope}
            \node[anchor=north west,inner sep=0] (image_a) at (0,0)
            {\includegraphics[width=0.49\columnwidth]{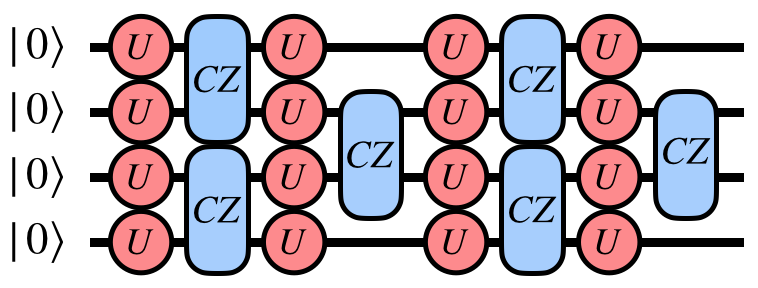}};
            \node [anchor=north west] (note) at (-0.4,0) {\small{\textbf{a)}}};
        \end{scope}
        
    \end{tikzpicture}
    \begin{tikzpicture}
        \begin{scope}
            \node[anchor=north west,inner sep=0] (image_a) at (0.2,0)
            {\includegraphics[width=0.63\columnwidth]{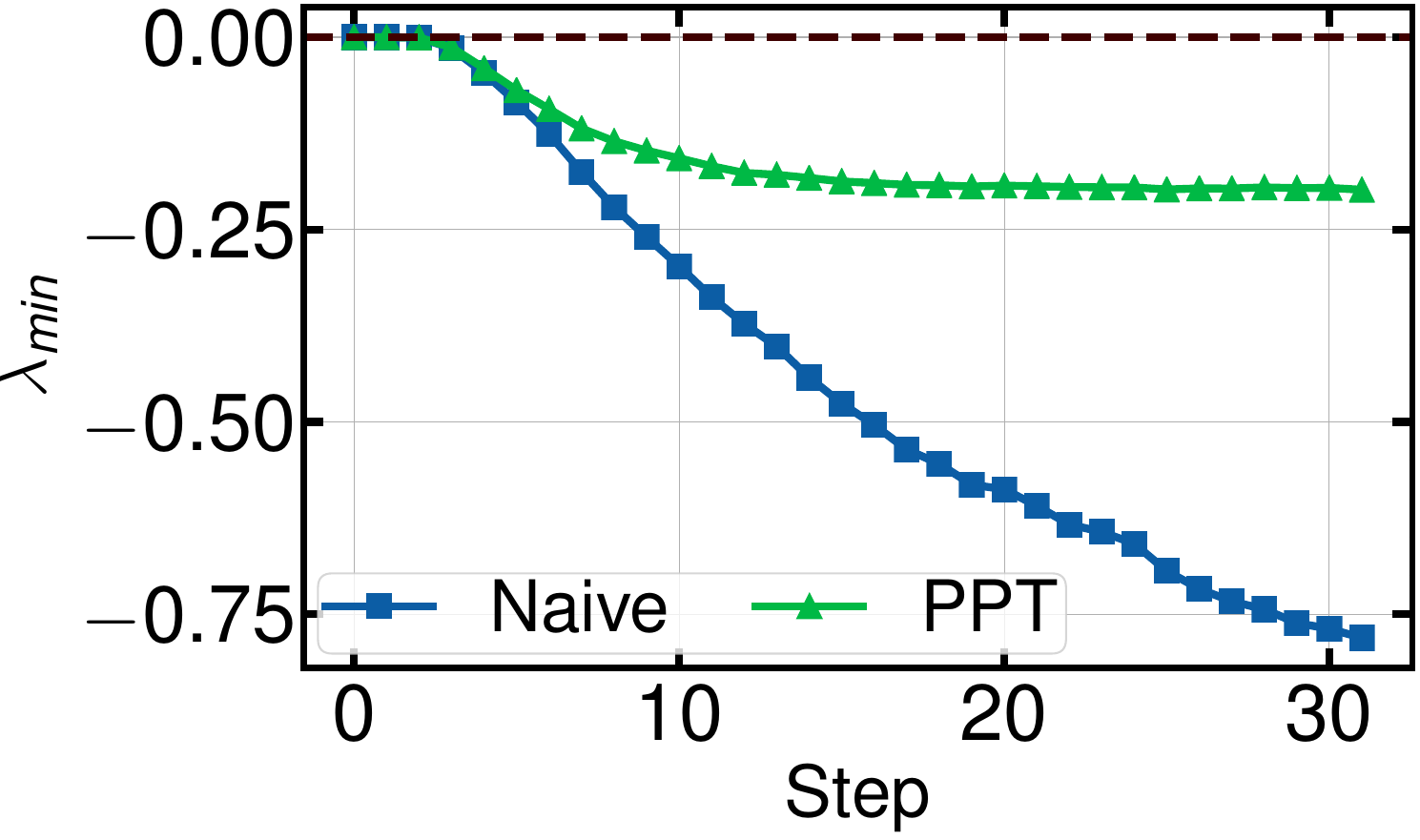}};
            \node [anchor=north west] (note) at (-0.0,0) {\small{\textbf{b)}}};
        \end{scope}
        \begin{scope}[xshift=0.67\columnwidth]
            \node[anchor=north west,inner sep=0] (image_a) at (0,0)
            {\includegraphics[width=0.63\columnwidth]{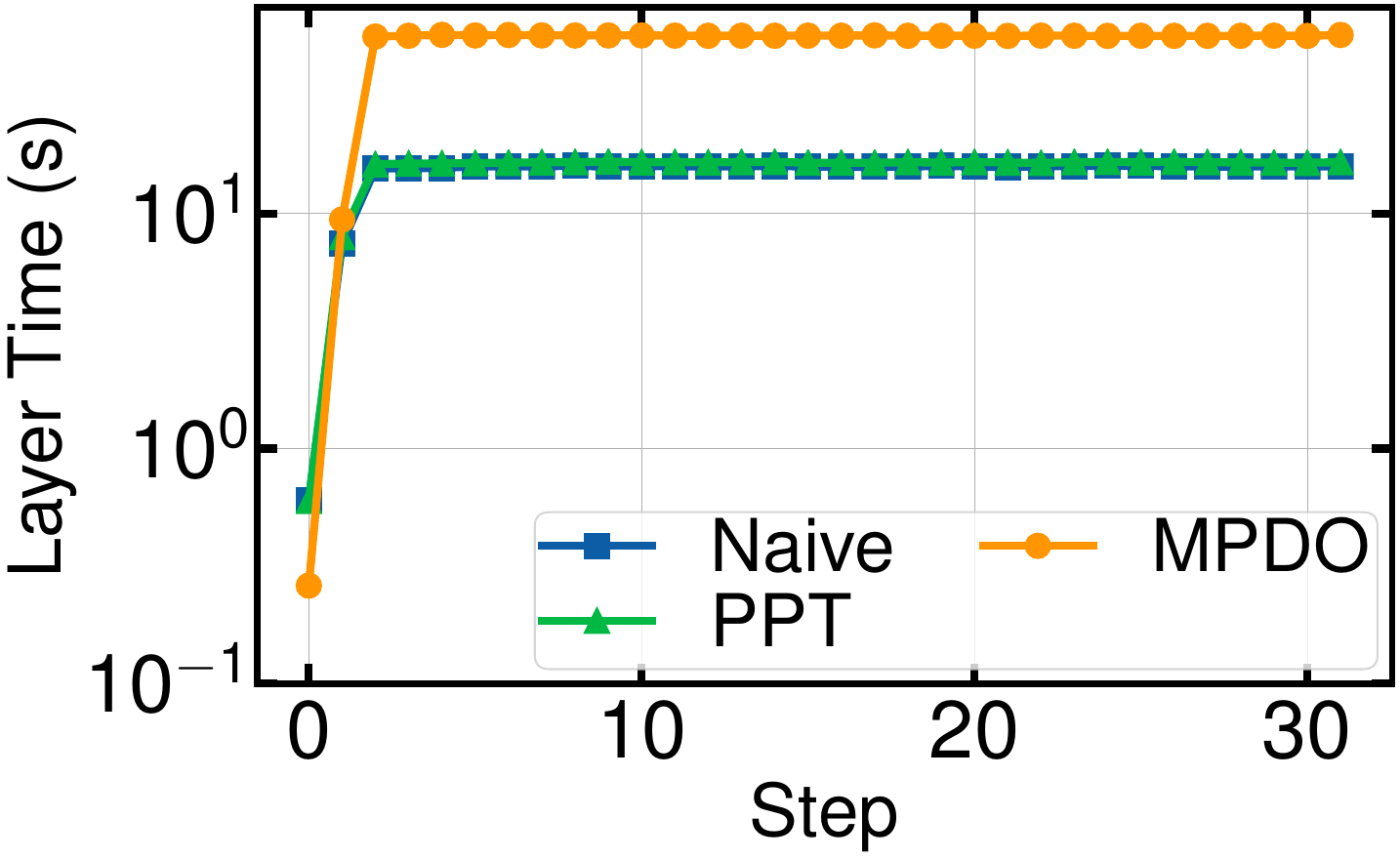}};
            \node [anchor=north west] (note) at (0.2,0.1) {\small{\textbf{c)}}};
        \end{scope}
        \begin{scope}[xshift=1.32\columnwidth]
            \node[anchor=north west,inner sep=0] (image_a) at (0.2,0)
            {\includegraphics[width=0.63\columnwidth]{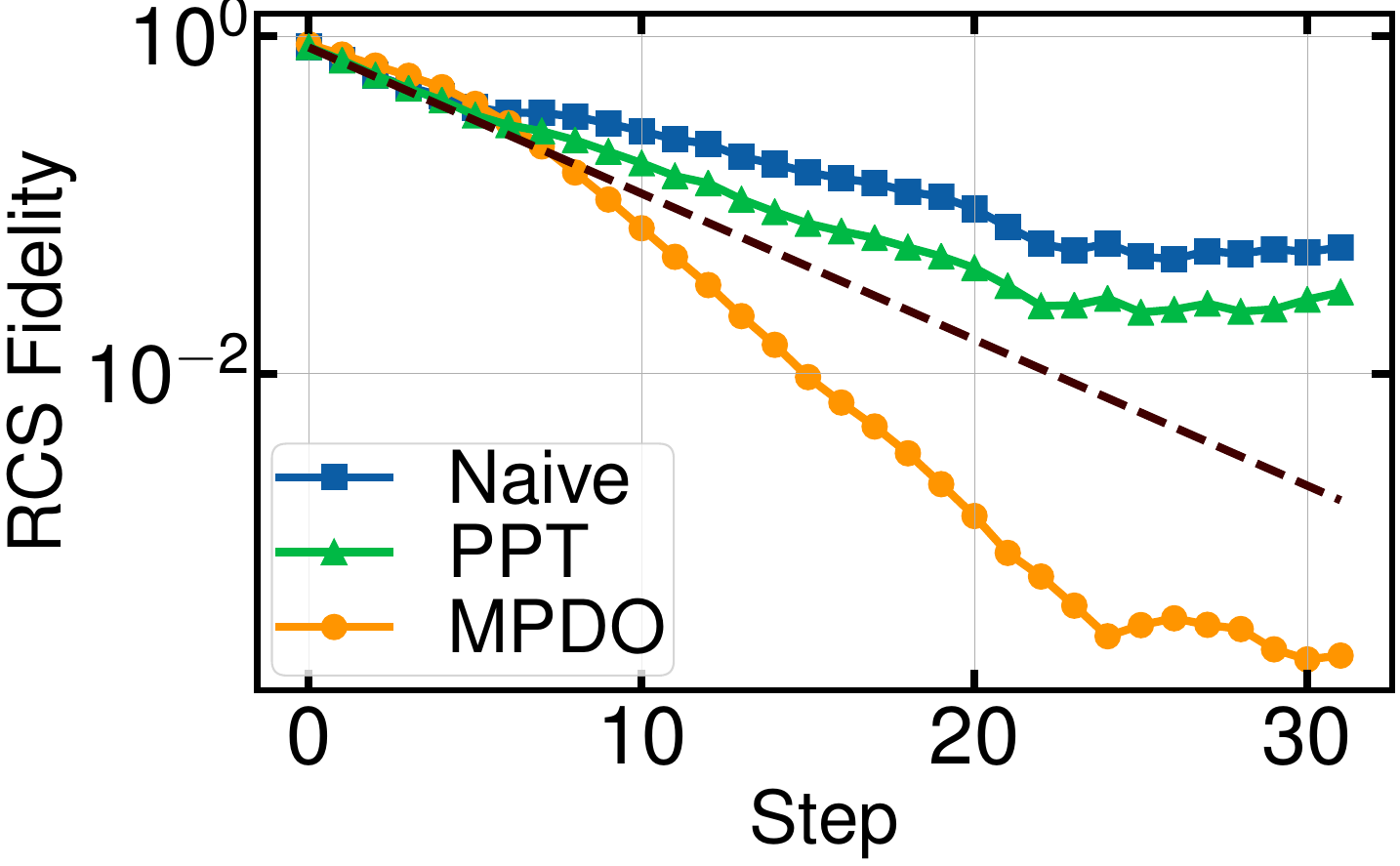}};
            \node [anchor=north west] (note) at (-0.0,0.1) {\small{\textbf{d)}}};
        \end{scope}
    \end{tikzpicture}
    \caption{Simulated random circuit fidelity over layers of a 12 site quantum circuit over 20 samples, under a Rydberg atom dissipation source $\gamma_{diss}=0.001$, with a maximum bond dimension of 64 and a maximum inner dimension of 4, if applicable. (a): Diagram of the random circuit. Each gate $U$ is an independent random 2x2 unitary chosen from the Haar measure, acting on the qubit states. (b): 
    Minimum eigenvalue comparisons of the non-positive MPO with and without PPT. The quantity $\lambda_{min}$ is the most negative eigenvalue of the density matrix after each layer of the random circuit, determined through DMRG. Without PPT, these eigenvalues increase rapidly, while with PPT they are bounded in absolute value from above. (c): Average wall time required to compute each layer of the random circuit with and without PPT and MPDOs on a single node personal computer. Using PPT introduces a small overhead to the time cost that becomes insignificant for MPS bond dimensions of 64 or above. Given that MPS's typically only become difficult to run at bond dimensions in the thousands, this is a minor cost for most circuits. (d): Random circuit fidelity between systems with and without PPT, including MPDO results. The fidelity is expected to maintain a consistent exponential decay, which is obeyed most closely by the highest bond dimension MPO with PPT -- in other circuits we see a sharp deviation from the initial exponential once the bond dimension saturates. }
    \label{fig:dm_ptt_comparison}
\end{figure*}

\subsection{Purity-Preserving Truncation}
There is a compromise between the simplicity of the vectorized MPO and the representational faithfulness of the MPDO -- we keep the density matrix as a vectorized MPO but ensure that all truncations of the density matrix do not change its purity $\tr(\rho^2)$. Specifically, we define 
\begin{align}
    \xi = \frac{\tr(\rho^2)}{\tr(\rho)^2}.
\end{align}
In Purity-Preserving Truncation (PPT), after the truncation of the bond's singular values $\Lambda_i$ to a smaller set $\Lambda_i' \equiv P_T\Lambda_i$ (where $P_T$ is the partial projection that truncates the least significant values), we modify the truncated singular values to $\tilde{\Lambda}_i$, the closest set of values that keeps $\xi$ constant. This is not significantly harder than regular truncation because all terms in the fraction can be represented as polynomial functions of $\Lambda_i$ and the environment tensors $T_i$, $P_{ij}$ (Fig.~\ref{fig:ansatzes_positive}b), 
\begin{gather}
    \tr(\rho) = \sum_i T_i \Lambda_i,\\
    \tr(\rho^2) = \sum_i P_{ij} \Lambda_i \Lambda_j.
\end{gather} 

Once we have determined $T_i$ and $P_{ij}$, finding the vector $\tilde{\Lambda}_i$ which is closest to the vector of original singular values $\Lambda_i$ becomes a constrained optimization problem. If the density matrix is canonized, this problem becomes even simpler, because $P_{ij}$ is the identity matrix. Then $\tilde{\Lambda}_i$ must satisfy
\begin{gather}
    \xi = \frac{|\Lambda|^2}{\left(\sum_i T_i \Lambda_i \right)^2} \overset{!}{=} \frac{|\tilde{\Lambda}|^2}{\left(\sum_i T_i' \tilde{\Lambda}_i\right)^2} 
\end{gather}
where $T' \equiv P_T T$ is the truncated trace environment and $|\tilde{\Lambda}|^2$ is the squared norm of $\tilde{\Lambda}$ treated as a vector. If we define $\phi$ as the angle between $\Lambda$ and $T$, and $\theta$ as the angle between $\tilde{\Lambda}$ and $T'$, then we must have
\begin{gather}
    \frac{\sec^2 \phi}{|T|^2} = \frac{\sec^2 \theta}{|T'|^2},\\
    \theta = \cos^{-1}\left(\frac{|T|}{|T'|} \cos \phi \right).\label{eq:xiformula}
\end{gather}
To satisfy this bound while maximizing the overlap with both the original singular values $\Lambda_i$ and its truncation $\Lambda'_i$, the new singular values $\tilde{\Lambda}_i$ should become the orthogonal projection of $\Lambda_i'$ onto the cone of constant angle $\theta$ around $T_i'$. If $\sigma$ is the angle between the original truncated values $\Lambda_i'$ and $T_i'$, we have (see Appendix~\ref{app:ppt_derivation} for the full derivation of this solution)
\begin{gather}
    \tilde{\Lambda}_i = \Lambda_i' + \bigg(\frac{\tan \theta}{\tan \sigma} - 1\bigg)\bigg(\Lambda_i' - \frac{|\Lambda'|}{|T'|} T_i' \cos \sigma \bigg). \label{eq:ppt}
\end{gather}
One caveat is that this bound is not necessarily always achievable. We see from Eq.~\eqref{eq:xiformula} that if $|T'| < |T| \cos \phi$, then there is no angle $\theta$ to satisfy the condition. This is equivalent to the possibility that the original trace environment $T$, when projected onto the span of the single vector $\Lambda$, is longer than its projection $T'$ onto the $D$ environment components corresponding to the largest singular values, where $D$ is the maximum bond dimension. As $D$ increases, this becomes more unlikely. In practice, we only find this occurring under heavy truncation. In these cases, setting $\theta=0$ is the best that we can achieve.

\begin{figure*}
    \centering
    \begin{tikzpicture}
        \begin{scope}
            \node[anchor=north west,inner sep=0] (image_a) at (0,-0.2)
            {\includegraphics[width=0.61\columnwidth]{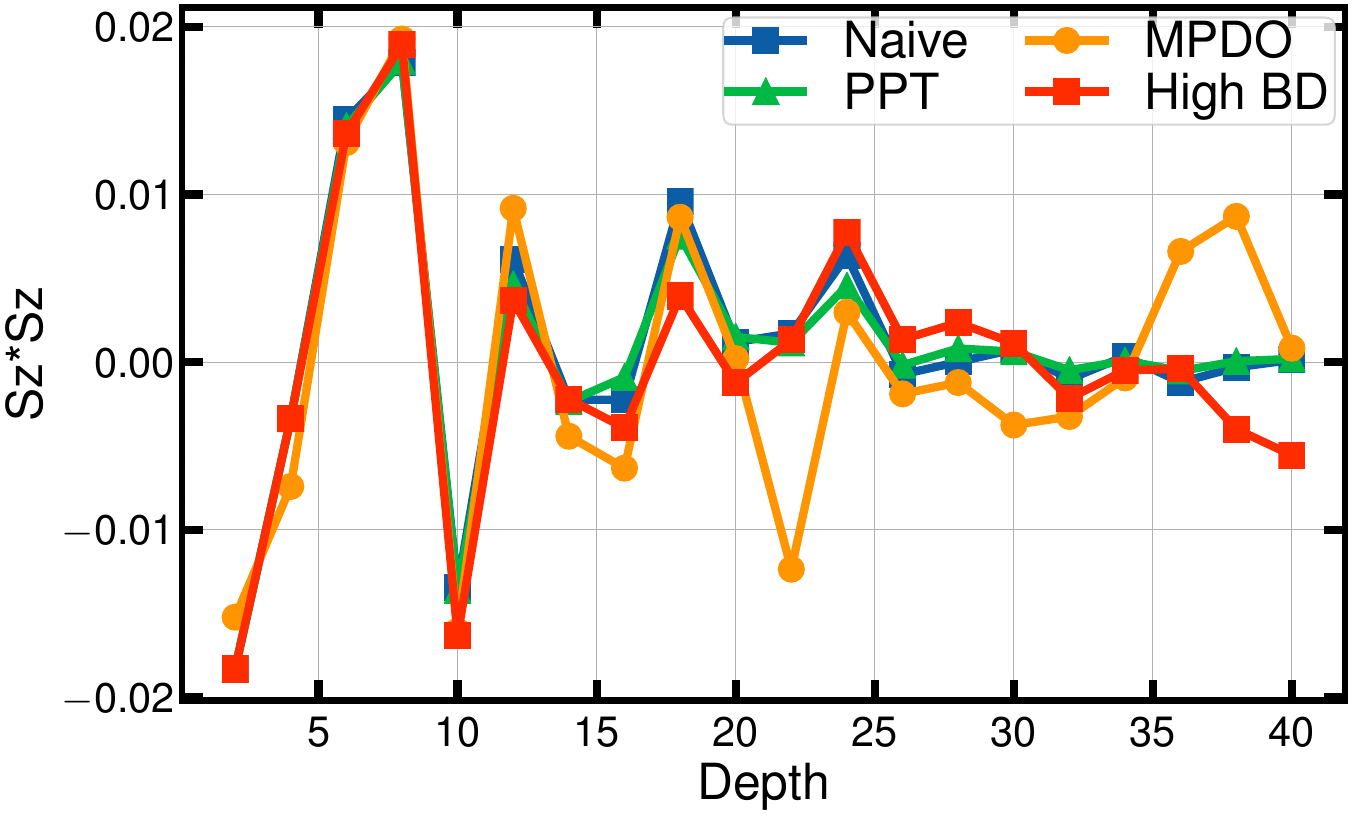}};
            \node [anchor=north west] (note) at (-0.25,0) {\small{\textbf{a)}}};
        \end{scope}
        \begin{scope}[xshift=0.66\columnwidth]
            \node[anchor=north west,inner sep=0] (image_a) at (0,-0.2)
            {\includegraphics[width=0.61\columnwidth]{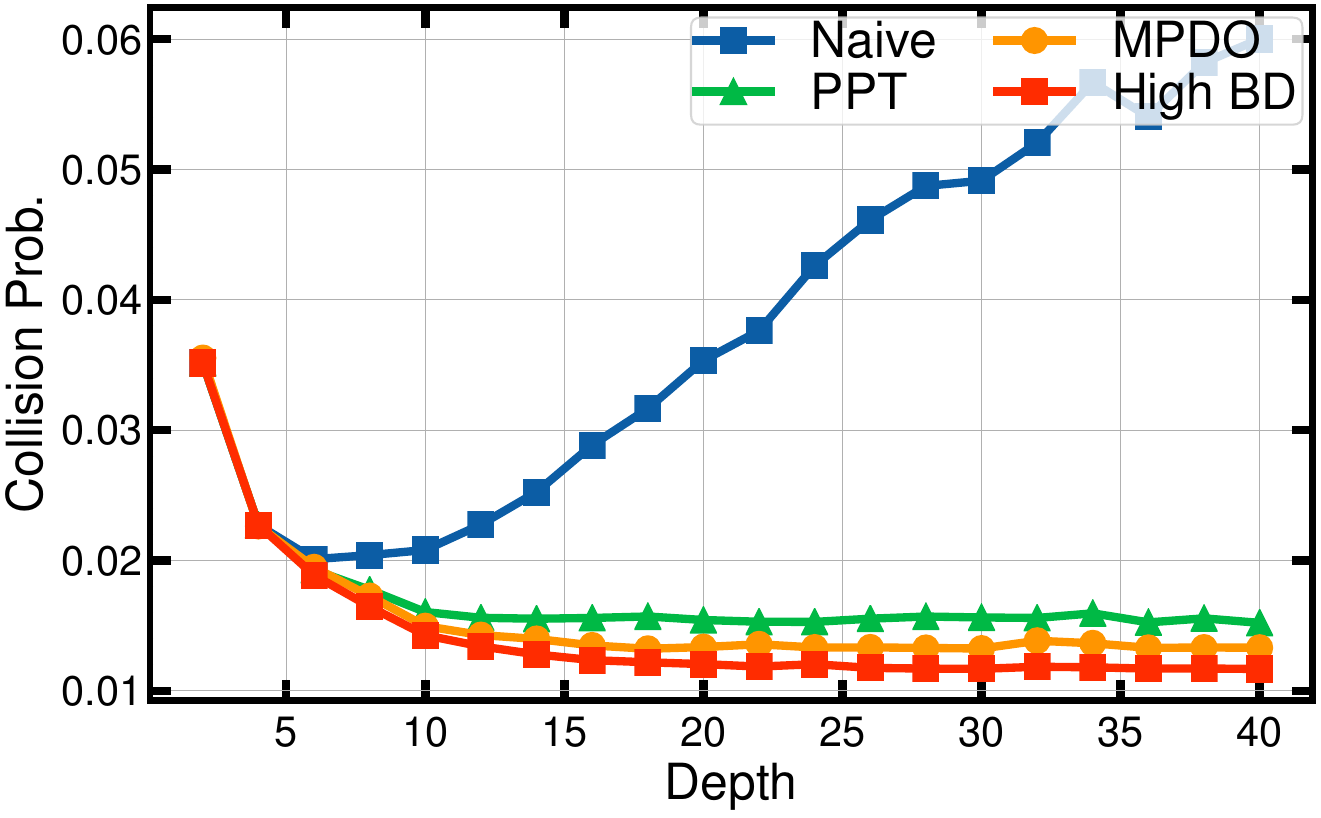}};
            \node [anchor=north west] (note) at (-0.25,0) {\small{\textbf{b)}}};
        \end{scope}
        \begin{scope}[xshift=1.32\columnwidth]
            \node[anchor=north west,inner sep=0] (image_a) at (0,-0.2)
            {\includegraphics[width=0.61\columnwidth]{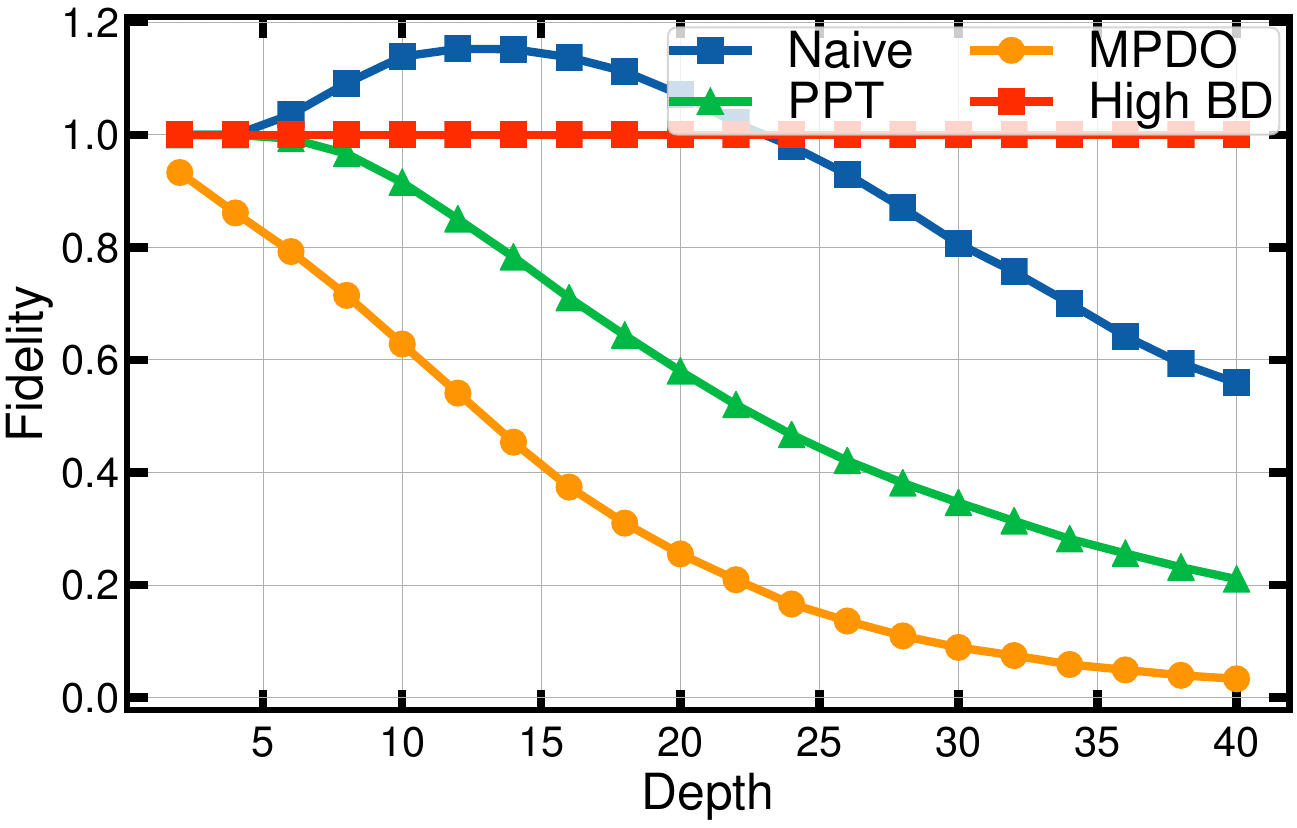}};
            \node [anchor=north west] (note) at (-0.25,0) {\small{\textbf{c)}}};
        \end{scope}
    \end{tikzpicture}
    \begin{tikzpicture}
        \begin{scope}
            \node[anchor=north west,inner sep=0] (image_a) at (0,-0.2)
            {\includegraphics[width=0.61\columnwidth]{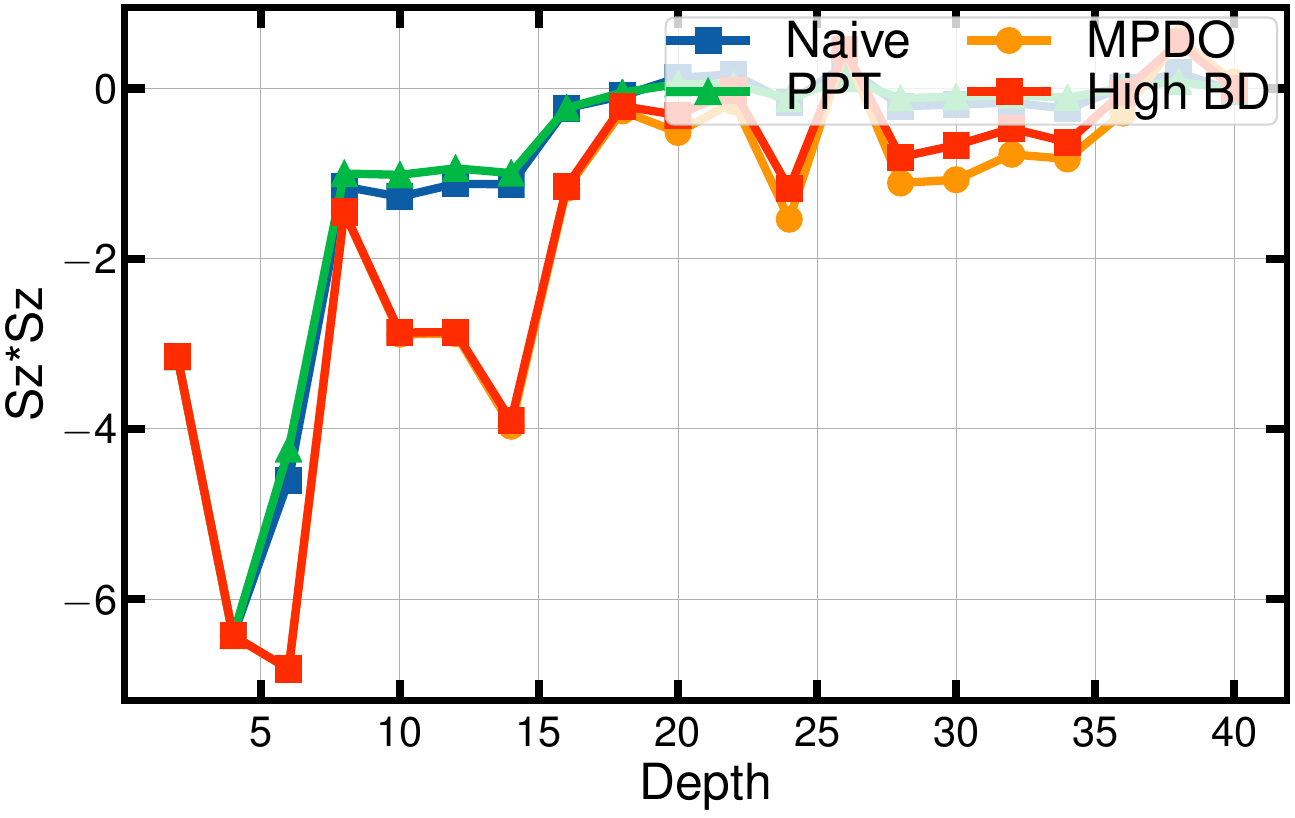}};
            \node [anchor=north west] (note) at (-0.25,0) {\small{\textbf{d)}}};
        \end{scope}
        \begin{scope}[xshift=0.66\columnwidth]
            \node[anchor=north west,inner sep=0] (image_a) at (0,-0.2)
            {\includegraphics[width=0.61\columnwidth]{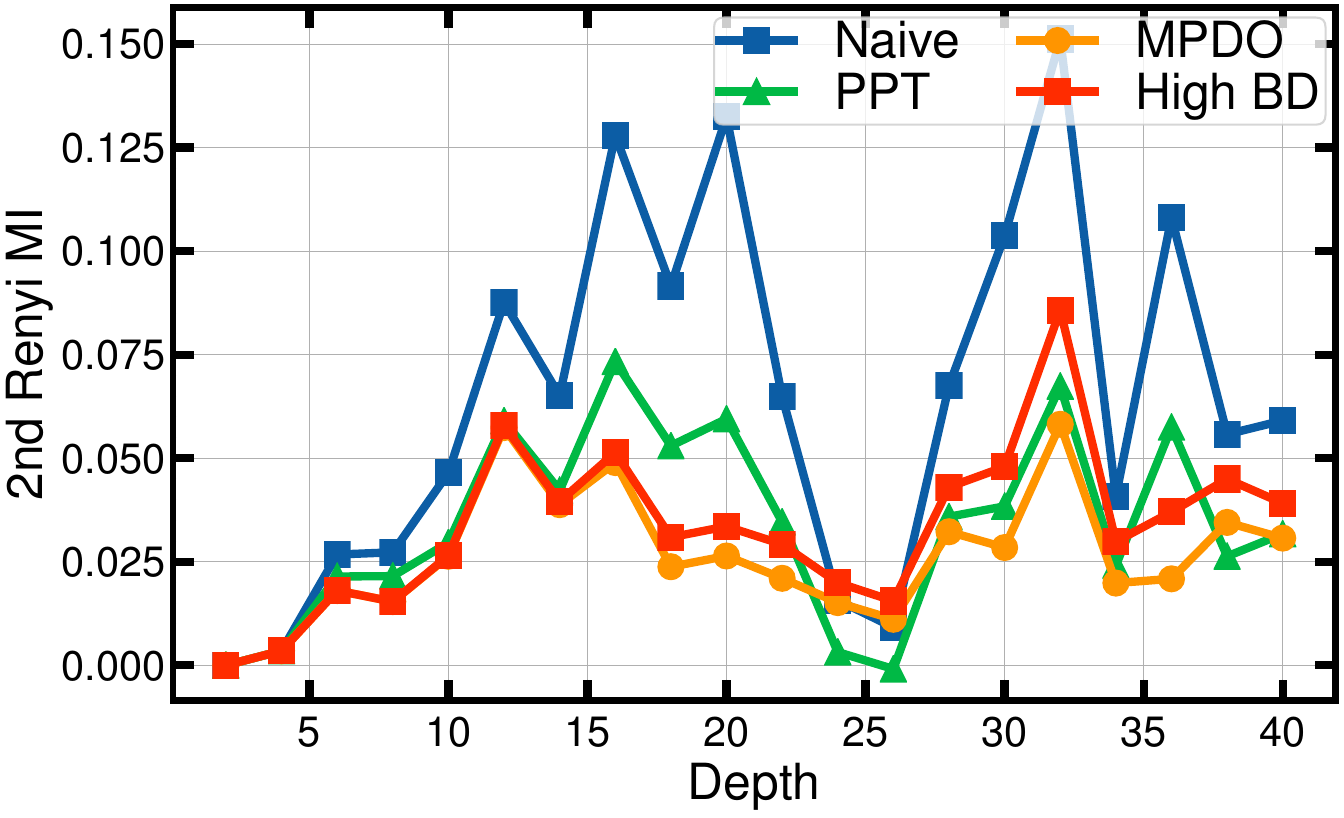}};
            \node [anchor=north west] (note) at (-0.25,0) {\small{\textbf{e)}}};
        \end{scope}
    \end{tikzpicture}
    
    \caption{(a) Average nearest-neighbor $S^z$ correlation on a 32 bond dimension and 500 bond dimension density matrix MPO, averaged over 40 samples of 8-site RCS. Introducing PPT had little effect on this observable. (b) The collision probability of the same 8-site RCS ensemble, comparing a 32 bond dimension density matrix with or without PPT to a 500 bond dimension density matrix, as well as an MPDO with 32 bond dimension and 16 Kraus dimension. (c) Average RCS fidelity in the absence of noise. The fidelity is measured with respect to a wavefunction simulated without truncation. Due to the lack of noise, all deviations from the high bond dimension limit come solely from truncation errors, although the MPO approaches both have negative eigenvalues which make interpreting the fidelity subtle and not necessarily bounded by one. (d) Average nearest-neighbor $S^z$ correlation on a 40 layer, 20 site QAOA iteration of the TFIM circuit with random parameters. (e) Average 2nd Renyi mutual information between sites 2 and 5 on the same circuit.}
    \label{fig:rcs_ptt_comparisons}
\end{figure*}

\subsection{Benchmarking of Density Matrix Ansatzes}

\subsubsection{Random circuit sampling}
We first use a random circuit architecture to compare the performance of each ansatz. Each layer consists of CZ gates surrounded by Haar random 1-site unitaries applied to every site in an even-odd pattern (Fig.~\ref{fig:dm_ptt_comparison}a). Given a sufficiently small noise, this circuit should eventually be intractible for any classical simulation, with a complexity growing exponentially in depth. 

We first test the positivity of the system with and without PPT under a heavily truncated RCS iteration. We see that PPT does not force the density matrix to be positive, but it does result in less negative eigenvalues for a minor computational cost (Fig.~\ref{fig:dm_ptt_comparison}b-c). In particular, for the 12-site random circuit of Fig.~\ref{fig:dm_ptt_comparison}, the minimum eigenvalue of the tensor network with PPT appears bounded at $\lambda_{min} \approx -0.2$. Given a mostly pure density matrix with maximum eigenvalue $\lambda_{max}$, we can have a general bound of $\lambda_{\min} \geq -\sqrt{\tr{\rho^2}-\lambda_{max}^2}$, however this does not appear to be enough to explain the plateau behavior of the minimum eigenvalue.

We then compare the random circuit fidelity between the simulated output of the circuit under noise and the ideal output without noise. This should take the form of an exponential decay in the circuit depth with a coefficient dependent on the noise factors\cite{Arute2019,Liu2021}. Deviations in this exponential decay represent the breakdown of the classical ansatz as too much information gets truncated. Fig.~\ref{fig:dm_ptt_comparison}d demonstrates the differences between the performance of the non-positive MPO, with and without PPT, as well as the MPDO ansatz. The MPO with PPT appears to be the most stable ansatz under RCS fidelity. The MPDO fidelity decreases much more rapidly than the expected exponential decay once its bond dimension becomes saturated, with the size of the inner dimension having little effect on this deviation (Appendix \ref{app:mpdo_rule}). On the other hand, the MPO without PPT drifts above the expected exponential once its bond dimension saturates. This is because it is no longer reporting a fidelity -- the negative density matrix eigenvalue increases the value of $\tr(\rho)^2$, which in turn increases the overlap of $\rho$ with the ideal wavefunction. This can even cause the reported fidelity to exceed 1 for deeper circuits. In Fig.~\ref{fig:rcs_ptt_comparisons}c, we repeat these observations for 8-site RCS without noise.

Without a guarantee of positivity, the RCS fidelity is not a completely reliable quantity to judge accuracy, as different sources of error can either increase or decrease the fidelity. However, there are other metrics by which we can measure the accuracy of a system. We also compare different observables of the density matrices to a density matrix with the same noise parameters, but much higher bond dimension. For example, we can consider the average nearest-neighbor $S^z$ correlation
\begin{gather}
    \frac{1}{N-1} \sum_{i=1}^{N-1} \langle S^z_i S^z_{i+1}\rangle.
\end{gather}
We perform 40 samples of 8-site RCS over a depth of 40, comparing two MPS's with bond dimension 32, one with PPT and one without. We also simulate an MPS of bond dimension 729 with the same gates, as this is an adequate bond dimension to simulate the 8-site density matrix over qubit and Rydberg states, to any depth, without truncation. We see (Fig.~\ref{fig:rcs_ptt_comparisons}a) that using a PPT truncation scheme does not significantly affect the nearest-neighbor $S^z$ correlation, with the discrepancies between either truncation method and the high bond dimension MPS being approximately the same for depths 30 and above.

While PPT does not affect the closeness of the density matrix to its exact, untruncated form, it keeps the density matrix in a subspace that more accurately reflects its physical properties. This can be seen most clearly by looking at the anticoncentration~\cite{Dalzell2022} of the system. Anticoncentration is a key quantity in random quantum circuits, and for a given random circuit ensemble $\varepsilon$ is equal to the inverse of the collision probability
\begin{gather}
    Z_C(\varepsilon) = E_{U \in \varepsilon}\left[\sum_{x \in [q]^N} p_U(x)^2\right].
\end{gather}
In Fig.~\ref{fig:rcs_ptt_comparisons}b, we measure the collision probability of an 8-site random circuit ensemble over depth. We see that using PPT allows the low bond dimension system to more accurately replicate the true decay of the collision probability, as given by the high bond dimension system. 

\subsubsection{QAOA iteration of time evolution under TFIM}
Next, we compare systems over a different circuit - namely, a QAOA iteration of the TFIM circuit (a more detailed description of which will be given in Section IV) with 20 sites and 40 layers using random phase parameters. Like before, we compared the systems using one linear observable, the average nearest-neighbor $S^z$ correlation, and one nonlinear observable. In this case, we chose the 2nd Renyi mutual information
\begin{gather}
    S_2(A) + S_2(B) - S_2(AB)
\end{gather}
where $S_2(X)$ is the 2nd Renyi entropy of subsystem $X$, and $A,B$ are single-site subsystems consisting of the 2nd and 5th sites of the system respectively. A similar picture (Fig.~\ref{fig:rcs_ptt_comparisons}d,e) emerges here - linear observables are roughly the same whether we use PPT or not, while PPT provides a noticeable improvement for nonlinear observables. 

\begin{figure}
    \centering
    \begin{tikzpicture}
        \begin{scope}
            \node[anchor=north west,inner sep=0] (image_a) at (0,-0.2)
            {\includegraphics[width=0.7\columnwidth]{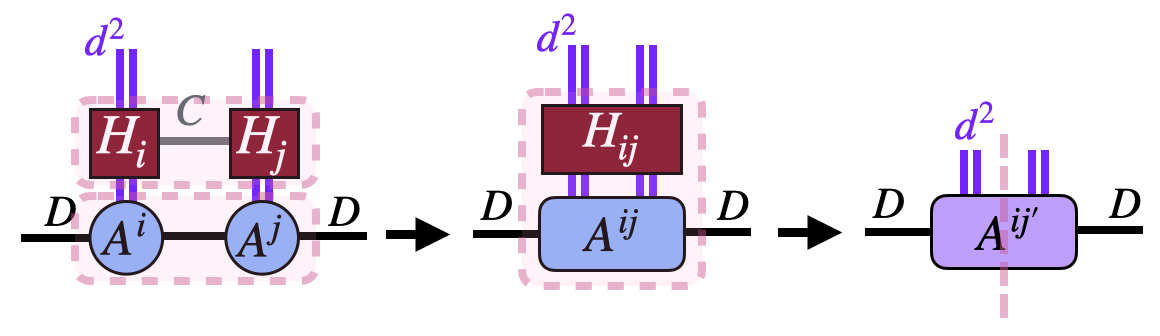}};
            \node [anchor=north west] (note) at (-0.4,0) {\small{\textbf{a)}}};
        \end{scope}
    \end{tikzpicture}
    \begin{tikzpicture}
        \begin{scope}[xshift=0.5\columnwidth]
            \node[anchor=north west,inner sep=0] (image_a) at (0.2,-0.2)
            {\includegraphics[width=0.9\columnwidth]{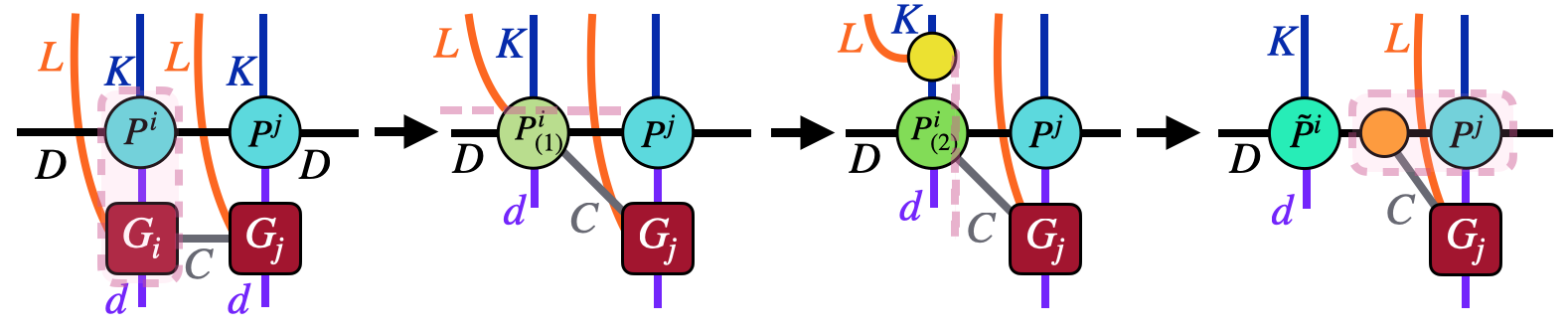}};
            \node [anchor=north west] (note) at (-0.25,0) {\small{\textbf{b)}}};
        \end{scope}
    \end{tikzpicture}
    \begin{tikzpicture}
        \begin{scope}
            \node[anchor=north west,inner sep=0] (image_a) at (0,-0.5)
            {\includegraphics[width=0.25\columnwidth]{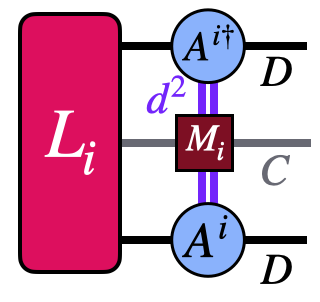}};
            \node [anchor=north west] (note) at (-0.4,-0.5) {\small{\textbf{c)}}};
        \end{scope}
        \begin{scope}[xshift=0.35\columnwidth]
            \node[anchor=north west,inner sep=0] (image_a) at (0,-0.2)
            {\includegraphics[width=0.28\columnwidth]{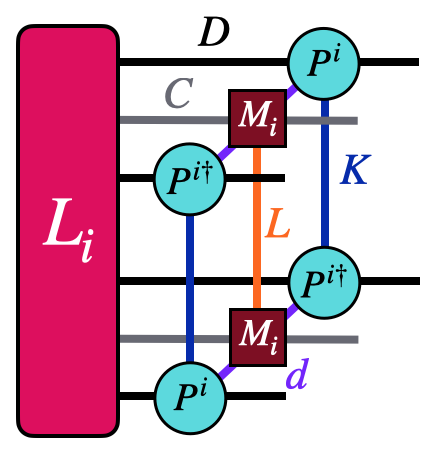}};
            \node [anchor=north west] (note) at (-0.5,-0.2) {\small{\textbf{d)}}};
        \end{scope}
    \end{tikzpicture}
    \caption{(a) The key steps of updating a density matrix, represented by MPO tensors $A^i$, with a two-site gate $H_i, H_j$. First the tensors corresponding to site $i$ and $j$, on both the MPO and the gate, are contracted together (pink dotted boxes), then combined into each other to form a single tensor, and finally is split back into MPO tensors with SVD (pink dotted line), possibly using PPT to change the singular values. The highest order cost of this operation is $O(d^8 D^2 + d^6 D^3)$. (b) The key steps of updating an MPDO, represented by tensors $P^i$, using a gate split into $G_i, G_j$ using a four-tensor split, by updating each site individually. The highest order cost is $O(dCK^2L^2D^2+dC^2KD^3)$. (c) One step of computing a generic second order measurement (i.e. a linear function of $\rho^{\otimes 2}$, such as the anticoncentration or the purity) on a density matrix represented by an MPO. At each step, we contract the sites $A_i, A_i^{\dagger}$ as well as the generic measurement operator $M_i$ into the left environment $L_i$. The highest order cost is $O(d^2 D^3 C + d^4 D^2 C^2)$. (d) One step of computing a generic second order measurement on a density matrix represented by an MPDO. The highest order cost is $O(d^2 K C^2 D^5+d^3 L C^3 D^4)$.}
    \label{fig:mpo_mpdo_update_comparisons}
\end{figure}

\subsubsection{Comparing MPO and MPDO}

\begin{figure}
    \centering
    \begin{tikzpicture}
        \begin{scope}
            \node[anchor=north west,inner sep=0] (image_a) at (0,-0.2)
            {\includegraphics[width=0.9\columnwidth]{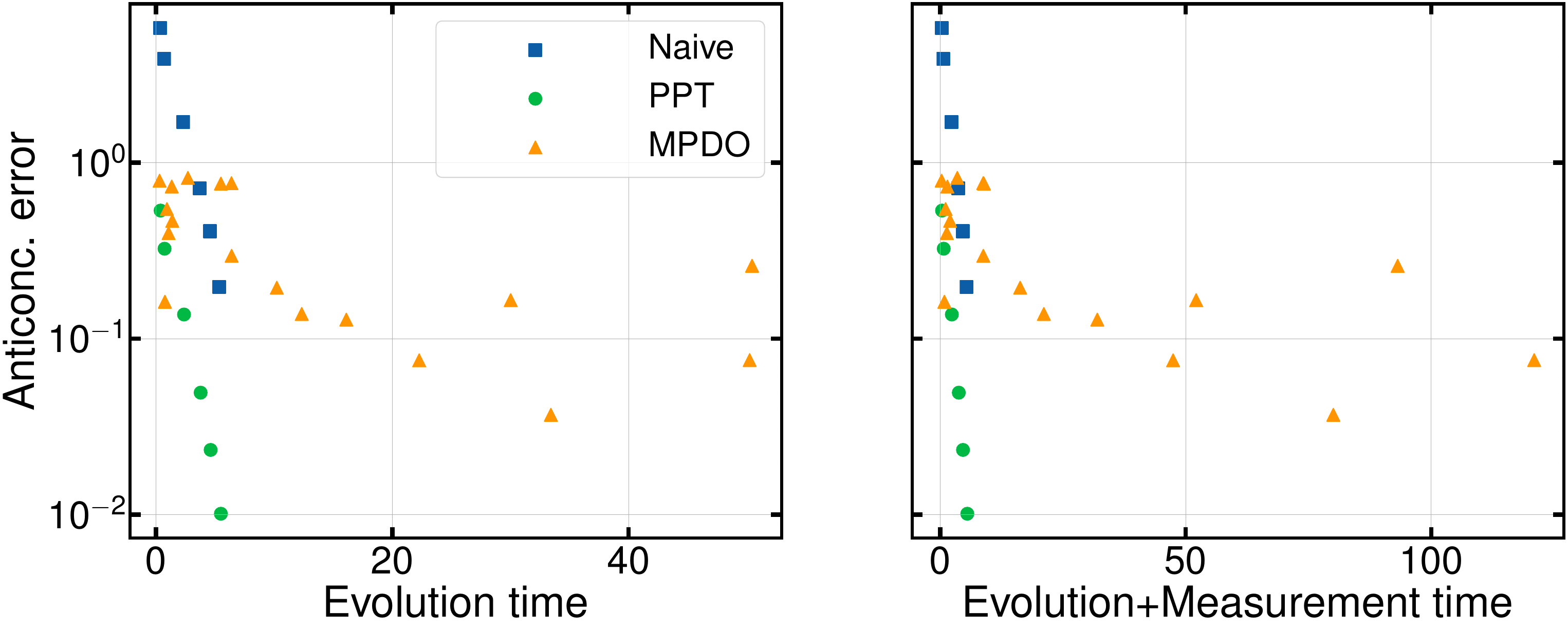}};
        \end{scope}
    \end{tikzpicture}
    \caption{The accuracy of the anticoncentration measurement of the 8-site RCS system with $\gamma_{diss}=0.1$, using each algorithm (naive updates without PPT, updates with PPT, and MPDO), for different maximum bond and inner dimensions. On the x-axis, we graph the time cost of the algorithm to update a 2-layer period of the circuit, assuming no measurements (\textit{left}) and one measurement at the end of the 20th period (\textit{right}). All times and accuracies are given as averages over the last 5 periods of the system. Naive and PPT updates have bond dimension 16-192 (a bond dimension otf 256 does not require truncations) and MPDOs have bond dimension 8-64 with inner dimension 1-16.}
    \label{fig:accuracy_time_comparisons}
\end{figure}

For many measurements, an MPDO at similar bond dimension and modest inner dimension generally gives more accurate measurements than either MPO-based approaches.
However, the MPDO is generally more costly to update, given similar bond dimensions and inner dimensions. From Fig.~\ref{fig:mpo_mpdo_update_comparisons}a-b, the relative cost of updating the MPDO to the MPO, assuming $D$ is the largest variable by far, is of order $\frac{C^2 K}{d^5}$. For a generic two-site gate, $C = O(d^2)$, so the relative cost of updating the MPDO is approximately $\frac{K}{d}$. While an MPDO with $K=d$ can theoretically represent any density matrix (Appendix~\ref{app:mpdo_rule}), at more realistic bond dimensions we would expect $K$ to be greater than $d$ in order to represent the system properly.

A more significant penalty arises when we attempt to take any nonlinear function of the system that requires the squared tensor product of the density matrix. These functions include the purity and anticoncentration. From Fig.~\ref{fig:mpo_mpdo_update_comparisons}c-d, the cost of computing such a measurement is far more expensive for the MPDO than the MPO, by roughly two factors of the bond dimension. This difference is so great that it can potentially outweigh the cost of creating the time-evolved MPDO in the first place, making otherwise easily simulable systems far more difficult to measure.

To quantify this comparison, we plot the accuracy versus wall-clock time for identical simulations using Naive, PPT, and MPDO approaches (see Fig.~\ref{fig:accuracy_time_comparisons}) at various different bond dimensions (and inner dimensions for MPDO) for time evolution both with and without a final measurement of the anticoncentration.  We find that for this particular example, at a fixed accuracy, the PPT algorithm is faster then the other approaches in the large plurality of cases. This gives evidence that (especially for non-linear observables) the PPT is the best choice amongst the various options.

\section{A Candidate Circuit: QAOA Iteration}

We study the ability of the quantum system to use QAOA to generate the ground state of a transverse field Ising model,
\begin{gather}
    H_{TFIM} = -J\underbrace{(\sum_{\langle i j \rangle} S^z_i S^z_j + b\sum_i S^z_i)}_{H_z} - g\underbrace{\sum_i S^x_i}_{H_t}\label{eq:tfim_hamiltonian}
\end{gather}
as one needs to do in a variational quantum eigensolver. Specifically, we will consider a system with a small bias field $b = 0.2$ and large, near-critical transverse field $g=1.2J$.

The QAOA ansatz \cite{Ho2019} alternates $k$ times between time evolution of the various terms of Hamiltonian $H_z$ and $H_t$ with weights $\alpha_k$ and $\beta_k$ respectively. With exact noiseless gates the state $|\psi\rangle$ would evolve under the quantum circuit as
\begin{align}
    |\psi\rangle \rightarrow &\prod_{k=1}^K e^{i \alpha_k H_z} e^{i \beta_k H_t} |\psi_0\rangle \label{eq:tfim_model}.
\end{align}
The final energy of $|\psi\rangle$ is measured with Hamiltonian in Equation~\eqref{eq:tfim_hamiltonian}. 

\begin{figure}
    \centering
    \begin{tikzpicture}
        \begin{scope}
            \node[anchor=north west,inner sep=0] (image_a) at (0,0)
            {\includegraphics[width=0.7\columnwidth]{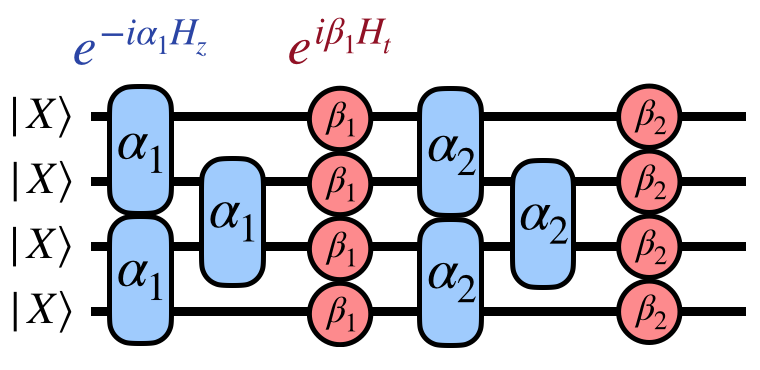}};
            \node [anchor=north west] (note) at (-0.5,0) {\small{\textbf{a)}}};
        \end{scope}
    \end{tikzpicture}
    \begin{tikzpicture}
        \begin{scope}
            \node[anchor=north west,inner sep=0] (image_a) at (0,0)
            {\includegraphics[width=0.9\columnwidth]{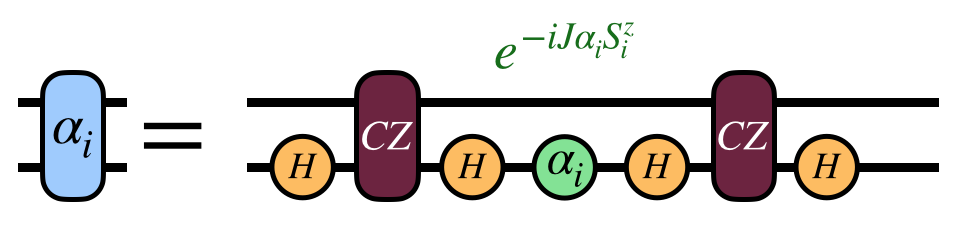}};
            \node [anchor=north west] (note) at (-0.5,-0.3) {\small{\textbf{b)}}};
        \end{scope}
    \end{tikzpicture}
    \caption{(a): A circuit diagram of the first two layers of a QAOA iteration for a 4-qubit system. In each layer, blocks of the entangling component $H_z$ of the TFIM Hamiltonian are applied to the circuit for each nearest-neighbor pair, followed by single site operators representing the transverse field component $H_t$. (b): Each two-site block is composed of two CZ gates simulated on the pulse level, as well as multiple single site operators, following Equation~\eqref{eq:HZ_block_representation}.}
    \label{fig:qaoa_circuti}
\end{figure}

A QAOA iteration therefore consists of single site $e^{i \beta_k H_t}$ gates and two site $e^{i \alpha_k H_z}$ gates. We assume the single site gates are comparably easy to perform in a noiseless manner and focus on simulating the two site gates. We can create an arbitrary $e^{i \alpha_k H_z}$ operation from pairs of CZ gates as follows:
\begin{gather}
    \text{CNOT}_{ij} = I_i \otimes \text{Had}_j \cdot  \text{CZ}_{ij} \cdot I_i \otimes \text{Had}_j\\
    e^{-iJ\alpha_k S^z_i S^z_j} = \text{CNOT}_{ij} \cdot  I_i \otimes e^{-iJ\alpha_k S^z_j } \cdot \text{CNOT}_{ij}\label{eq:HZ_block_representation} \\
    e^{-iJ\alpha_k H_z } = \bigg(\prod_{\langle ij\rangle} e^{-iJ\alpha_k S^z_i S^z_j }\bigg) \prod_l e^{-ibJ \alpha_k S^z_l }. \label{eq:HZ_representation}
\end{gather}
Each CZ gate is a copy of the one constructed by the LME
, Equation~\eqref{eq:LME}. Since all terms in the final product (\ref{eq:HZ_representation}) commute with each other, our simulation of the QAOA iteration is as follows (Fig.~\ref{fig:qaoa_circuti}). We first assemble each two-site term in Equation~\eqref{eq:HZ_representation} using Equation~\eqref{eq:HZ_block_representation}. We then apply each term sequentially to the circuit, as well as the single site terms. Finally we apply the layer of transverse field gates $e^{i \beta_k H_t}$ for each site $l$.

We simulate the QAOA iteration using realistic noise sources and Rydberg blockade 
using a vectorized MPO with Purity-Preserving Truncation and a maximum bond dimension of 768, over different system sizes, with $K=8$ layers, with the initial state
$|\psi_0\rangle = \bigotimes_{i=1}^N |+_X\rangle$ as a product of positive eigenstates of the Pauli X operator. 
The most expensive calculations were run on the Argonne National Laboratory Computing Resource Center (LCRC) using the distributed-memory Cyclops Tensor Framework (CTF, Appendix~\ref{app:distributed_memory}). 
\setlength{\tabcolsep}{16pt}
\begin{table}[]
    \centering
    \begin{tabular}{|c| c c|} \hline
        \thead{$k$} & \thead{$\alpha_k$} & \thead{$\beta_k$} \\ \hline
        1 & 0.11076513 & 0.75428624 \\ \hline
        2 & 0.2771272 & 0.73016842 \\ \hline
        3 & 0.36282021 & 0.7096901 \\ \hline
        4 & 0.40618171 & 0.68739375 \\ \hline
        5 & 0.43256044 & 0.65733871 \\ \hline
        6 & 0.44492256 & 0.60978220 \\ \hline
        7 & 0.42887337 & 0.51570246 \\ \hline
        8 & 0.3225842 & 0.19145101 \\ \hline
    \end{tabular}
    \caption{Classically pre-optimized time evolution parameters from Eq.~\eqref{eq:tfim_model} for a 10-site, 8-layer TFIM.}
    \label{tab:tfim_params}
\end{table}

\begin{table}
    \centering
    \begin{tabular}{|c||c|c|}\hline 
    $N$ & $\varepsilon_{\text{QAOA}}$ & $\varepsilon_{\text{True GS}}$\\ \hline
    20 & -1.51726 & -1.51836\\ 
    40 & -1.52980 & -1.53102\\
    60 & -1.53398 & -1.53524\\
    80 & -1.53607 & -1.53735 \\ \hline
    \end{tabular}
    \caption{Comparison of the noiseless energy density returned by near-optimized QAOA and the true ground state energy density of the TFIM with the same Hamiltonian parameters, over different $N$. This true ground state energy was computed through DMRG, with a bond dimension of 250.}
    \label{tab:tfim_energy_comparison}
\end{table}

\begin{figure}
    \centering
    \begin{tikzpicture}
        \begin{scope}[xshift=0.0\columnwidth]
            \node[anchor=north west,inner sep=0] (image_a) at (0.0,-0.12)
            {\includegraphics[width=0.44\columnwidth]{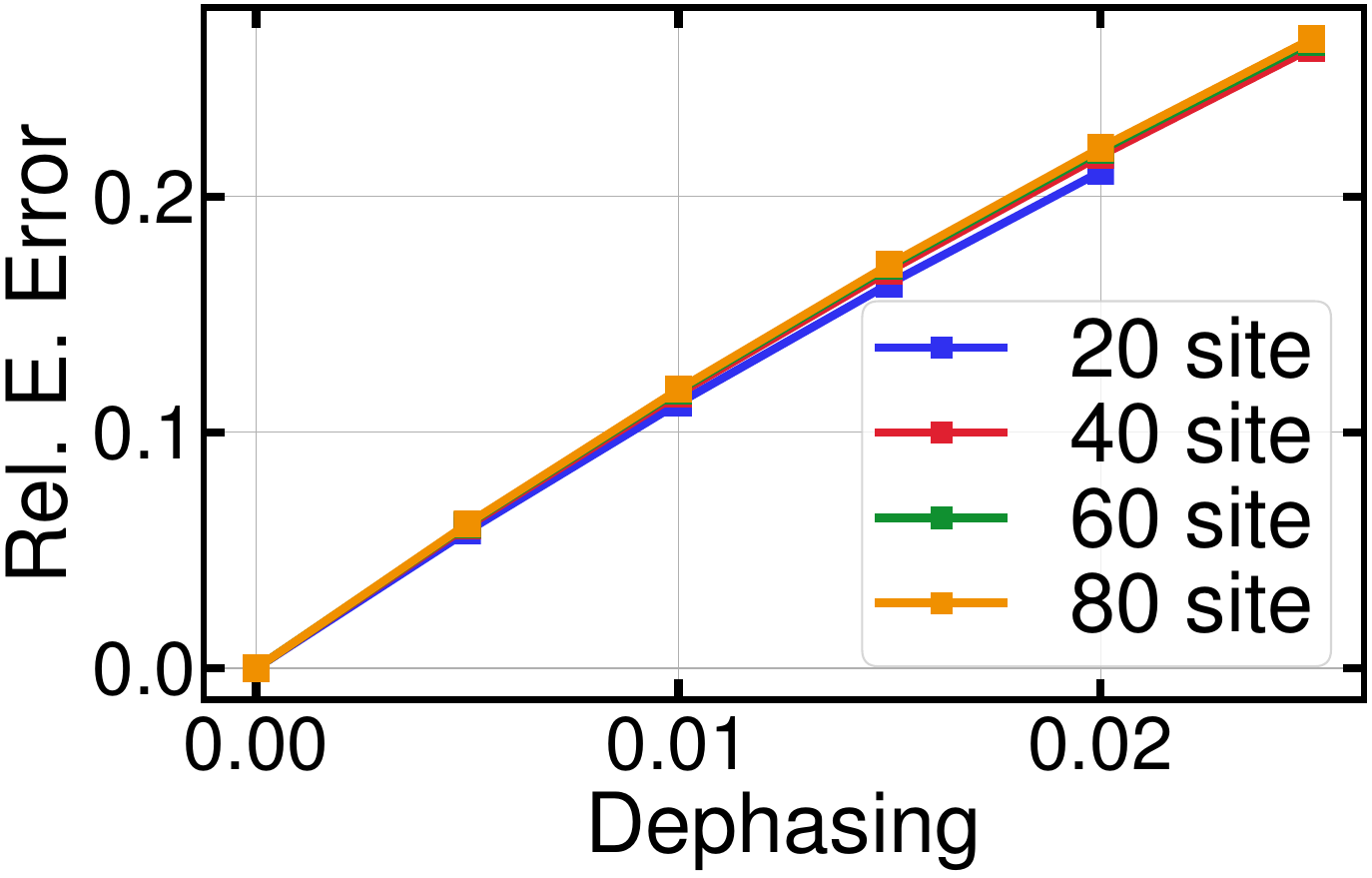}};
            \node [anchor=north] (note) at (-0.2,-0.2) {\small{a)}};
        \end{scope}
        \begin{scope}[xshift=0.5\columnwidth]
            \node[anchor=north west,inner sep=0] (image_a) at (0.0, -0.12)
            {\includegraphics[width=0.44\columnwidth]{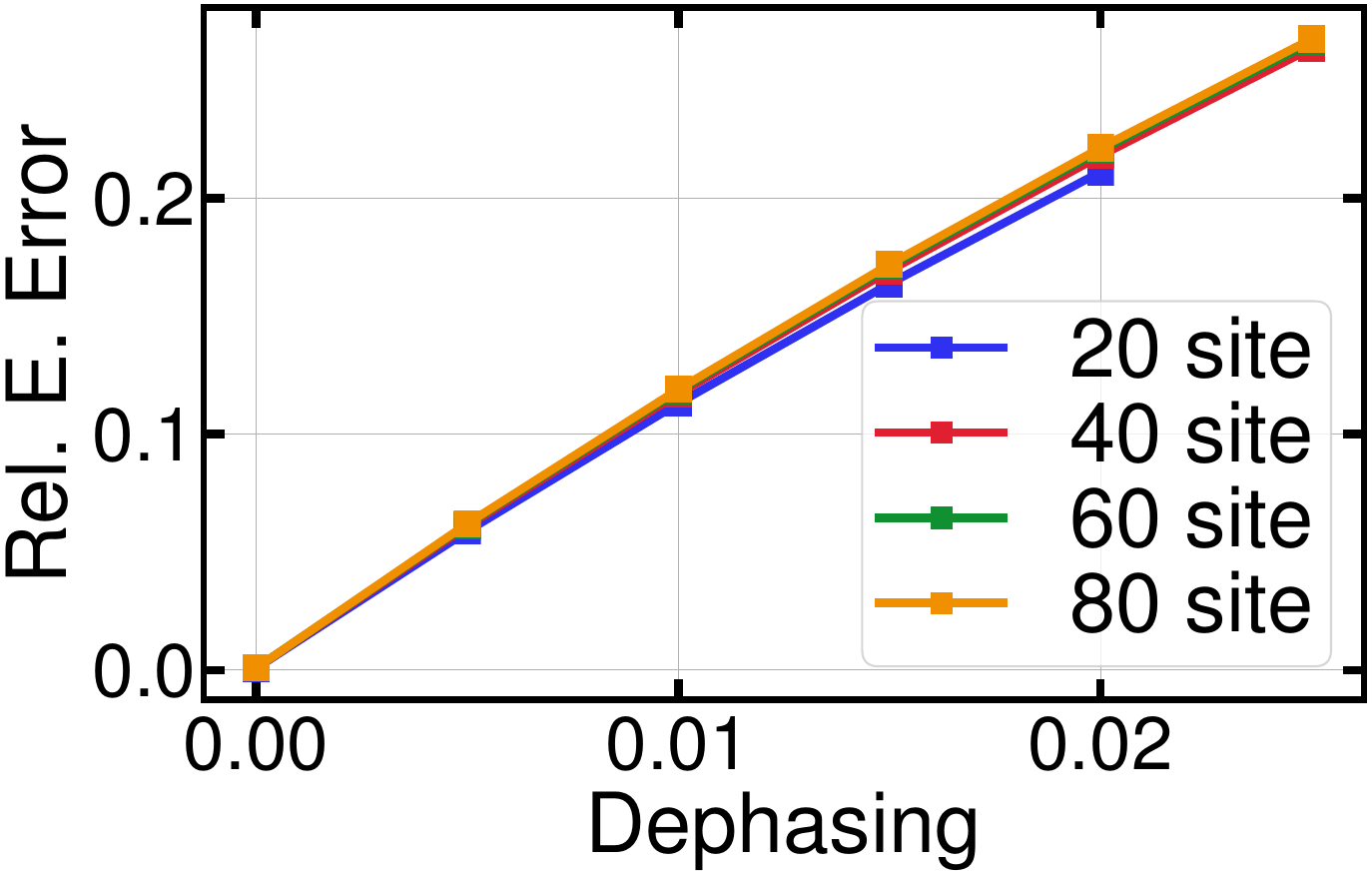}};
            \node [anchor=north] (note) at (-0.2,-0.2) {\small{b)}};
        \end{scope}
    \end{tikzpicture}
    \vspace*{-0.1cm}
    \caption{Comparison of the relative energy error over dephasing of a near-optimized QAOA on the TFIM, using either the near-optimized energy (\textit{a}) or the true ground state energy (\textit{b}) as a baseline. The two graphs are very similar because the energy differences are small. }
    \label{fig:stat_mech_process}
\end{figure}

For each system size, we use the same parameters $\alpha_j, \beta_j$ optimized classically on a 10-site system (Table \ref{tab:tfim_params}). On a larger number of sites, these parameters will produce an energy per site that is within $1.4\times 10^{-4}$ of the true ground state energy (Table \ref{tab:tfim_energy_comparison}). This difference is negligible compared to those caused by noise or change in parameters (Fig.~\ref{fig:stat_mech_process}).

\subsection{QAOA Energy in the Presence of Noise}

\begin{figure}
    \centering
    \begin{tikzpicture}
        \begin{scope}
            \node[anchor=north west,inner sep=0] (image_a) at (0,0)
            {\includegraphics[width=0.46\columnwidth]{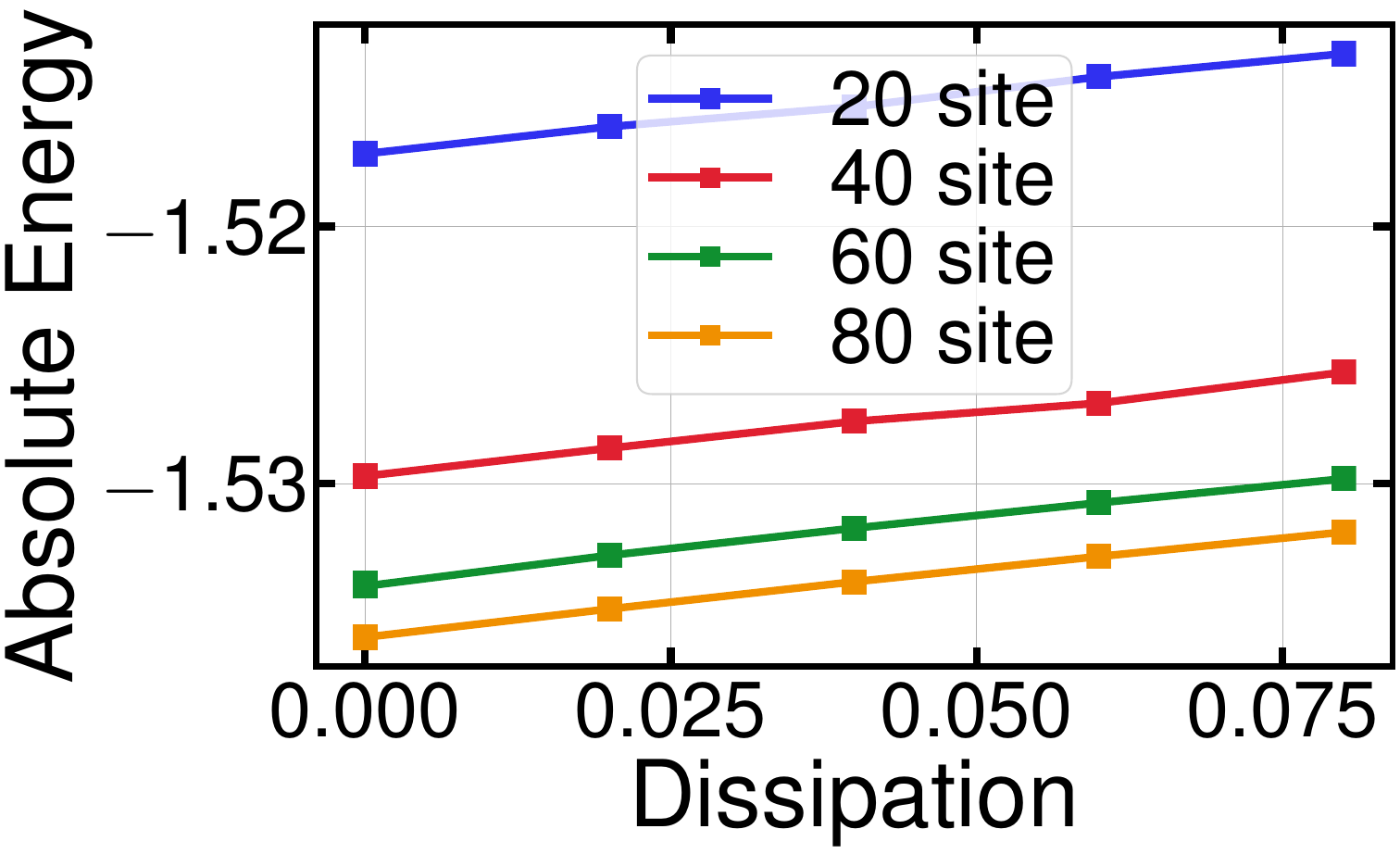}};
            \node [anchor=north west] (note) at (-0.4,0.35) {\footnotesize{\textbf{a)}}};
        \end{scope}
        
        \begin{scope}[xshift=0.51\columnwidth]
            \node[anchor=north west,inner sep=0] (image_b) at (0,0)
            {\includegraphics[width=0.46\columnwidth]{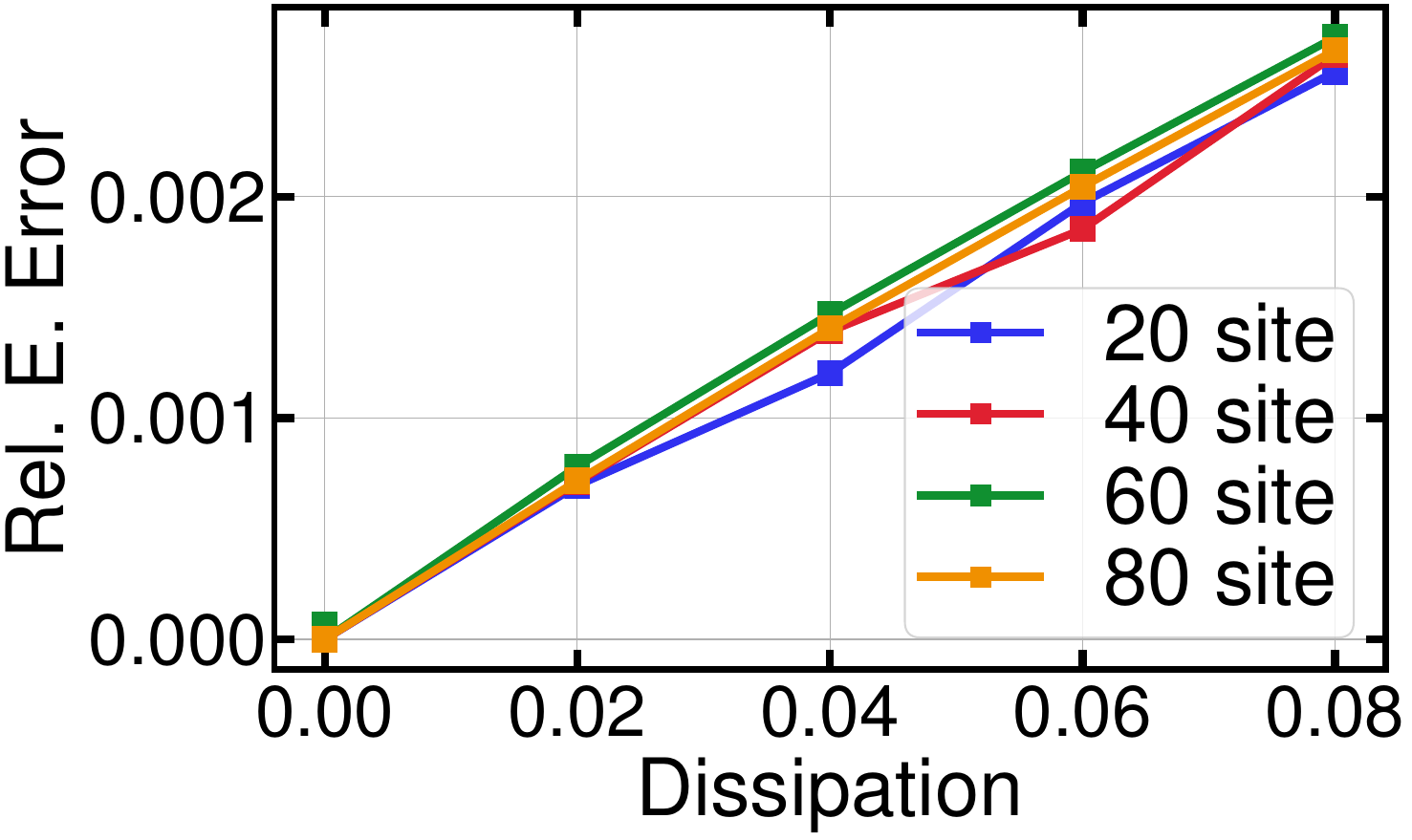}};
            \node [anchor=north west] (note) at (-0.33,0.35) {\footnotesize{\textbf{b)}}};
        \end{scope}
    \end{tikzpicture}
    \vspace*{0.01\textwidth}
    \begin{tikzpicture}
        \begin{scope}[xshift=0.0\columnwidth]
            \node[anchor=north west,inner sep=0] (image_d) at (0,-0.12)
            {\includegraphics[width=0.46\columnwidth]{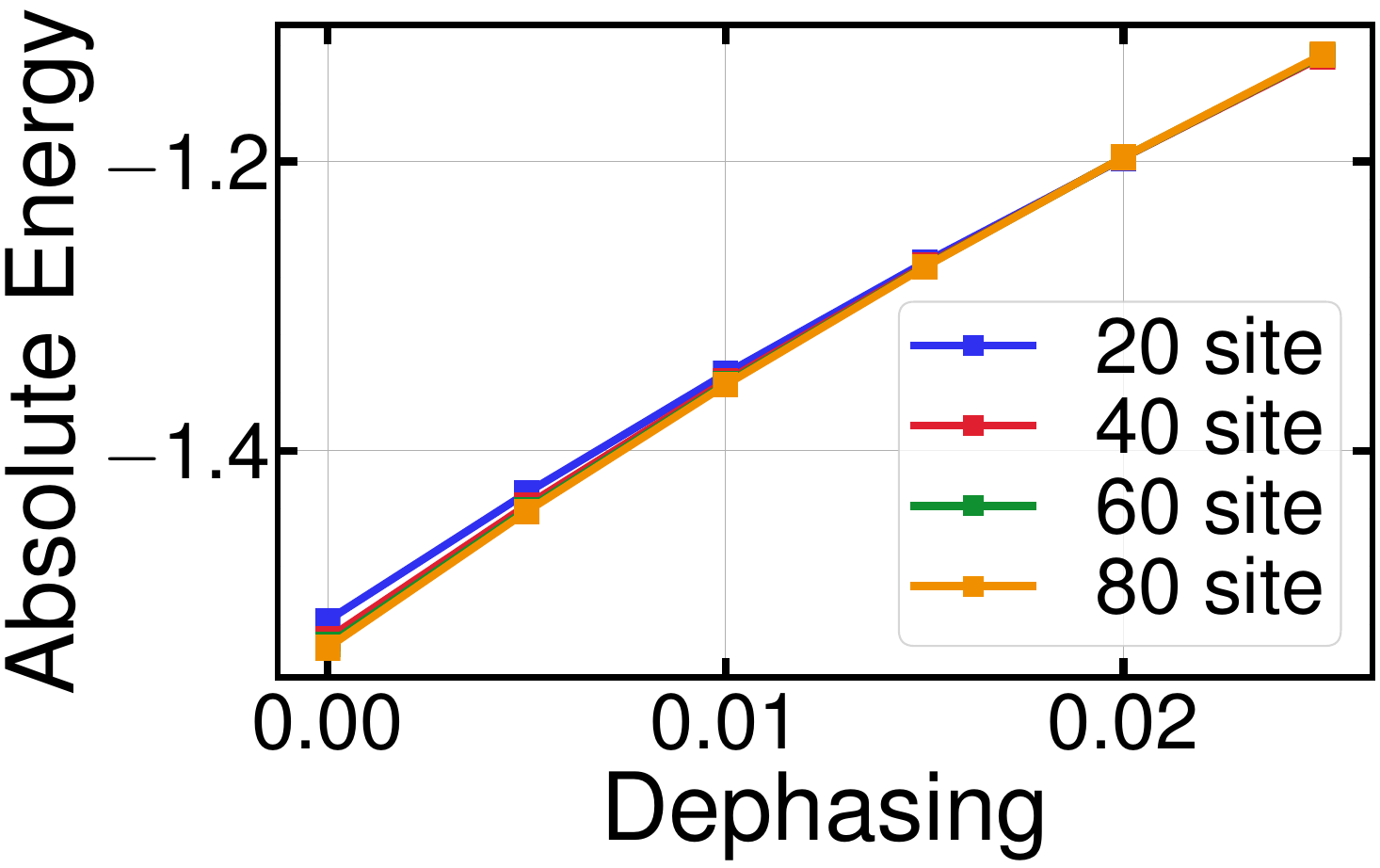}};
            \node [anchor=north west] (note) at (-0.4,0.20) {\footnotesize{\textbf{c)}}};
        \end{scope}
        \begin{scope}[xshift=0.51\columnwidth]
            \node[anchor=north west,inner sep=0] (image_d) at (0,-0.12)
            {\includegraphics[width=0.46\columnwidth]{plots/tfim_relative_energy_over_site_qdeph_ideal.pdf}};
            \node [anchor=north west] (note) at (-0.33,0.20) {\footnotesize{\textbf{d)}}};
        \end{scope}
    \end{tikzpicture}
    \begin{tikzpicture}
        \begin{scope}[xshift=0.0\columnwidth]
            \node[anchor=north west,inner sep=0] (image_c) at (0,0)
            {\includegraphics[width=0.46\columnwidth]{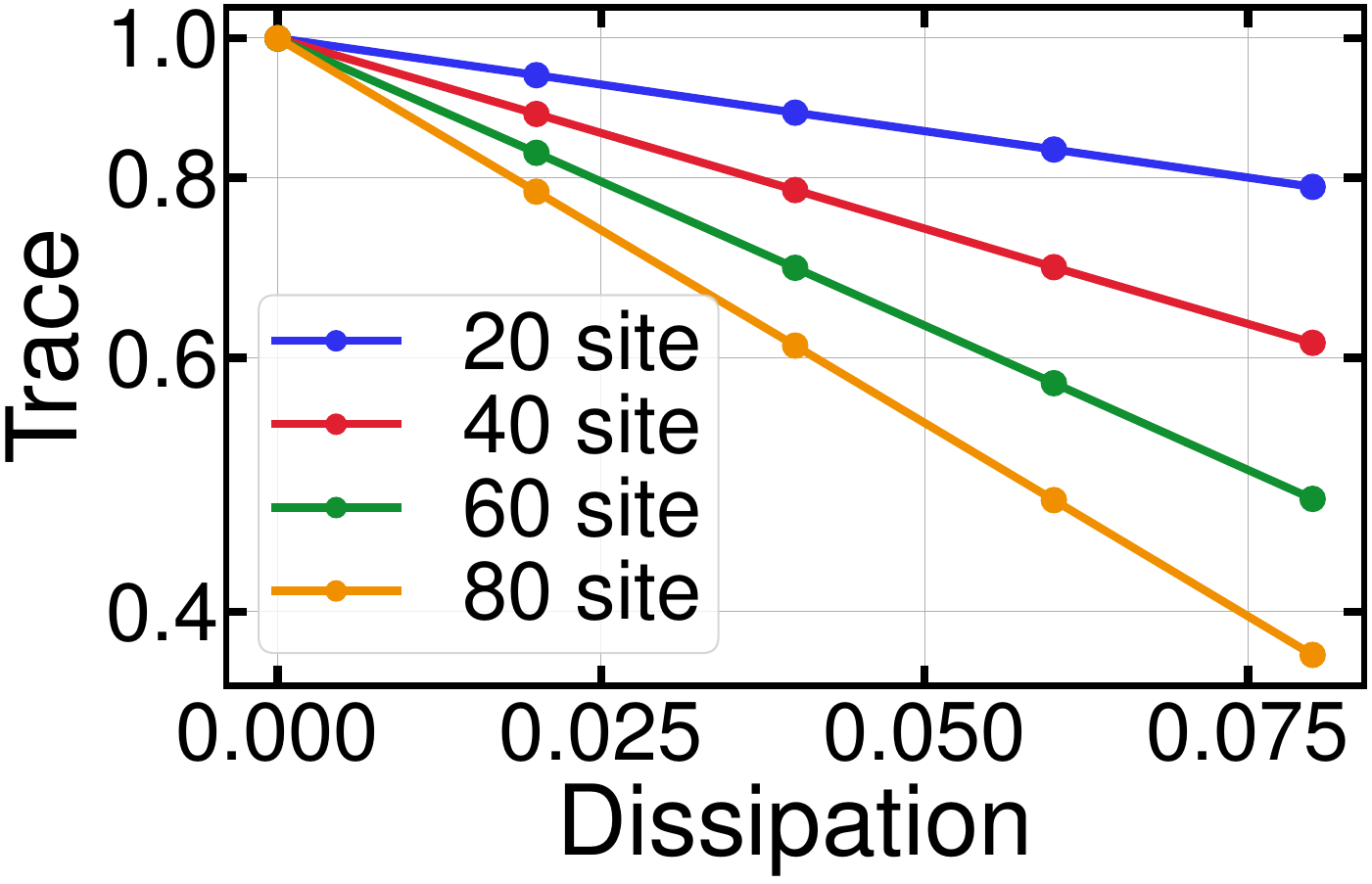}};
            \node [anchor=north west] (note) at (-0.3,0.2) {\footnotesize{\textbf{e)}}};
        \end{scope}
        \begin{scope}[xshift=0.51\columnwidth]
            \node[anchor=north west,inner sep=0] (image_c) at (0,0)
            {\includegraphics[width=0.46\columnwidth]{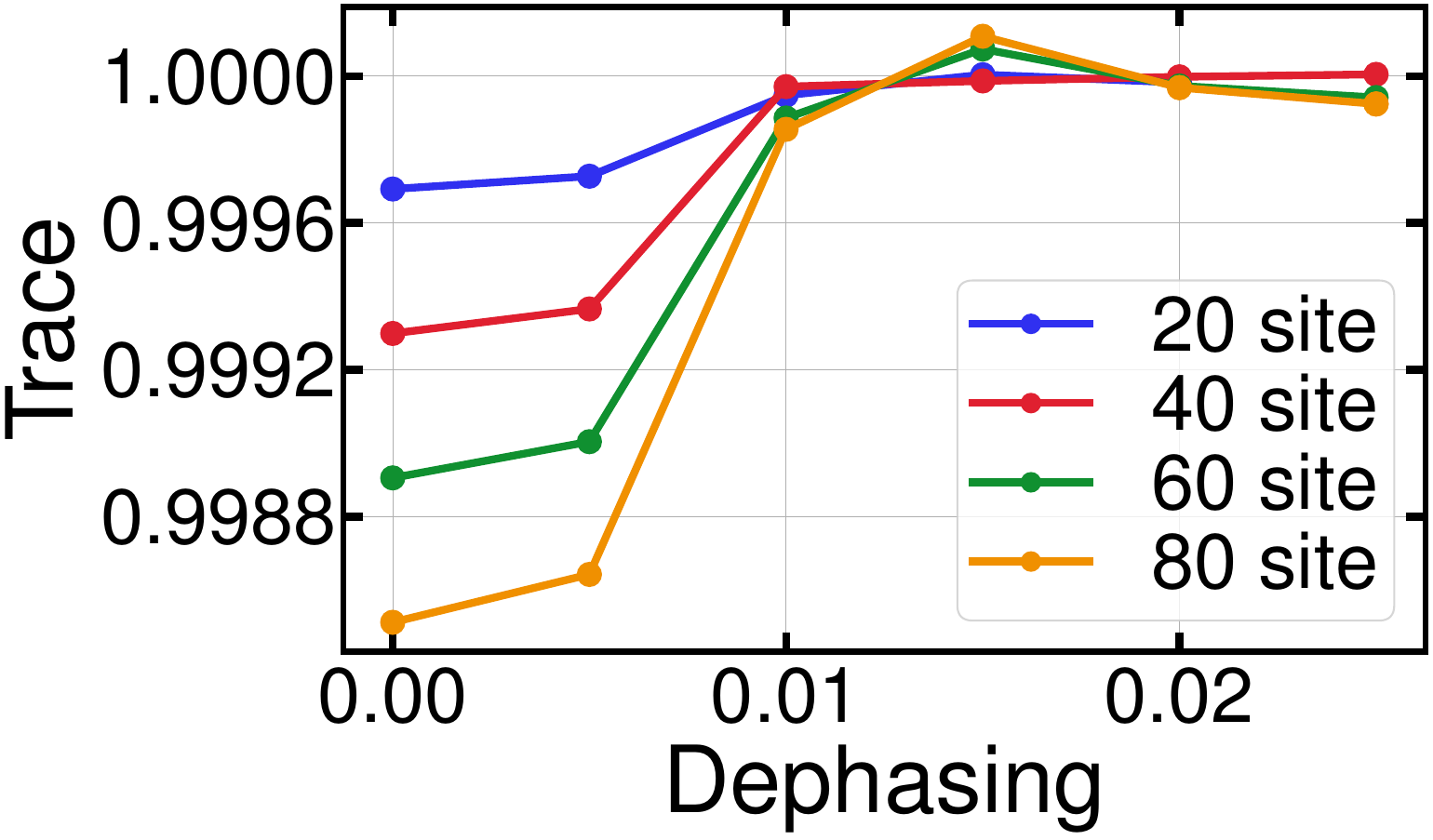}};
            \node [anchor=north west] (note) at (-0.3,0.2) {\footnotesize{\textbf{f)}}};
        \end{scope}
    \end{tikzpicture}
    \caption{Final energy of a QAOA iteration with optimized parameters under noise. (a): Absolute energy per site over Rydberg atom dissipation. The energy has very little dependence on dissipation.    (b): Relative energy error per site $(E-E_0)/E_0$, where $E_0$ is the energy per site of the noiseless circuit. The relative energy decreases with dissipation but has very little dependence on the system size. (c,d): Absolute energy (c) and relative energy error (d) per site over qubit dephasing; the system size dependence is very small. (e,f): Trace of the qubit component of the density matrix over system size and dissipation (e) and dephasing (f). }
    \label{fig:final_energy_over_noise}
\end{figure}

\begin{figure}
    \centering
    \begin{tikzpicture}
        \begin{scope}
            \node[anchor=north west,inner sep=0] (image_a) at (0,0)
            {\includegraphics[width=0.47\columnwidth]{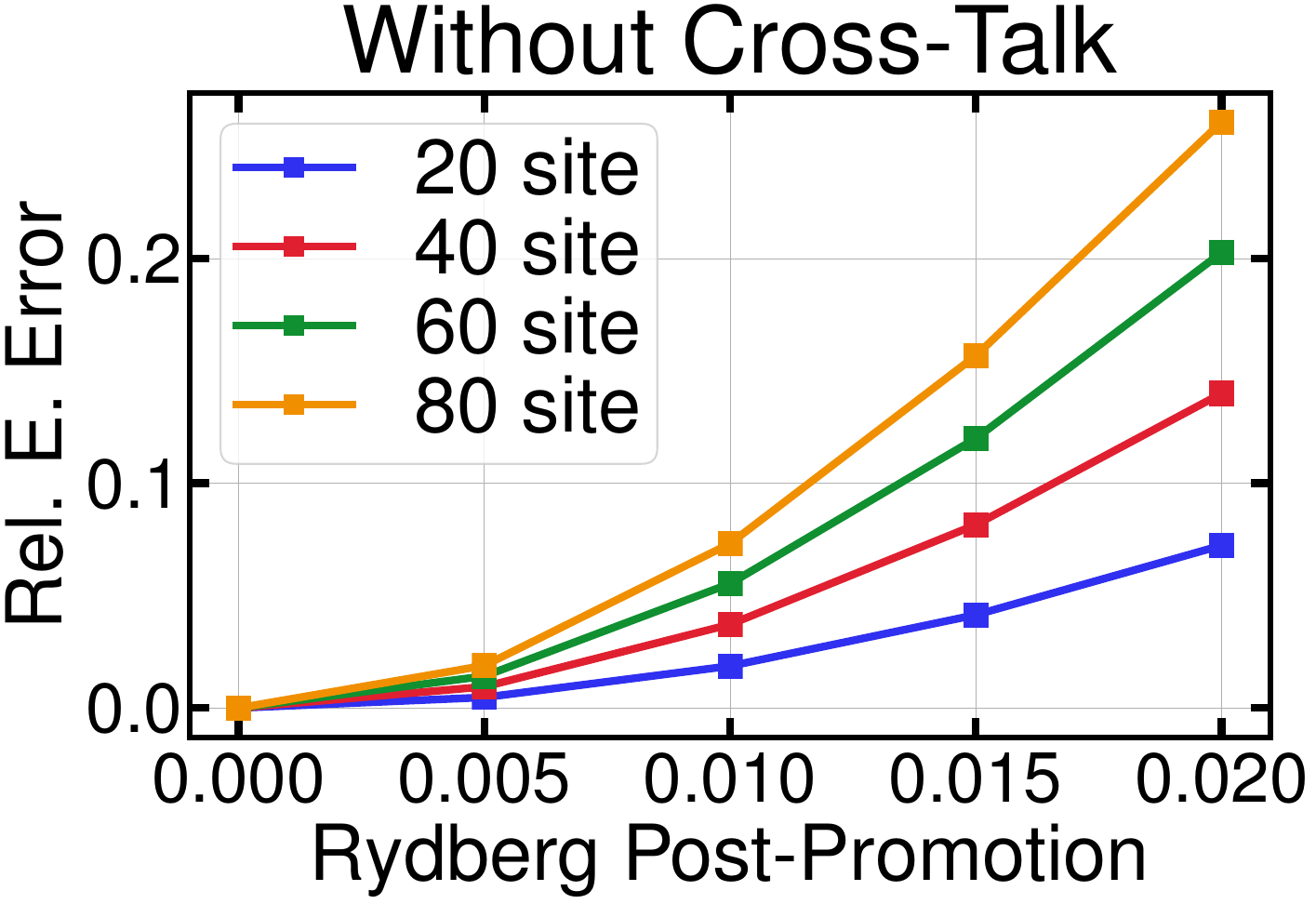}};
            \node [anchor=north west] (note) at (-0.35,0.1) {\footnotesize{\textbf{a)}}};
        \end{scope}
        \begin{scope}[xshift=0.51\columnwidth]
            \node[anchor=north west,inner sep=0] (image_b) at (0,0)
            {\includegraphics[width=0.47\columnwidth]{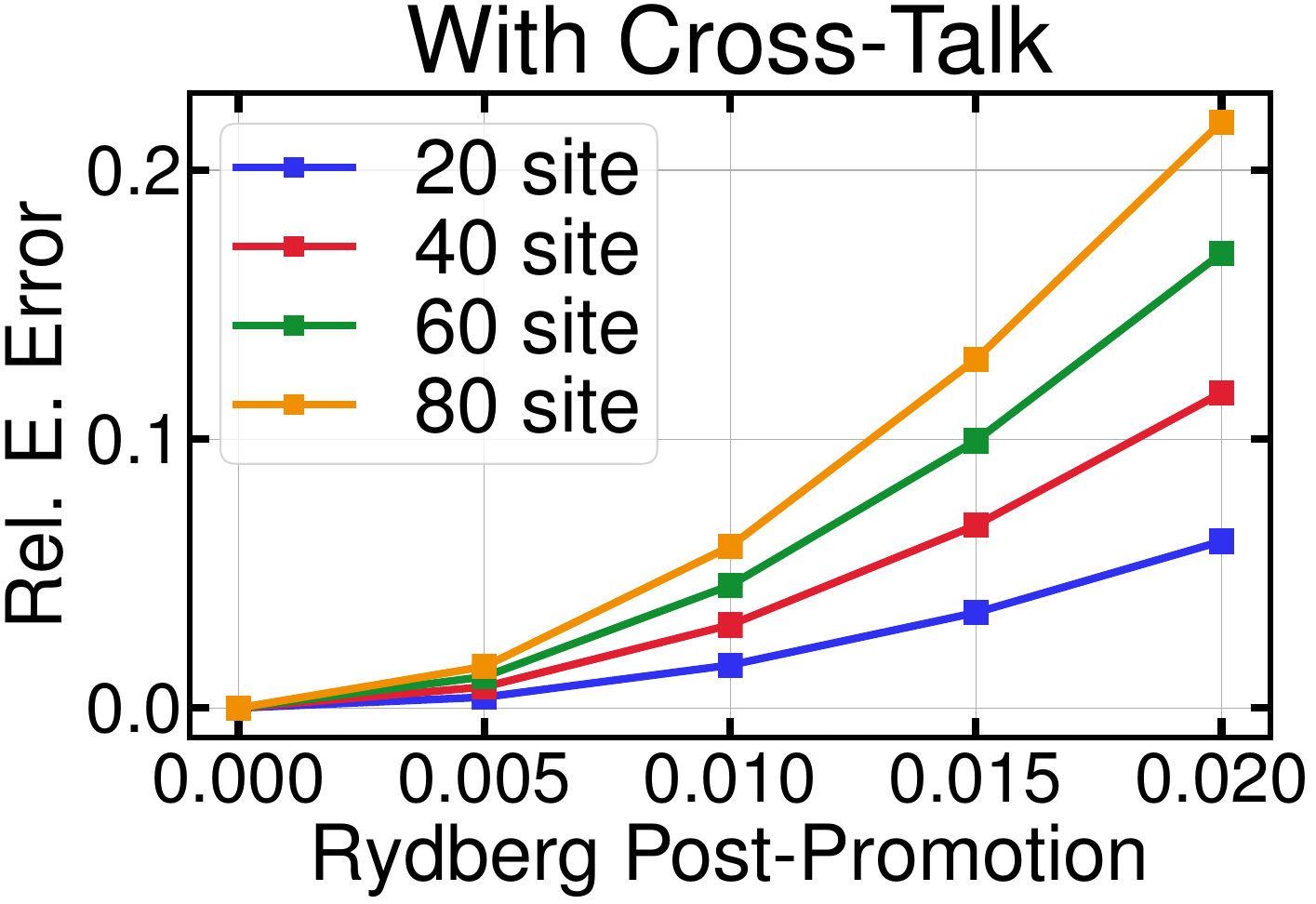}};
            \node [anchor=north west] (note) at (-0.35,0.1) {\footnotesize{\textbf{b)}}};
        \end{scope}
    \end{tikzpicture}
    \caption{Relative energy per site over Rydberg atom post-promotion, both without (a) and with (b) crosstalk. Introducing crosstalk in a sequential pattern of gates creates an error that is dependent on both post-promotion level and system size.}
    \label{fig:final_energy_over_rpp}
\end{figure}

We investigate the performance of the QAOA iteration over two types of incoherent noise: dissipation of the Rydberg population and dephasing of the individual qubit states (Fig.~\ref{fig:final_energy_over_noise}). We first fix the Rydberg blockade at $2\pi \times 60$MHz and adjust the dissipation from 0 to 0.1.   We note that the accuracy of the energy gets worse as we increase both the dissipation and dephasing although is more strongly affected by dephasing.  Interestingly, the energy per site has almost no system size dependence suggesting that the errors due to both noise sources do not accumulate as the system gets larger. This suggests that a neutral atom experiment could successfully measure the QAOA energy for large systems even in the face of significant dissipation. Unfortunately, while the energy eventually measured by the circuit is independent of system size, the number of circuit iterations required to measure the energy will increase for larger systems, due to the decrease in the qubit population of the density matrix caused by Rydberg atom dissipation.  
This creates a larger chance of errant population in the Rydberg atom and dark states, which would make the energy measurement invalid. For large enough system sizes, this makes the energy difficult to evaluate, even if the energy would theoretically be accurate if one were lucky enough to measure it.  We can see this effect by looking at the trace of the Rydberg/dark state components of the density matrix and its deviation from 0 as seen in Fig.~\ref{fig:final_energy_over_noise}e. The accumulation of non-qubit population is a site-wise independent process -- the overall qubit trace of the density matrix is an exponential in dissipation and system size,
\begin{gather}
    \tr_q(\rho) = e^{-0.1556\gamma_{diss} N}.
\end{gather}
Dephasing does not contribute to these trace errors as the corresponding time evolution operator does not affect the dark state population. As can be seen in Fig.~\ref{fig:final_energy_over_noise}f, the trace of a system under only dephasing has minimal errors. All fluctuations in the trace are due to truncation errors in the MPO, as the Linbladian itself does not have any trace-affecting terms - this is evident in how the trace error only reaches its maximum when the dephasing is as low as possible.
We also simulated the influence of a possible coherent error within the system, that of unwanted Rydberg atom crosstalk. The Rydberg blockade can be problematic if there is a residual Rydberg atom population on sites neighboring those where a gate is being applied, as they can interfere with the dynamics of the gate. These residual populations are normally too small to have an observable effect during the normal execution of a circuit, so we introduce a post-promotion term
\begin{gather}
    \hat{A}_{PP} = e^{-i\delta_{PP}(|1\rangle\langle r| + |r\rangle \langle 1|}
\end{gather}
to increase the Rydberg atom population after a gate is applied. Fig.~\ref{fig:final_energy_over_rpp} shows the effects on the relative energy error of this post-promotion, with and without any crosstalk between sites. Post-promotion introduces an error in the relative energy per site that increases with both post-promotion strength $\delta_{PP}$ and system size. However, the effects of introducing crosstalk are not only minor, the crosstalk even appears to cause a slight improvement in the energy (Fig.~\ref{fig:final_energy_over_rpp}b).

\subsection{QAOA Parameter Optimization over errors}

\begin{figure*}
    \centering
    \begin{tikzpicture}
        \begin{scope}
            \node[anchor=north west,inner sep=0] (image_a) at (0,0)
            {\includegraphics[width=0.63\columnwidth]{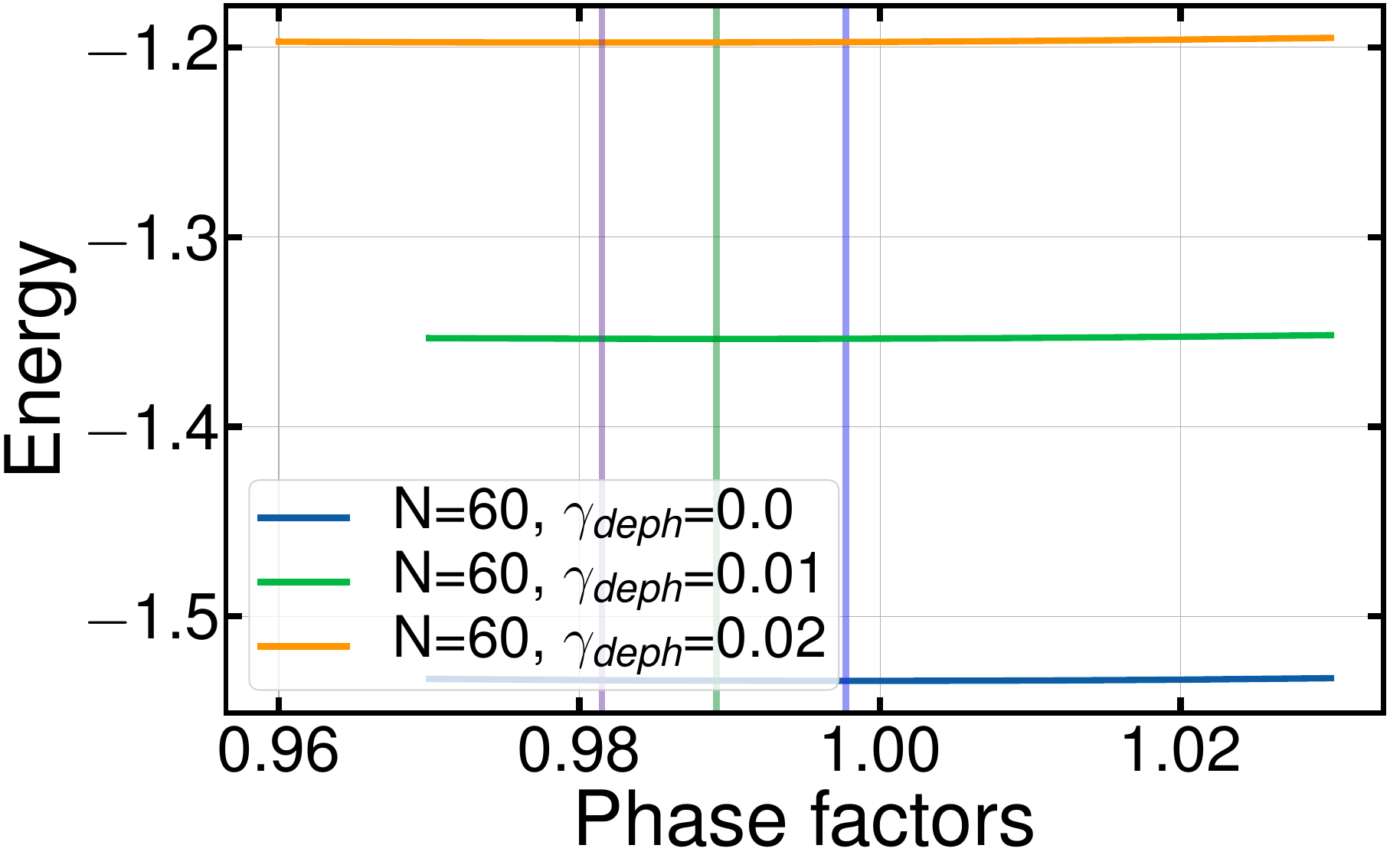}};
            \node [anchor=north west] (note) at (-0.1,0.1) {\footnotesize{\textbf{a)}}};
        \end{scope}
        \begin{scope}[xshift=0.67\columnwidth]
            \node[anchor=north west,inner sep=0] (image_b) at (0,0)
            {\includegraphics[width=0.63\columnwidth]{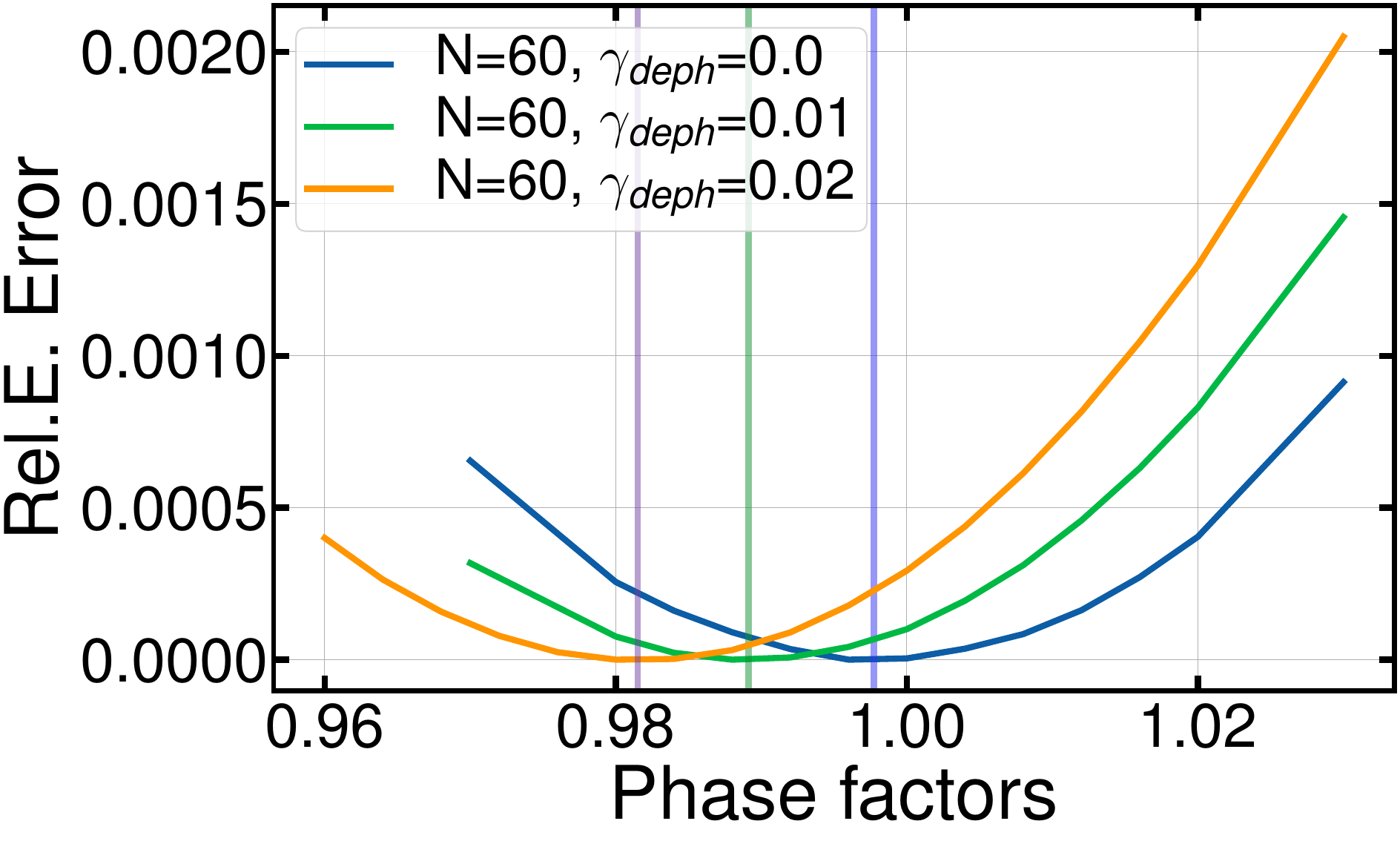}};
            \node [anchor=north west] (note) at (-0.2,0.1) {\footnotesize{\textbf{b)}}};
        \end{scope}
        \begin{scope}[xshift=1.32\columnwidth]
            \node[anchor=north west,inner sep=0] (image_a) at (0,0)
            {\includegraphics[width=0.63\columnwidth]{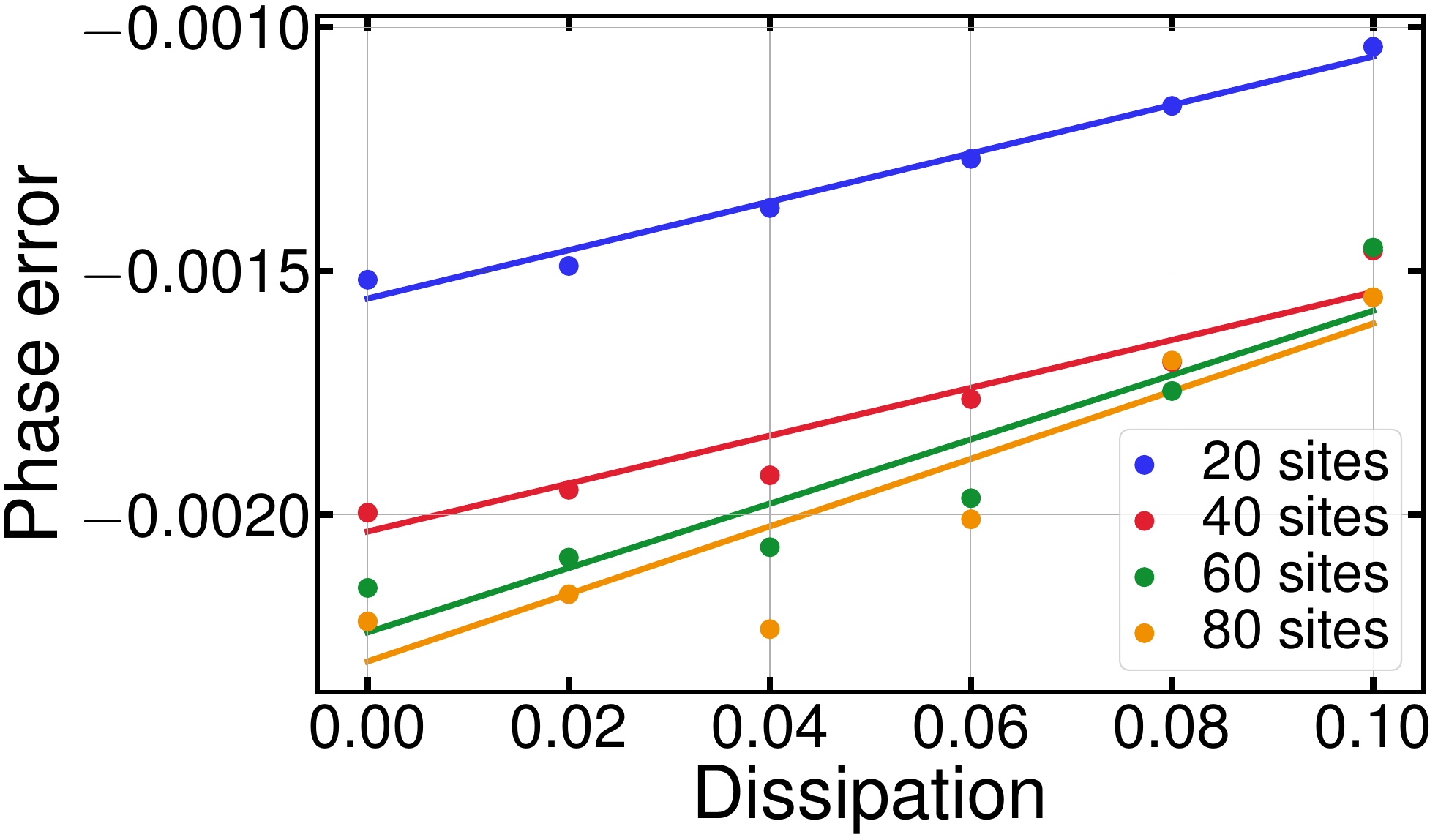}};
            \node [anchor=north west] (note) at (-0.1,0.1) {\footnotesize{\textbf{c)}}};
        \end{scope}
    \end{tikzpicture}
    \begin{tikzpicture}
        \begin{scope}
            \node[anchor=north west,inner sep=0] (image_b) at (0,0)
            {\includegraphics[width=0.63\columnwidth]{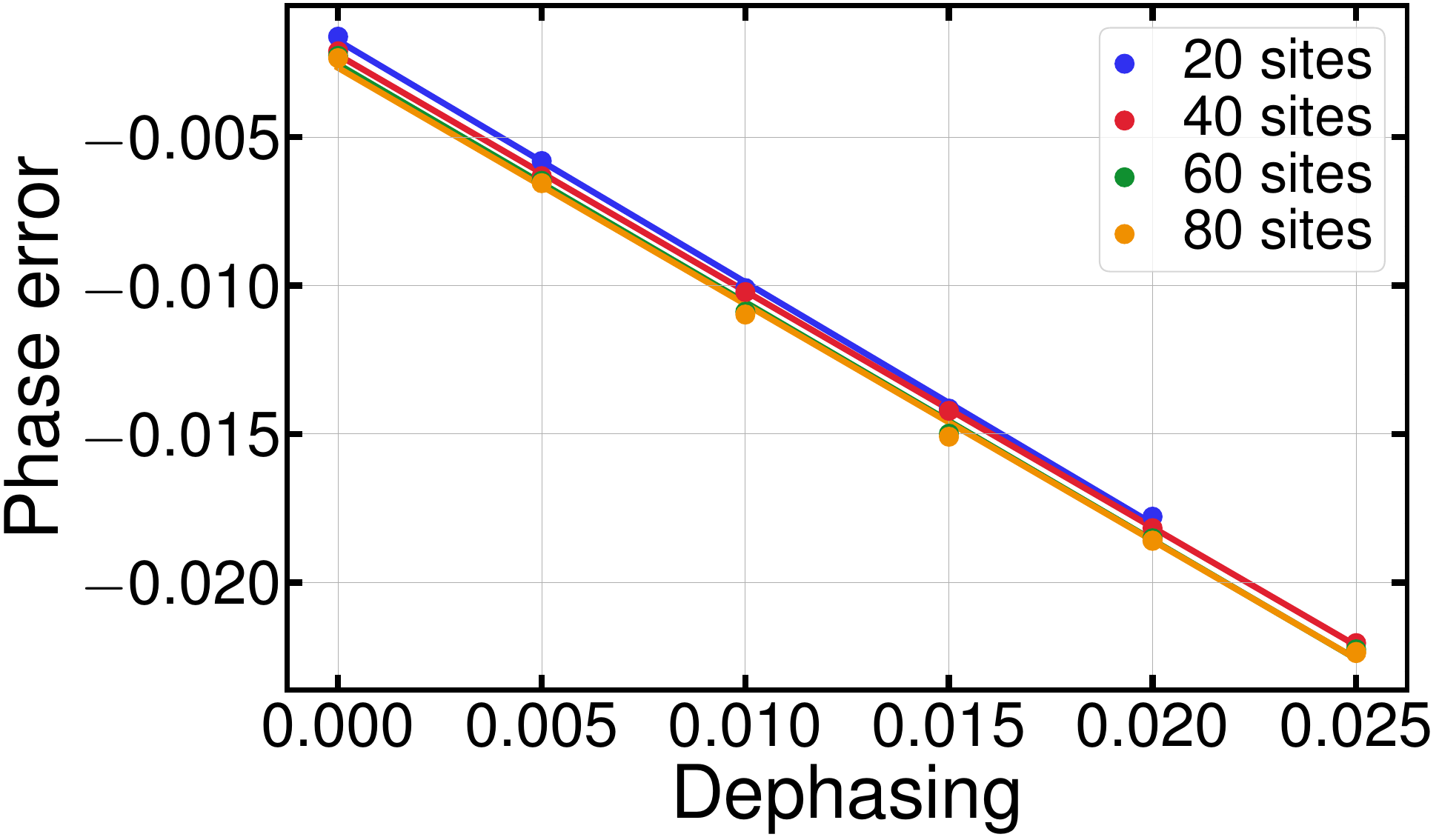}};
            \node [anchor=north west] (note) at (-0.1,0.1) {\footnotesize{\textbf{d)}}};
        \end{scope}
        \begin{scope}[xshift=0.67\columnwidth]
            \node[anchor=north west,inner sep=0] (image_c) at (0,0)
            {\includegraphics[width=0.63\columnwidth]{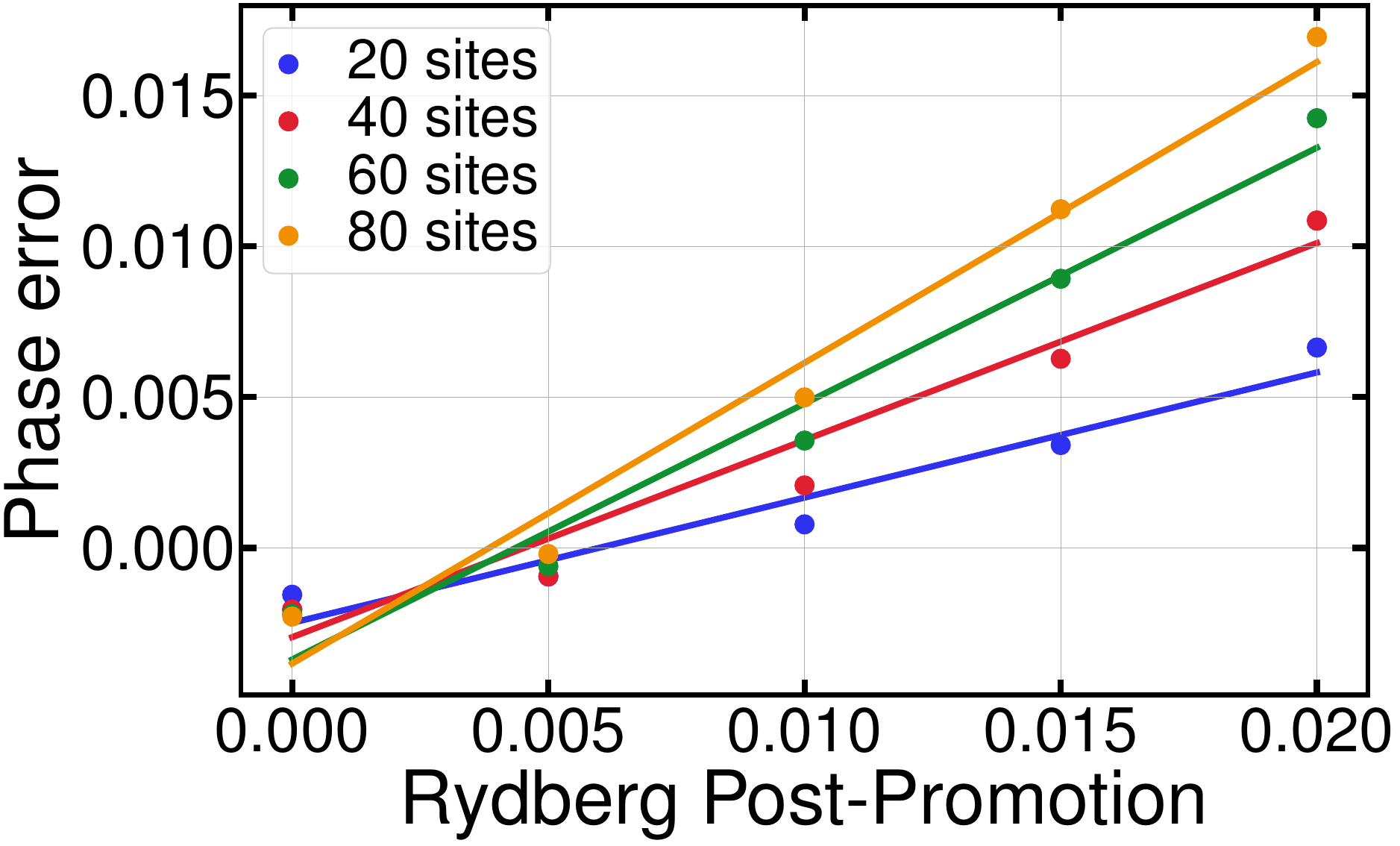}};
            \node [anchor=north west] (note) at (-0.2,0.1) {\footnotesize{\textbf{e)}}};
        \end{scope}
        \begin{scope}[xshift=1.32\columnwidth]
            \node[anchor=north west,inner sep=0] (image_c) at (0,0)
            {\includegraphics[width=0.63\columnwidth]{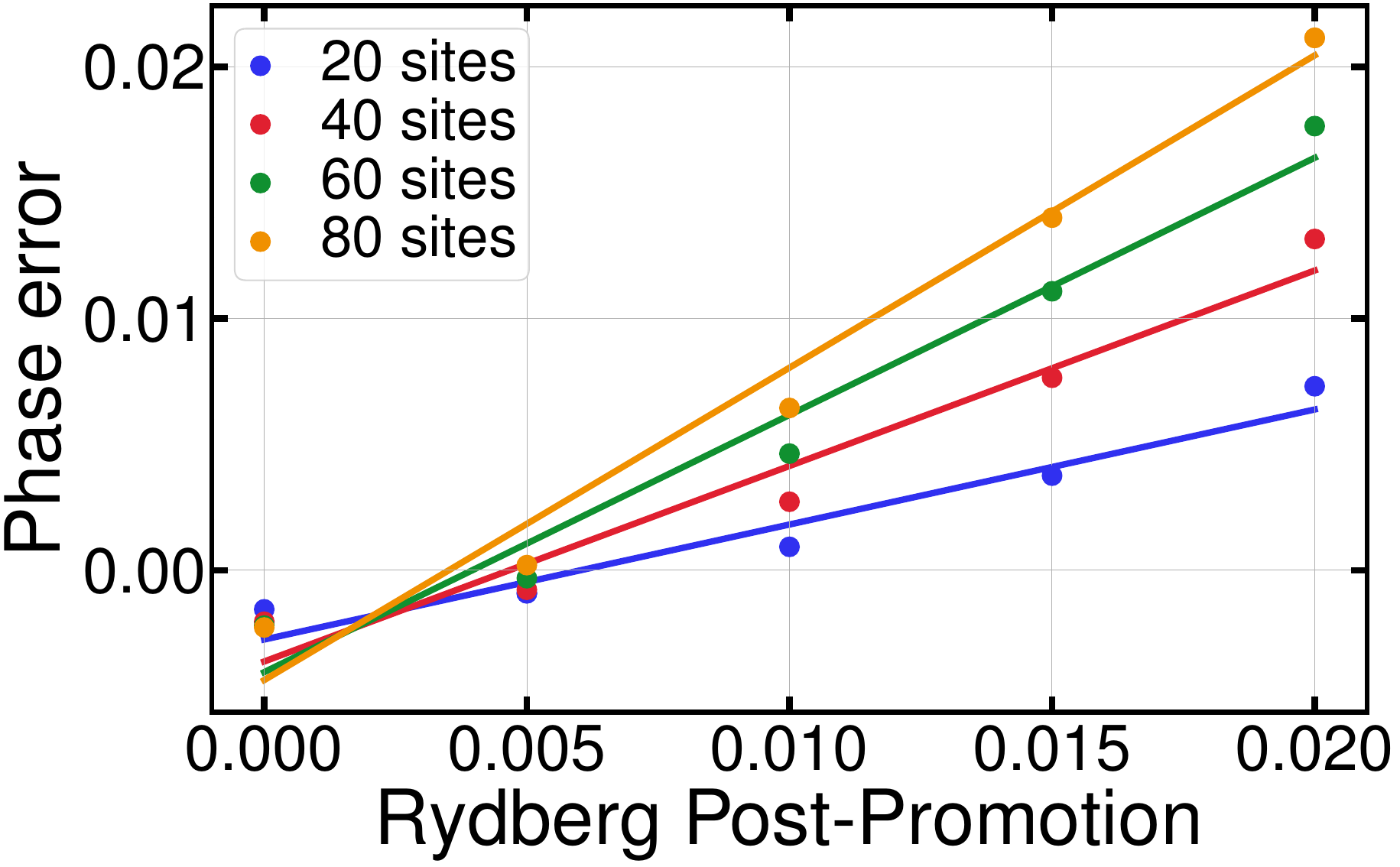}};
            \node [anchor=north west] (note) at (-0.1,0.1) {\footnotesize{\textbf{f)}}};
        \end{scope}
    \end{tikzpicture}
    \caption{Final absolute (a) or relative error (b) energy of a QAOA iteration at 60 sites and various levels of dephasing, where the circuit gates differ from the pre-optimized minimum parameters by a phase factor applied to $\alpha_3$. The vertical lines in the graph are located at the phase factor that gives the minimal energy for each circuit (the difference in the optimal phase factor over each circuit is not visible at this scale of the graph).  (c-f): Errors $p-1$ in the energy-minimizing phase factor $p$ of $\alpha_3$ estimated by noisy gates, over the relevant noise parameter and system size. The noise parameters are dissipation (c), dephasing (d), and Rydberg Post-Promotion without crosstalk (e) and with crosstalk (f). }
    \label{fig:final_energy_over_params}
\end{figure*}
\setlength{\tabcolsep}{10pt}
\begin{table}[]
    \centering
    \begin{tabular}{|c c c c c|} \hline
        \thead{$N$} & \thead{$\chi_{diss}$} & \thead{$\chi_{deph}$} & \thead{$\underset{\text{no x-talk}}{\chi_{PP}}$} & \thead{$\underset{\text{with x-talk}}{\chi_{PP}}$} \\ \hline
        20 & 4.963$\times 10^{-3}$ & -0.8133 & 0.4152 & 0.4569 \\ \hline
        40 & 4.910$\times 10^{-3}$ & -0.7955 & 0.6536 & 0.7768 \\ \hline
        60 & 6.594$\times 10^{-3}$ & -0.8008 & 0.8487 & 1.022 \\ \hline
        80 & 6.926$\times 10^{-3}$ & -0.7959 & 0.9976 & 1.241 \\ \hline
    \end{tabular}
    \caption{Error in the $\alpha_3$ parameter that minimizes the energy over noise for different system sizes. Here $\chi_i$ is the slope of the parameter error $p$ over noise source $\gamma_i$, or the coherent error $\delta_{PP}$ in the case of post-promotion.}
    \label{tab:param_slopes}
\end{table}

In addition to measuring the robustness of the noisy neutral atom device with respect to the energy, we would also like to understand if the parameters $\alpha_k,\beta_k$, one would find during optimization of the QAOA process are robust to noise. 
To study this, we start with a circuit that is near-optimized, by fixing all $\alpha_i$ and $\beta_i$ parameters at the pre-optimized values except for one $\alpha_3$, which we vary by a multiplicative phase factor $p$, and then measure the circuit energy over that phase factor (Fig.~\ref{fig:final_energy_over_params}a-b). As the noise factor $\gamma_i$ for source $i$ is increased, the optimal phase factor should vary according to some function $p(\gamma_i)$. In our observations, this function is linear, $p(\gamma_i) \approx \chi_i \gamma_i + p_0$. We can measure the overall degree $\chi_i$ to which this phase factor changes. 

The dissipation of the system changes the optimal value of $\alpha_3$ to a much lesser degree than it changes the final energy. For example, at a dephasing of 0.01 with 60 qubits, the optimal $\alpha_3$ parameter decreases by a factor of $8.6\times 10^{-3}$; this error in $\alpha_3$ would affect the final energy per site by $9.8\times 10^{-5}$. The effect on the energy from the changed parameter is much smaller than the energy error measured due to dephasing which, at this level of dephasing, is at 0.18. Therefore, the energy errors induced by selecting an improper optimization parameter are not the main source of inaccuracy. 

This suggests that if we only want to find the optimal parameters of this model, we do not need a very clean system. Therefore, we can consider a protocol where we use a cheap, noisy system to determine the optimal parameters of the circuit, then use a cleaner, more expensive circuit to measure the expectation value of operators over the optimized wavefunction. Provided that the higher noise does not reduce the qubit population of the density matrix to the extent that most circuits are immediately rejected, this can save time on the QAOA optimization process.

There is no discernible dependence of the optimization error $\frac{\dd p}{\dd \gamma}$ with system size for incoherent noise  types or environmental effects encoded by the LME (Fig.~\ref{fig:final_energy_over_params}c-f, Table ~\ref{tab:param_slopes}). For Rydberg post-promotion, however, the slope increases with system size without crosstalk and, unlike the relative energy measurements, becomes even worse with crosstalk. Therefore, larger systems will be proportionally more difficult to optimize parameters over in the presence of this error.

\setlength{\tabcolsep}{8pt}
\begin{table}[]
    \centering
    \begin{tabular}{|c|c c|c c|} \hline
        \thead{$k$} & \thead{$\alpha_k^p$} & \thead{$\beta_k^p$} & \thead{$\alpha_k^r$} & \thead{$\beta_k^r$} \\ \hline
        1 & 0.1759028 & 0.8370010 & 0.4481228 & 0.9891198 \\ \hline
        2 & 0.1924228 & 0.7371920 & 0.9514315 & 0.4154509 \\ \hline
        3 & 0.4384482 & 0.7426397 & 0.3459991 & 0.6669658 \\ \hline
        4 & 0.4252176 & 0.7838991 & 0.2498503 & 0.4133823 \\ \hline
        5 & 0.3655380 & 0.6881939 & 0.6141819 & 0.8871989 \\ \hline
        6 & 0.4713333 & 0.6609786 & 0.7105851 & 0.7540628 \\ \hline
        7 & 0.4288353 & 0.4869830 & 0.9924587 & 0.4864159 \\ \hline
        8 & 0.2536605 & 0.0983591 & 0.2900677 & 0.7650869 \\ \hline
    \end{tabular}
    \caption{Perturbed time evolution parameters $\alpha_k^p, \beta_k^p$ and fully random time evolution parameters $\alpha_k^r, \beta_k^r$ from Eq.~\eqref{eq:tfim_model} for a 10-site, 8-layer TFIM. }
    \label{tab:tfim_params_alternate}
\end{table}
\begin{figure}
    \centering
    \begin{tikzpicture}
        \begin{scope}
            \node[anchor=north west,inner sep=0] (image_a) at (0,0)
            {\includegraphics[width=0.46\columnwidth]{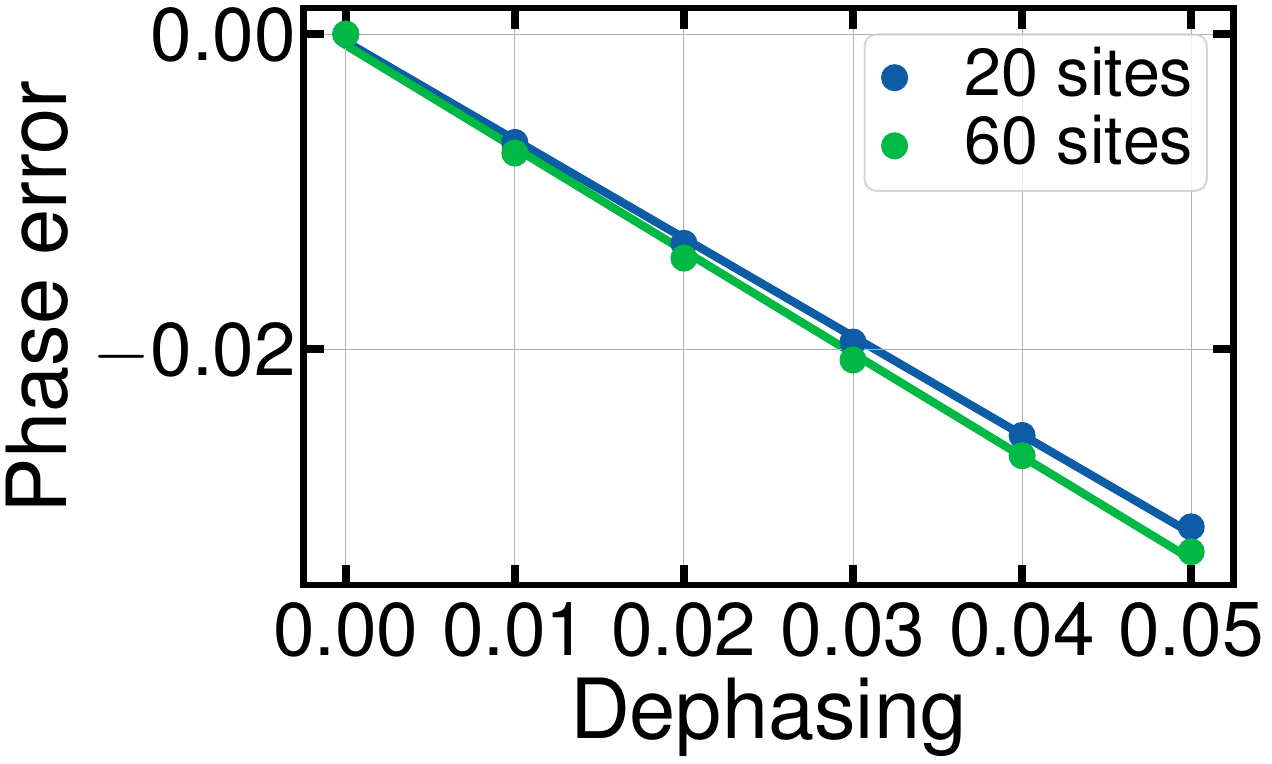}};
            \node [anchor=north west] (note) at (-0.4,0.35) {\footnotesize{\textbf{a)}}};
        \end{scope}
        
        \begin{scope}[xshift=0.5\columnwidth]
            \node[anchor=north west,inner sep=0] (image_b) at (0,0)
            {\includegraphics[width=0.46\columnwidth]{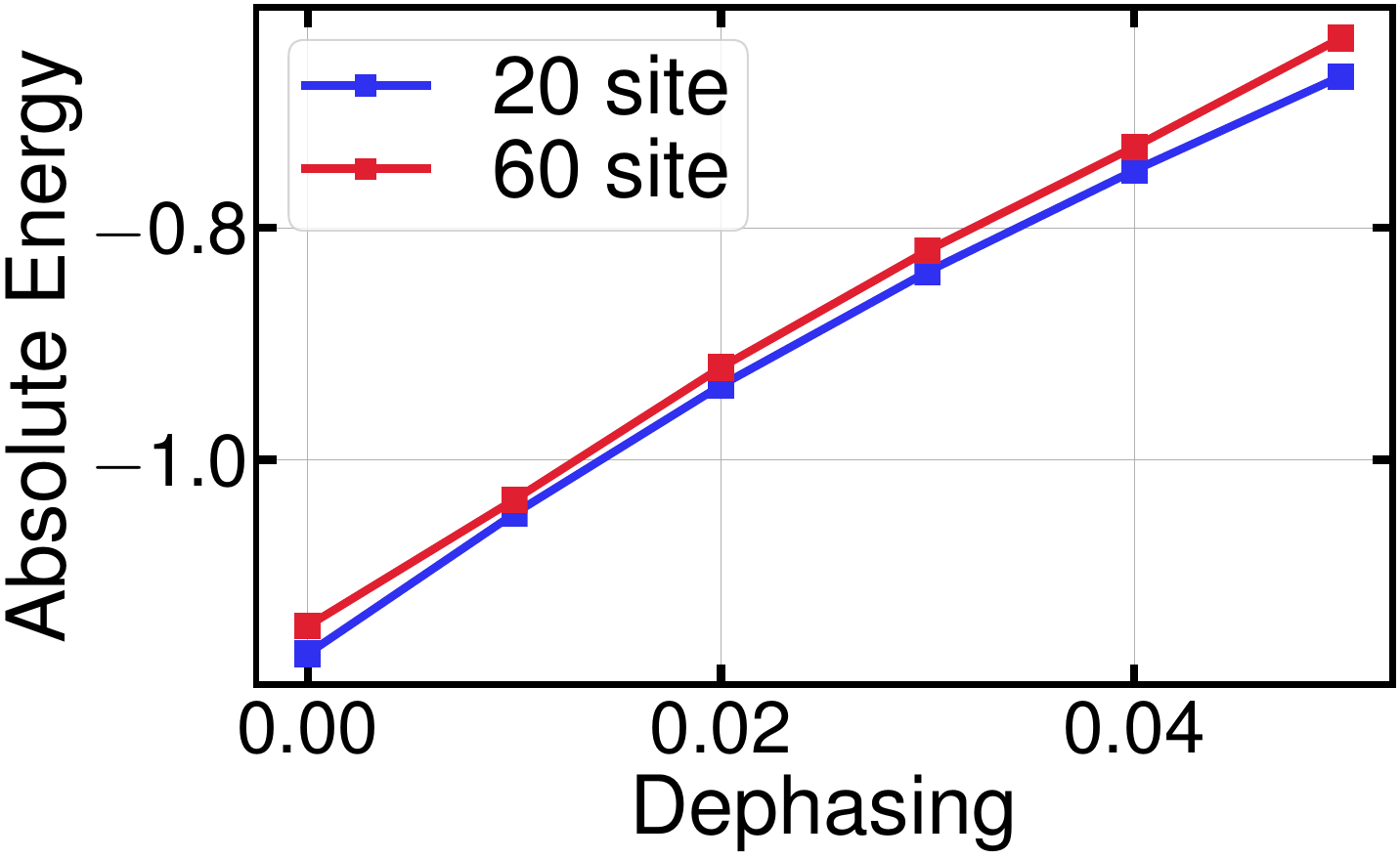}};
            \node [anchor=north west] (note) at (-0.33,0.35) {\footnotesize{\textbf{b)}}};
        \end{scope}
    \end{tikzpicture}
    \vspace*{0.01\textwidth}
    \begin{tikzpicture}
        \begin{scope}[xshift=0.0\columnwidth]
            \node[anchor=north west,inner sep=0] (image_d) at (0,-0.12)
            {\includegraphics[width=0.46\columnwidth]{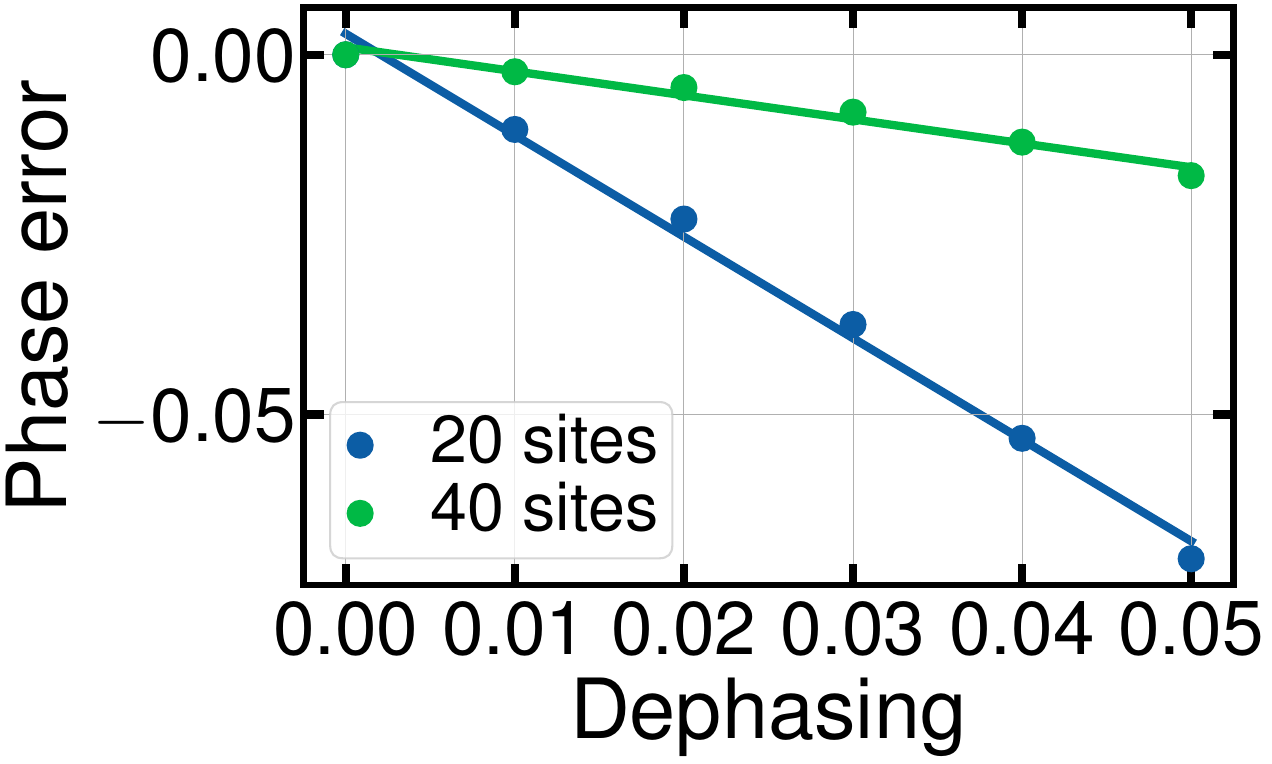}};
            \node [anchor=north west] (note) at (-0.4,0.20) {\footnotesize{\textbf{c)}}};
        \end{scope}
        \begin{scope}[xshift=0.5\columnwidth]
            \node[anchor=north west,inner sep=0] (image_d) at (0,-0.12)
            {\includegraphics[width=0.46\columnwidth]{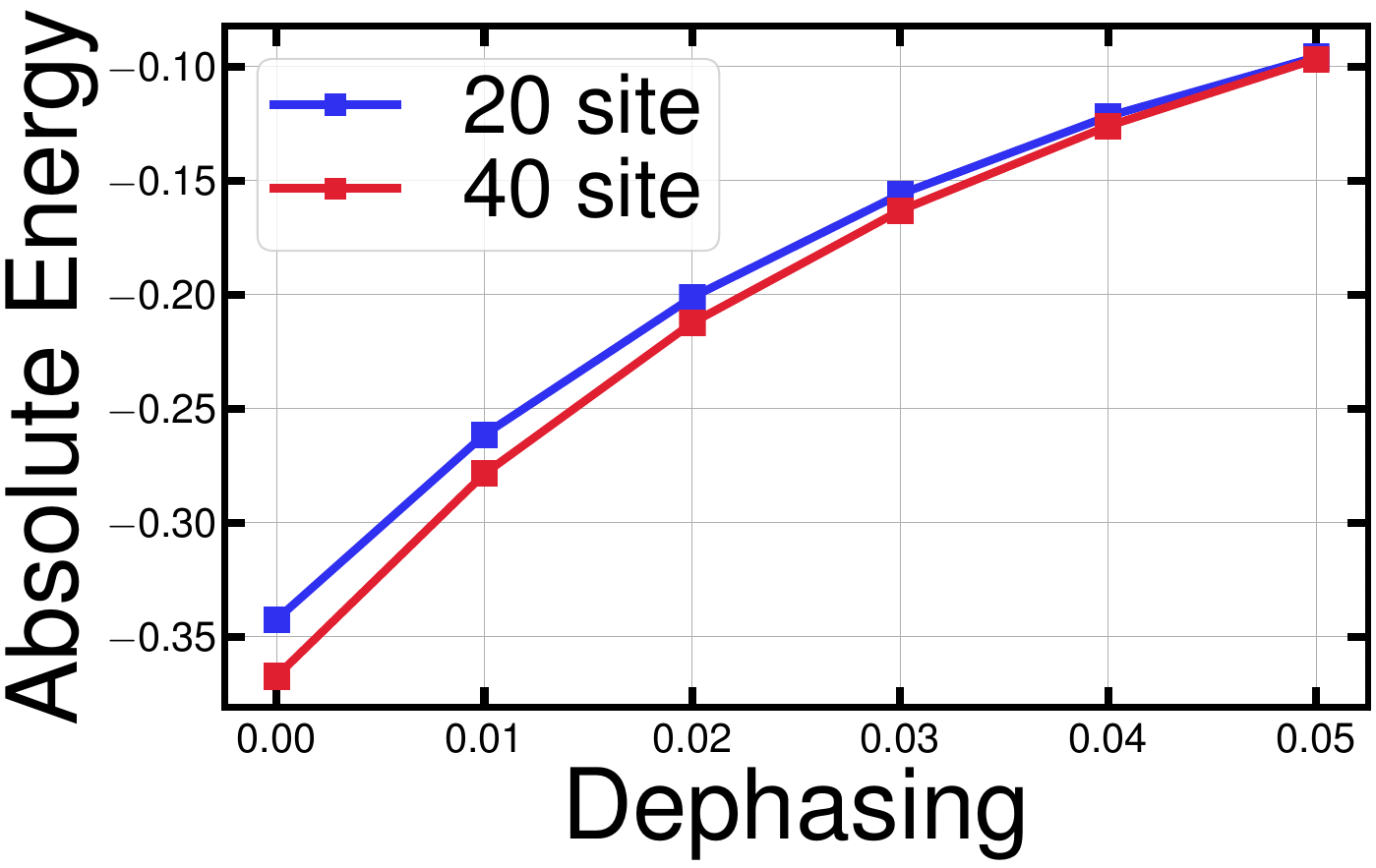}};
            \node [anchor=north west] (note) at (-0.33,0.20) {\footnotesize{\textbf{d)}}};
        \end{scope}
    \end{tikzpicture}
    \caption{(a,b): Minimum $\alpha_3$ phase errors (a) and absolute energy (b) of a QAOA iteration given perturbed parameters $\alpha_k^p, \beta_k^p$. (c,d): Minimum phase errors (c) and absolute energy (d) of a QAOA iteration given completely random parameters $\alpha_k^r, \beta_k^r$.}
    \label{fig:random_param_optimization}
\end{figure}

\subsubsection{QAOA with Random Parameters}
Besides testing QAOA from a near-optimized parameter position, we also measured the effect of noise and system size on optimized parameters selected by the circuit when the parameters are not optimized. Specifically, we considered starting parameters $\alpha_k^p, \beta_k^p$ which were perturbed from the pre-optimized parameters by a random value in the range $[-0.1, 0.1]$, as well as parameters $\alpha_k^r, \beta_k^r$ which were completely random values between 0 and 1 (Table~\ref{tab:tfim_params_alternate}). With both sets of parameters, we chose the dephasing as the noise parameter to vary, due to the relatively minor effects of Rydberg atom dissipation on the resulting energy.

For the perturbed parameters, the errors in the minimum $\alpha_3$ phase of the optimization over dephasing are given in Fig.~\ref{fig:random_param_optimization}a. There does not appear to be any difference in the change of this phase error over dephasing between the 20 site and 60 site systems. This is despite the perturbed parameters being different enough from the pre-optimized parameters to account for an absolute energy difference of around 0.4 per site, as seen in Fig.~\ref{fig:random_param_optimization}b. For the fully randomized parameters, given in Fig.~\ref{fig:random_param_optimization}c, the phase error does change its relationship, but in the negative, with the 40 site system accruing \textit{less} phase error as the dephasing is increased. Therefore, it does not appear that the 40 site system would produce a more inaccurate optimization compared to the 20 site system under large dephasing.

\section{Outlook}

We have constructed a tensor-network based pulse-level simulation of large-scale, one-dimensional neutral atom arrays, using vectorized Matrix Product Operators (MPOs) to represent the density matrix and two-site gate created by integrating a Linblad master equation including Rydberg blockade.
We have developed a new algorithmic approach, Purity-Preserving Truncation (PPT), to help maintain the physicality of the density matrix as a quantum circuit is acted on it. We have benchmarked the PPT and found that it is more efficient than Matrix Product Density Operators (MPDOs) while having only a minimal number of unphysical (negative)  eigenvalues in the resulting density matrix.  In practice, because we can go to larger bond dimensions than an MPDO at similar computational cost, we are closest to the ground truth using PPT. We proceeded to then use this machinery to simulate the pulse-level dynamics of open quantum circuits on a neutral atom array for the quantum approximate optimization algorithm (QAOA) on a transverse field Ising model.  We find that at fixed depth there is little to no dependence on the system size of the accuracy of the circuit under noisy errors, although there is a non-trivial increase in the failure rate under such errors. Under coherent errors, there is a possible decrease in accuracy (specifically final energy and the correct optimization parameters) as the number of qubits is increased. 

If we extrapolate our findings to systems of arbitrary size, the failure rate of a QAOA iteration under a dissipation of $0.001 T^{-1}$ such as in \cite{Saffman2020} will yield a qubit trace of $\tr_q(\rho) = e^{-1.556\times 10^{-4} N}$. At $N=200$ this would result in a 3\% trace error. The expected error in relative energy is extremely minor at only $-3.3\times 10^{-5}$, and a near-optimized system would find a parameter error of order $10^{-5}$. Dephasing is a much more significant error - under a dephasing of $0.001T^{-1}$, the relative energy error becomes $0.011$, and the parameter error becomes order $10^{-3}$. This is possibly due to the limited amount of time in which Rydberg atom populations are large enough that dissipation is allowed to act. This suggests that it would be possible to run QAOA iterations under this noise model with relatively low error and failure rates.

Given the harsher scaling of coherent errors, an interesting open question is whether coherent errors are generically the dominant error source in open quantum devices.  In fact, even for noise sources such as dissipation there is a coherent and incoherent piece and it is plausible that even in dissipation, the coherent piece is driving errors (see Appendix~\ref{app:coherent_parts} for a discussion of this).

While we mainly focused on parameters that were close to the optimal value, it remains to determine how the QAOA behaves under noise at any stage of optimization, including completely random starting parameters and semi-optimized parameters. This would mainly be useful for problems where even the roughest optimization is classically unfeasible, which does not apply to the current TFIM.

The ability to determine the effect of errors on realistic quantum devices at scale with respect to interesting algorithms is important to making progress in the field of quantum computers.  We have demonstrated a particular example of this in neutral atoms systems and believe our new techniques are an important step forward for future applications. 

\section{Acknowledgements}
We acknowledge useful discussions with Mark Saffman, Martin Suchara and Xiaoyu Jiang. This material is based upon work supported by the U.S. Department of Energy, Office of Science, National Quantum Information Science Research Centers. Work performed at the Center for Nanoscale Materials, a U.S. Department of Energy Office of Science User Facility, was supported by the U.S. DOE, Office of Basic Energy Sciences, under Contract No. DE-AC02-06CH11357. We gratefully acknowledge the computing resources provided on Bebop, a high-performance computing cluster operated by the Laboratory Computing Resource Center at Argonne National Laboratory.  

\bibliographystyle{unsrt}

\begin{appendices}

\section{The Four-Tensor Split}\label{app:4tsplit}

\begin{figure}
    \centering
    \begin{tikzpicture}
        \begin{scope}
            \node[anchor=north west,inner sep=0] (image_a) at (0,0)
            {\includegraphics[width=0.8\columnwidth]{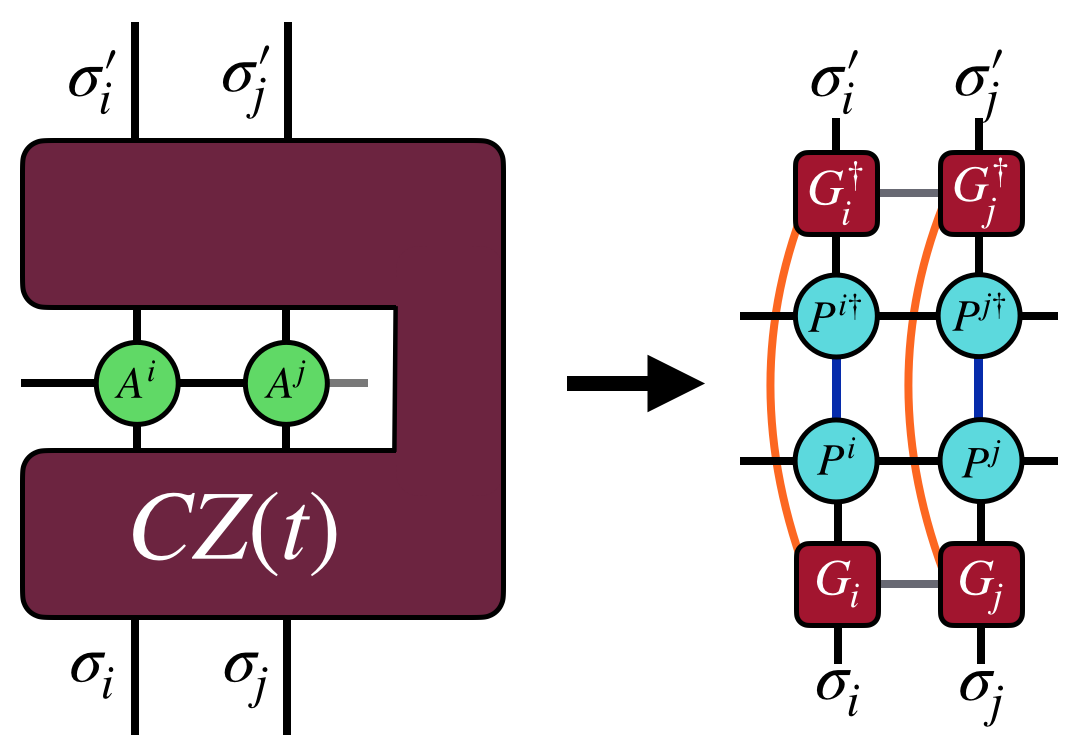}};
        \end{scope}
    \end{tikzpicture}
    \caption{A two-site operator acting on a density matrix in MPDO form must be split along both the bond and inner dimension before being applied to the MPDO, creating a four-tensor split.}
    \label{fig:four_tensor_split}
\end{figure}

An operator with both entangling and mixing terms needs to be split along both directions before being applied to the MPDO. For a two-site operator $\hat{O}_{i_f, j_f, i_b, j_b}^{i_f', j_f', i_b', j_b'}$, where $i_f, j_f, i_b, j_b$ are the forward and backward local Hilbert space indices at sites $i$ and $j$, we must identify two bond dimensions $l,m$ and two Kraus dimensions $k_i, k_j$, with a split of the form
\begin{gather}
    \hat{O}_{i_f, j_f, i_b, j_b}^{i_f', j_f', i_b', j_b'} \rightarrow \hat{A}_{i_f, l, k_i}^{i_f'} \hat{B}_{j_f, l, k_j}^{j_f'}\hat{A}_{i_b, m, k_i}^{\dagger i_b'} \hat{B}_{j_b, m, k_j}^{\dagger j_b'}.
\end{gather} 
$\hat{A}$ and $\hat{B}$ are the operators that will be applied to sites $i$ and $j$ respectively of the MPDO, with $\hat{A}^\dagger$ and $\hat{B}^\dagger$ implicitly applied to the complex conjugated half of the MPDO (Fig.~\ref{fig:four_tensor_split}). We first separate $\hat{O}$ along the Kraus direction, creating a combined Kraus index $k$ for both sites with maximum dimension $K^2$, before splitting this index into two indices $k_i$, $k_j$ with maximum dimension $K$ each. Initially, these indices have little relation to the physical qualities of the site they are supposed to represent. As a result, there is potentially a large amount of entanglement between the sites. We use an entanglement reduction algorithm between the indices of sites $i$ and $j$, by applying unitary operators that act on $k_i$ and $k_j$. This is the same algorithm that enabled the Moses move\cite{Zaletel2020}, and can be found in Ref.~\onlinecite{Hauschild2018}. We attempt to reduce the entanglement by minimizing the second Renyi entropy across the sites, $S_2 = -\log(\tr(\rho_G^2))$, where $\rho_G$ is an effective density matrix across the gate. The term in the log is a quartic function of the unitary, so its derivative over that unitary can be easily obtained, and from that derivative we can iteratively optimize the entropy.

\section{Derivation of the PPT Formula}\label{app:ppt_derivation}

To update the truncated singular values $\Lambda_i'$ to the new singular values $\tilde{\Lambda}_i$ such that their angle with $T_i'$ is exactly $\theta$, we need to find the closest point on the cone of constant angle $\theta$ around $T_i'$. This point will lie on the plane spanned by $\Lambda_i'$ and $T_i'$, which intersects the cone at some target ray. That is, we should make $\tilde{\Lambda}_i$ some linear combination of $\Lambda_i'$ and $T_i'$.

Set a vector
\begin{gather}
    \Delta_i' = \Lambda_i' - |\Lambda'| \cos \sigma \hat{T}_i',
\end{gather} 
that is, the component of $\Lambda'$ which is orthogonal to $T'$. Up to some normalization constant, $\tilde{\Lambda}$ is the point on the ray $T' +x\Delta'$ which is at an angle $\theta$ with $T'$. That is, $x$ must satisfy
\begin{gather*}
    (\Lambda' + x\Delta') \cdot T' = \cos \theta |T'| |\Lambda' + x\Delta'|\\
    |\Lambda'| |T'| \cos \sigma = \cos \theta |T'| \sqrt{|\Lambda'|^2 + x^2 |\Delta'|^2 + 2x\left(\Lambda' \cdot \Delta'\right)}\\
    |\Lambda'| \frac{\cos \sigma}{\cos \theta} = \sqrt{|\Lambda'|^2 + x^2 |\Delta'|^2 + 2x\Lambda' \cdot \Delta'}.
\end{gather*}
Because
\begin{gather*}
    |\Delta'|^2 = |\Lambda'|^2(1+\cos^2 \sigma) -2 \cos^2 \sigma |\Lambda'|^2= |\Lambda'|^2\sin^2\sigma\\
    \Lambda' \cdot \Delta' = |\Lambda'|^2\sin^2\sigma
\end{gather*}
we have
\begin{gather*}
    |\Lambda'| \frac{\cos \sigma}{\cos \theta} = \sqrt{|\Lambda'|^2  + (x^2+2x)\left(|\Lambda'|^2\sin^2\sigma\right)}\\
    \frac{\cos^2 \sigma}{\cos^2 \theta} = 1 + (x^2+2x)\sin^2 \sigma  = (x+1)^2 \sin^2 \sigma + \cos^2 \sigma\\
    \cot^2\sigma(\sec^2\theta-1) = (x+1)^2\\
    x = \frac{\tan\theta}{\tan \sigma} - 1
\end{gather*}
where we choose the positive solution because the negative solution is on the opposite site of $T'$ from $\Lambda'$.

\section{Enforcing the Inner Dimension of an MPDO}\label{app:mpdo_rule}

\begin{figure}
    \centering
    \begin{tikzpicture}
        \begin{scope}
            \node[anchor=north west,inner sep=0] (image_a) at (0,0)
            {\includegraphics[width=0.54\columnwidth]{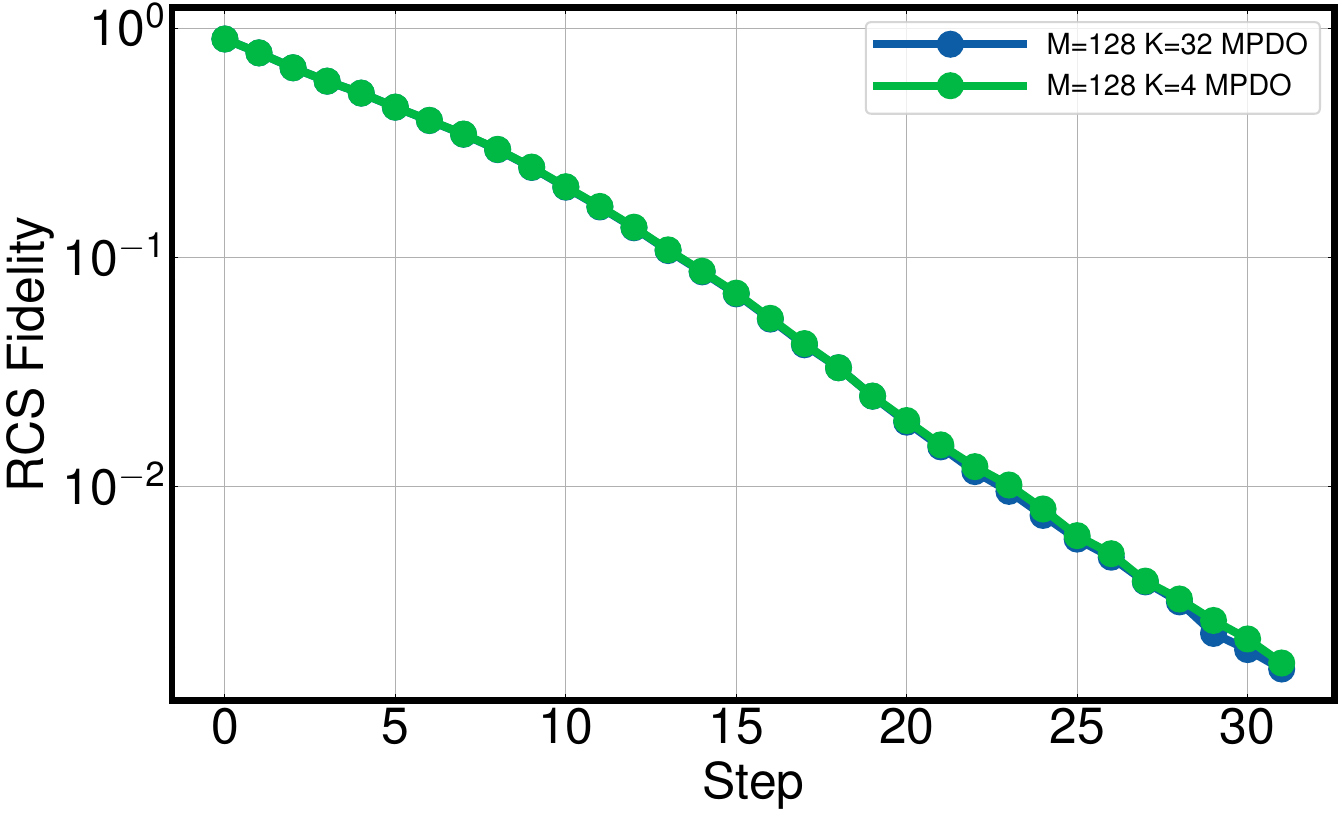}};
        \end{scope}
    \end{tikzpicture}
    \caption{12-site RCS fidelity of two MPDO's with the same bond dimension $M=128$ but different inner dimensions $K=4, 32$. The two fidelities only deviate slightly for very large time steps.}
    \label{fig:rcs_kraus_test}
\end{figure}

We have both strong numerical evidence and an intuitive suggestion that a high inner dimension is not required to simulate the MPDO. The latter is a constructive proof that any density matrix can be represented as an MPDO with low inner dimension.
Starting with the standard decomposition of a density matrix into orthogonal wavefunctions
\begin{gather}
    \rho = \sum_{i=0}^{d^N-1} p_i |\psi_i\rangle \langle \psi_i|,
\end{gather}
each wavefunction $|\psi_i\rangle$ can be represented as an MPS with tensors $A^i_j$ on site $j$. We can construct an MPS for the sum of wavefunctions
\begin{gather}
    |\Psi\rangle = \sum_{i=0}^{d^N-1} \sqrt{p_i} |\psi_i\rangle
\end{gather}
by turning each site tensor $B_j$ into a block-diagonal composition of the individual site tensors $A^i_j$. 
If we consider the projection
\begin{gather}
    |\Psi\rangle \langle \Psi| = \sum_{i,j=0}^{d^N-1} \sqrt{p_i p_j} |\psi_i\rangle \langle \psi_j|
\end{gather}
we obtain the original density matrix $\rho$ as long as we discard all cross-terms $|\psi_i\rangle \langle \psi_j|,\; i \neq j$. We can achieve this by using the inner indices at each site of the MPDO.
For each $\psi_i$, we consider the base-$d$ decomposition of $i$,
\begin{gather}
    i \rightarrow k_{i1} k_{i2} ... k_{iN}.
\end{gather}

We can then construct the positive TNS of $\rho$ with inner dimension $d$ at every site, by starting with $B_j$ and setting each block $A^i_j$ to 0 unless the $j$th digit $k_{ij}$ matches the inner index. The total contraction of this TNS therefore only keeps the terms $|\psi_i\rangle \langle \psi_j|$ where $i$ and $j$ have the exact same digits at every site, i.e. when $i=j$ exactly.

The bond dimension of such a MPDO would generally be quite large at $O(d^{2N})$. This is expected, as we had not made any assumptions about the form of the density matrix. While we do not have a general scheme for reducing the inner dimension while keeping the bond dimension relatively low, we do observe that the maximum allowed inner dimension of a MPDO can be as low as 4 without significantly changing the RCS fidelity (Fig.~\ref{fig:rcs_kraus_test}).

\section{Distributed-Memory Speedup}\label{app:distributed_memory}

\begin{figure}
    \centering
    \begin{tikzpicture}
        \begin{scope}
            \node[anchor=north west,inner sep=0] (image_a) at (0,0)
            {\includegraphics[width=0.67\columnwidth]{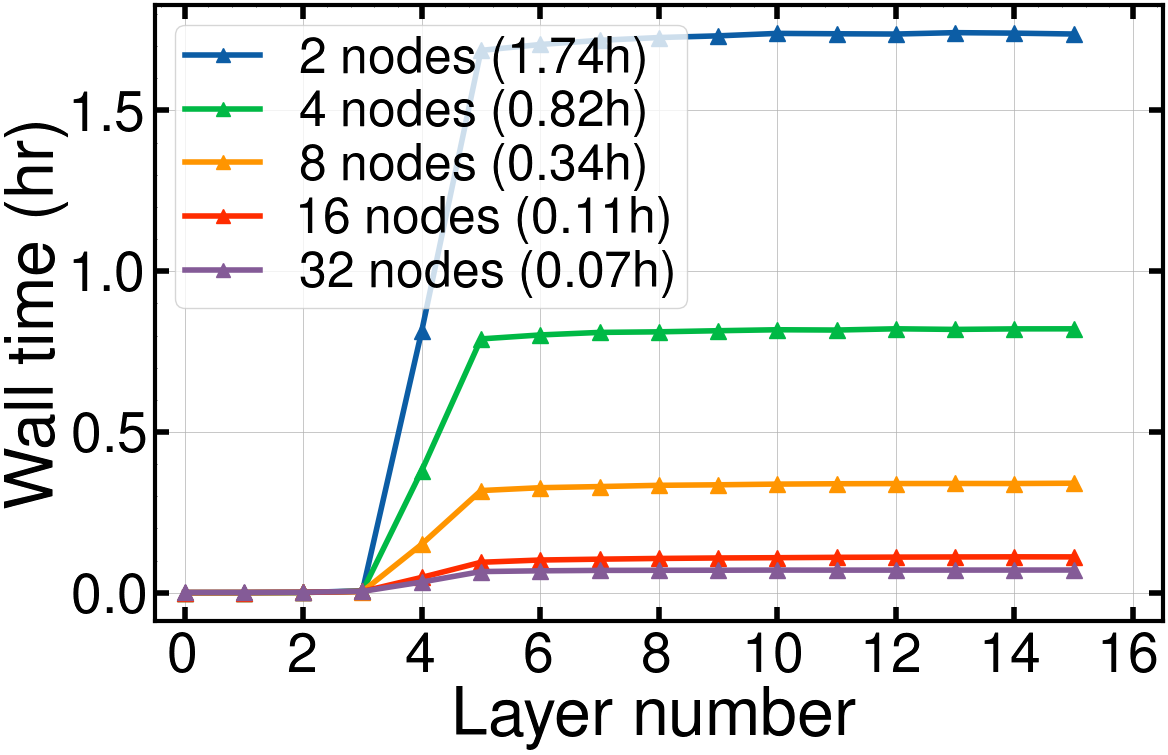}};
            \node [anchor=north west] (note) at (-0.3,-0.1) {\footnotesize{\textbf{a)}}};
        \end{scope}
        \end{tikzpicture}
        \begin{tikzpicture}
        \begin{scope}
            \node[anchor=north west,inner sep=0] (image_b) at (0,0)
            {\includegraphics[width=0.67\columnwidth]{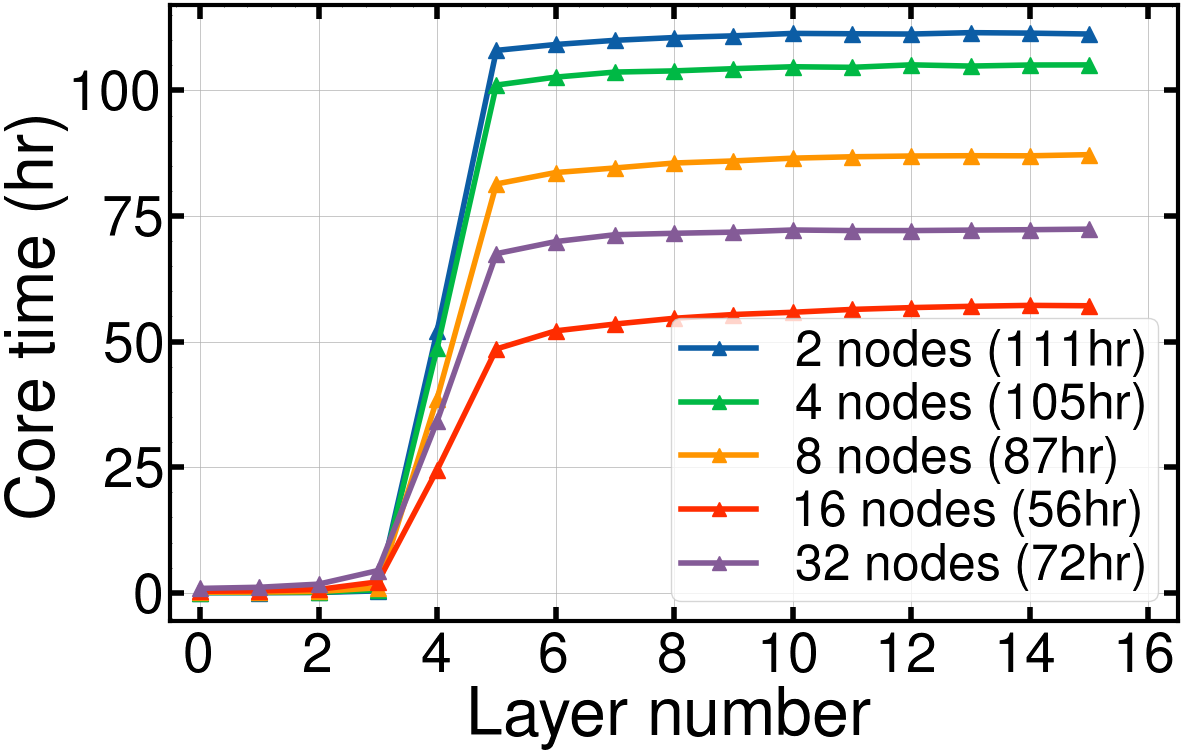}};
            \node [anchor=north west] (note) at (-0.3,-0.1) {\footnotesize{\textbf{b)}}};
        \end{scope}
    \end{tikzpicture}
    \caption{Stress test of a 32-site, 16-layer random circuit with 1024 bond dimension. (a): Wall time of each process over each layer. (b): Total cost in core-hours of each process. Due to the easier handling of large memories, the program is the most efficient at 16 nodes.}
    \label{fig:rcs_stress_test}
\end{figure}

For computationally difficult circuits like RCS\cite{Bouland2019a}, the ideal dimensions required to maintain accuracy grow exponentially with depth, resulting in increasingly costly operations. We use the tensor-tools library\cite{Levy2020} built on top of the Cyclops Tensor Framework (CTF)\cite{Solomonik2014} to reduce the time cost of these simulations by distributing the data of individual tensors over multiple nodes. 

We tested this framework by performing a 12-site, 20-layer RCS simulation on one sample with maximum bond dimension 1024 over multiple nodes (Fig.~\ref{fig:rcs_stress_test}a). As we allot more nodes to the program, the overall wall time of the operations decreases. This speedup was better than proportional at some node sizes, resulting in an overall reduction of total core hours as well as wall time (Fig.~\ref{fig:rcs_stress_test}b). This may be due to the distributed-memory nature of the program reducing the size of the tensors each node has to manipulate, removing costs associated with tensors that are too large for the node memory to properly handle. On the other hand, the performance worsens at 32 nodes because the increased cost of communicating between more nodes outweights the benefits of working with smaller individual tensors. 

\section{Coherent Components of Incoherent Errors}\label{app:coherent_parts}

An interesting open question is to better understand 
whether the coherent components of incoherent errors like Rydberg atom dissipation are the main way they reduce the accuracy of the TFIM. Specifically, a pulse under Rydberg atom dissipation $\gamma_{diss}$ experiences single and double site population depletion
\begin{gather}\small
    \begin{pmatrix}
    \rho_{|00\rangle\langle00|} & \rho_{|00\rangle\langle01|} & \rho_{|01\rangle\langle00|} & \rho_{|01\rangle\langle01|}\\
    \rho_{|00\rangle\langle10|} & \rho_{|00\rangle\langle11|} & \rho_{|01\rangle\langle10|} & \rho_{|01\rangle\langle11|}\\
    \rho_{|10\rangle\langle00|} & \rho_{|10\rangle\langle01|} & \rho_{|11\rangle\langle00|} & \rho_{|11\rangle\langle01|}\\
    \rho_{|10\rangle\langle10|} & \rho_{|10\rangle\langle11|} & \rho_{|11\rangle\langle10|} & \rho_{|11\rangle\langle11|}
    \end{pmatrix} \quad \rightarrow \n \small
    \begin{pmatrix}
    \rho_{|00\rangle\langle00|} & -\alpha  \rho_{|00\rangle\langle01|} & -\alpha  \rho_{|01\rangle\langle00|} & -\beta^*  \rho_{|01\rangle\langle01|}\\
    -\alpha  \rho_{|00\rangle\langle10|} &  \alpha^2 \rho_{|00\rangle\langle11|} & \alpha^2  \rho_{|01\rangle\langle10|} & \alpha \beta^*  \rho_{|01\rangle\langle11|}\\
    -\alpha  \rho_{|10\rangle\langle00|} & -\alpha^2  \rho_{|10\rangle\langle01|} & \alpha^2 \rho_{|11\rangle\langle00|} & \alpha \beta^*  \rho_{|11\rangle\langle01|}\\
    -\beta  \rho_{|10\rangle\langle10|} & \alpha \beta  \rho_{|10\rangle\langle11|} & \alpha \beta  \rho_{|11\rangle\langle10|} & |\beta|^2 \rho_{|11\rangle\langle11|}
    \end{pmatrix}
\end{gather}\normalsize
where
\begin{gather}
    \alpha = e^{-0.012291\gamma_{diss}}\\
    \beta = e^{-0.0185454\gamma_{diss}}.
\end{gather}
This in turn causes the Ising phase gates $e^{iS_z^i S_z^j}$ to experience extra phase shifts of the form
\begin{gather}
    e^{i\theta S_z^i S_z^j} \rightarrow e^{i\theta (S_z^i S_z^j + \Delta \theta_1 (S_z^i + S_z^j)+\Delta \theta_2 S_z^i S_z^j)}\\
    \Delta \theta_1 = \left(\frac{1-\alpha}{1+\alpha}\right)^2\qquad
    \Delta \theta_2 = \left(\frac{\alpha-\beta}{\alpha+\beta}\right)^2.
\end{gather}
Which suggests that we can mimic this part of the error by modifying these phases directly, without implementing any noisy pulse-level simulations in the first place. If the energy and parameter profiles of this system match the respective system with Rydberg atom dissipation, we can deduce that the main source of error in the dissipative system comes from this coherent component. We can also separate this error from the rest of the dissipation by reversing all the phase shifts afterwards, creating a channel which only contains the incoherent part of the dissipation error.

\end{appendices}

\end{document}